\def\VersionLong{}
\def\VersionProofAppendix{}
\ifdefined\VersionLong%
	\newcommand{\longVersion}[1]{#1}
	\newcommand{\shortVersion}[1]{}
\else
	\newcommand{\longVersion}[1]{}
	\newcommand{\shortVersion}[1]{#1}
\fi
\newcommand{\journalVersion}[1]{#1}
\newcommand{\versionProofOut}[1]{\ifdefined\VersionProofAppendix{\color{black}#1}\fi}
\newcommand{\versionProofIn}[1]{\ifdefined\VersionProofInPaper{\color{green!50!black}#1}\fi}
\documentclass[a4paper,10pt]{lmcs} %
\pdfoutput=1

\usepackage{lastpage}
\lmcsdoi{17}{2}{13}
\lmcsheading{}{\pageref{LastPage}}{}{}%
{Oct.~01,~2019}{May~10,~2021}{}

\usepackage[utf8]{inputenc}
\usepackage[english]{babel}

\usepackage{mathpartir}

\usepackage{wrapfig}

\usepackage{caption}
\usepackage{framed}
\usepackage{subcaption}
\newenvironment{ienumeration}
	{\ifdefined\VersionLong\begin{enumerate}} %
	{\ifdefined\VersionLong\end{enumerate}} %

	{\ifdefined\VersionLong\end{enumerate}} %

\usepackage{amsmath} %
\usepackage{amssymb} %
\usepackage{stmaryrd}
\usepackage{amsfonts}
\usepackage{algorithmic}

\usepackage{tikz}
\usetikzlibrary{arrows,positioning,automata,shadows}

\tikzstyle{every node}=[initial text=]
\tikzstyle{location}=[rectangle, rounded corners, minimum size=12pt, draw=black, fill=blue!10, inner sep=2pt]
\tikzstyle{invariant}=[draw=black, dotted, inner sep=1pt] %
\tikzstyle{final}=[double]

	\definecolor{coloract}{rgb}{0.50, 0.70, 0.30}
	\definecolor{colorclock}{rgb}{0.4, 0.4, 1}
	\definecolor{colordisc}{rgb}{1, 0, 1}
	\definecolor{colorloc}{rgb}{0.4, 0.4, 0.65}
	\definecolor{colorparam}{rgb}{1, 0.6, 0.0}

\newcommand{\styleact}[1]{\ensuremath{\textcolor{coloract}{\mathrm{#1}}}}
\newcommand{\styleclock}[1]{\ensuremath{\textcolor{colorclock}{#1}}}

\newcommand{\styleloc}[1]{\ensuremath{\mathrm{#1}}}
\newcommand{\styleparam}[1]{\ensuremath{\textcolor{colorparam}{#1}}}

\newcommand{\paraminline}[1]{\ensuremath{\colorbox{yellow}{\ensuremath{#1}}}}

\ifdefined\VersionWithComments%
	\usepackage{draftwatermark}
	\SetWatermarkText{draft}
	\SetWatermarkScale{15}
	\SetWatermarkColor[gray]{0.9}
\fi
\usepackage{scalerel}
\usetikzlibrary{svg.path}
\makeatletter
\def\orcidID#1{\smash{\href{https://orcid.org/#1}{\protect\raisebox{-1.25pt}{\protect\includegraphics{orcid}}}}}
\makeatother

\newlength{\ORCIDidheight}
\newlength{\ORCIDidunit}
\definecolor{ORCIDgreen}{HTML}{A6CE39}

\definecolor{darkgreen}{rgb}{0, 0.5, 0}
\definecolor{darkpurple}{rgb}{0.7, 0, 0.7}
\definecolor{darkblue}{rgb}{0, 0, 0.7}

\usepackage{hyperref}

\newenvironment{preuve}
	{\begin{proof}}
	{\end{proof} } %

\newcommand{\refClaim}[1]{\cref{#1}}

\usepackage{multirow}
\usepackage{tikz}
\usetikzlibrary{automata, arrows, positioning}

\tikzstyle{location}=[minimum size=12pt, circle, fill=blue!20, draw=black, text=black, inner sep=1.5pt, initial text={}] %
\tikzstyle{invariant}=[yshift=3, rectangle, draw=black, fill=white, text=black, inner sep=1pt]

\tikzstyle{decidable}=[green!80!black,-]
\tikzstyle{undecidable}=[red!80!black, dashed,-]
\tikzstyle{contribution}=[very thick]
\tikzstyle{legende}=[draw=none, font=\scriptsize]
\tikzstyle{unknown}=[black,dotted,thick]

\usepackage[ruled,vlined,linesnumbered]{algorithm2e}

\usepackage[capitalise,english,nameinlink]{cleveref}
\usepackage{thmtools}
\makeatletter
\let\c@thm\undefined%
\makeatother

\theoremstyle{plain}
\declaretheorem[name=Theorem,numberwithin=section]{thm}
\newtheorem{clm}[thm]{Claim}
\newtheorem{lem}[thm]{Lemma}
\newtheorem{cor}[thm]{Corollary}
\newtheorem{prop}[thm]{Proposition}
\newtheorem{fact}[thm]{Fact}

\theoremstyle{thmC}
\newtheorem{lemC}[thm]{Lemma}

\theoremstyle{definition}
\newtheorem{exa}[thm]{Example}
\newtheorem{defi}[thm]{Definition}

\crefname{prop}{Proposition}{Propositions}
\crefname{thm}{Theorem}{Theorems}
\crefname{clm}{Claim}{Claims}
\crefname{lem}{Lemma}{Lemmas}
\crefname{lemC}{Lemma}{Lemmas}
\crefname{cor}{Corollary}{Corollaries}
\crefname{exa}{Example}{Examples}
\crefname{defi}{Definition}{Definitions}
\crefname{fact}{Fact}{Facts}

\crefname{section}{Section}{Sections}
\crefname{figure}{Figure}{Figures}
\newcommand{\crefDef}[1]{\cref{#1}}
\newcommand{\crefLemma}[1]{\texorpdfstring{\cref{#1}}{Lemma~\ref{#1}}}
\newcommand{\crefProp}[1]{\texorpdfstring{\cref{#1}}{Proposition~\ref{#1}}}

\newcommand{\refc}[1]{\ref{#1}}

\newcommand{\init}{_0}

\newcommand{\A}{\ensuremath{\mathcal{A}}}

\newcommand{\Actions}{\Sigma}
\newcommand{\action}{\ensuremath{a}}

\newcommand{\C}{C}

\newcommand{\Clock}{\mathbb{X}} %
\newcommand{\ClockCard}{H} %
\newcommand{\clock}{x} %
\newcommand{\clockk}{y} %
\newcommand{\clockval}{w} %
\newcommand{\ClocksZero}{\vec{0}}
\newcommand{\compOp}{\bowtie}
\newcommand{\compOpLeq}{\triangleleft}
\newcommand{\CONSTMAX}{K}
\newcommand{\D}{D}
\newcommand{\E}{E} %
\newcommand{\OPDBMs}{\ensuremath{\param\textsf{--}\mathcal{PDBM}_{\blacksquare}(\paramR)}}
\newcommand{\edge}{e}
\newcommand{\Edges}{\zeta}
\newcommand{\longuefleche}[1]{\stackrel{#1}{\longrightarrow}}
\newcommand{\longueflecheRel}[1]{\stackrel{#1}{\mapsto}}

\newcommand{\flecheRel}{{\rightarrow}}

\newcommand{\grandn}{\mathbb{N}}
\newcommand{\grandq}{\mathbb{Q}}
\newcommand{\grandqplus}{\grandq_{+}} %
\newcommand{\grandr}{\mathbb{R}}
\newcommand{\grandrplus}{\grandr_{+}} %
\newcommand{\grandz}{\mathbb{Z}}
\newcommand{\guard}{g}

\newcommand{\loc}{\ell} %
\newcommand{\locinit}{\loc\init}
\newcommand{\Loc}{L} %

\newcommand{\Param}{\mathbb{P}} %
\newcommand{\param}{p} %
\newcommand{\ptotal}{\ensuremath{\mathsf{max}}} 
\newcommand{\ParamCard}{M} %
\newcommand{\partieFrac}{\mathit{frac}} %

\newcommand{\pval}{v} %

\newcommand{\paramregions}{\mathcal{R}_{p}}

\newcommand{\styleTCTL}[1]{\ensuremath{\mathsf{#1}}}
\newcommand{\EFsynth}{\textsf{EFsynth}}

\newcommand{\clockR}{\ensuremath{\Pregion_{c}}}
\newcommand{\paramR}{\ensuremath{\Pregion_{p}}}
\newcommand{\Pregion}{R}
\newcommand{\CPr}{\ensuremath{\param\textsf{--}\mathcal{PDBM}_{\odot}(\paramR)}}
\newcommand{\PDBMRp}{\ensuremath{\param\textsf{--}\mathcal{PDBM}(\paramR)}}
\newcommand{\pointPDBM}{\ensuremath{\mathsf{point}}--\mPDBM{}}
\newcommand{\oPDBM}{\ensuremath{\mathsf{open}}--\mPDBM{}}
\newcommand{\pointPDBMs}{\ensuremath{\mathsf{point}}--\mPDBM{}s}
\newcommand{\oPDBMs}{\ensuremath{\mathsf{open}}--\mPDBM{}s}

\newcommand{\PLT}{\ensuremath{\mathcal{PLT}}}
\newcommand{\itsRegion}[2]{\ensuremath{{[#1]}_{#2}}}
\newcommand{\RA}{\ensuremath{\mathcal{R}(\A)}}

\newcommand{\de}{\ensuremath{\delta}} %

\newcommand{\Doo}{\ensuremath{D_{0,0}}}
\newcommand{\Dij}{\ensuremath{D_{i,j}}}
\newcommand{\Dji}{\ensuremath{D_{j,i}}}
\newcommand{\Dik}{\ensuremath{D_{i,k}}}
\newcommand{\Dkj}{\ensuremath{D_{k,j}}}
\newcommand{\Dki}{\ensuremath{D_{k,i}}}
\newcommand{\Djk}{\ensuremath{D_{j,k}}}
\newcommand{\Dio}{\ensuremath{D_{i,0}}}
\newcommand{\Doi}{\ensuremath{D_{0,i}}}
\newcommand{\Doj}{\ensuremath{D_{0,j}}}
\newcommand{\Djo}{\ensuremath{D_{j,0}}}
\newcommand{\mPDBM}{\ensuremath{\mathsf{p}}--PDBM}
\newcommand{\mPDBMs}{\mPDBM{}s}

\newcommand{\Dix}{\ensuremath{D_{i,\clock}}}
\newcommand{\Dxi}{\ensuremath{D_{\clock,i}}}
\newcommand{\Djx}{\ensuremath{D_{j,\clock}}}
\newcommand{\Dxj}{\ensuremath{D_{\clock,j}}}
\newcommand{\Dxo}{\ensuremath{D_{\clock,0}}}
\newcommand{\Dox}{\ensuremath{D_{0,\clock}}}
\newcommand{\Dyo}{\ensuremath{D_{\clockk,0}}}
\newcommand{\Doy}{\ensuremath{D_{0,\clockk}}}
\newcommand{\Dxy}{\ensuremath{D_{\clock,\clockk}}}
\newcommand{\Dyx}{\ensuremath{D_{\clockk,\clock}}}

\newcommand{\Dko}{\ensuremath{D_{k,0}}}
\ifdefined\Dok%
	\renewcommand{\Dok}{\ensuremath{D_{0,k}}}
\else
	\newcommand{\Dok}{\ensuremath{D_{0,k}}}
\fi

\newcommand{\Dkx}{\ensuremath{D_{k,\clock}}}
\newcommand{\Dxk}{\ensuremath{D_{\clock,k}}}
\newcommand{\Dii}{\ensuremath{D_{i,i}}}
\newcommand{\Djj}{\ensuremath{D_{j,j}}}

\newcommand{\dij}{\ensuremath{d_{i,j}}}
\newcommand{\dji}{\ensuremath{d_{j,i}}}
\newcommand{\dzeroi}{\ensuremath{d_{0,i}}}
\newcommand{\dio}{\ensuremath{d_{i,0}}}
\newcommand{\doj}{\ensuremath{d_{0,j}}}
\newcommand{\djo}{\ensuremath{d_{j,0}}}
\newcommand{\dxo}{\ensuremath{d_{x,0}}}
\newcommand{\dox}{\ensuremath{d_{0,x}}}

\newcommand{\doy}{\ensuremath{d_{0,y}}}
\newcommand{\dxy}{\ensuremath{d_{\clock,\clockk}}}
\newcommand{\djk}{\ensuremath{d_{j,k}}}
\newcommand{\dkj}{\ensuremath{d_{k,j}}}
\newcommand{\dik}{\ensuremath{d_{i,k}}}
\newcommand{\dki}{\ensuremath{d_{k,i}}}
\newcommand{\dok}{\ensuremath{d_{0,k}}}

\newcommand{\LFP}{\ensuremath{\mathsf{LFP}}} %
\newcommand{\LFPd}[1]{\ensuremath{\mathsf{LFP}_{\paramR}(#1)}} %

\newcommand{\resetF}{\mathit{update}} %
\newcommand{\resetpF}{\overline{\mathit{update}}} %
\newcommand{\TEF}{\mathit{TE}} %

\newcommand{\guardF}{\ensuremath{\mathit{guard}_{\forall}}} %
\newcommand{\guardpF}{\ensuremath{\mathit{p\mbox{-}guard}_{\exists}}} %

\newcommand{\resetfun}{u}
\newcommand{\resetfunnp}{u_{np}}

\renewcommand{\r}{\ensuremath{:=}}
\newcommand{\RtoPPTA}{\textcolor{colorok}{U2P-PTA}}

\newcommand{\RPGRtoPPTA}{\textcolor{colorok}{R-U2P-PTA}}
\newcommand{\RPGRtoPPTAs}{\textcolor{colorok}{R-U2P-PTAs}}
\newcommand{\SRPGRtoPPTA}{\textcolor{colorok}{S-R-U2P-PTA}}
\newcommand{\SPGRtoPPTAs}{\textcolor{colorok}{S-R-U2P-PTAs}}

\newcommand{\sinit}{s\init} %
\newcommand{\state}{\ensuremath{s}} %
\newcommand{\States}{S} %
\newcommand{\Succ}{\mathsf{Succ}}

\newcommand{\floor}[1]{\lfloor#1\rfloor}

\newcommand{\reset}[2]{\ensuremath{{[#1]}_{#2}}}

\newcommand{\valid}{is valid for~$\paramR$}
\newcommand{\valuate}[2]{\ensuremath{#2(#1)}}
\newcommand{\stpw}{\ensuremath{\mathsf{stop}}} 
\newcommand{\imitator}{\textsf{IMITATOR}}

\newcommand{\recallResult}[2]
{%
	\smallskip

	\noindent\fcolorbox{black}{green!15}{
		\begin{minipage}{.95\columnwidth}
			\noindent\textbf{\cref{#1} (recalled).}
			{\em{}#2} 
		\end{minipage}
	}
}

\newcommand{\defProblem}[3]
{%
\noindent\fcolorbox{black}{blue!15}{
	\begin{minipage}{.95\columnwidth}
		\textbf{#1 problem:}\\
		\textsc{Input}: #2\\
		\textsc{Problem}: #3
	\end{minipage}
}

	\smallskip

}

\usepackage[pagewise]{lineno} %
\ifdefined\VersionWithComments%
	\usepackage{marginnote}
	\newcommand{\marginX}{\marginnote{\huge{\quad\quad\textbf{!}\quad\quad}}}
	\newcommand{\instructions}[1]{{\color{red}\marginX{}\textbf{Instructions: #1}}}
	\newcommand{\ea}[1]{{\color{blue}\marginX{}[\textbf{\'Etienne}: #1]}}
	\newcommand{\dl}[1]{{\color{green!50!black}\marginX{}[\textbf{Didier}: #1]}}
	\newcommand{\mr}[1]{{\color{orange!90!black}\marginX{}[\textbf{Mathias}: #1]}}

	\newcommand{\reviewer}[2]{\mbox{}{\color{red}\marginX{}[\textbf{Reviewer #1}: ``#2'']}}
	\newcommand{\todo}[1]{\mbox{}{\color{red}{\marginX{}\textbf{TODO}\ifx#1\\\else:\ \fi #1}}} %
\else
	\newcommand{\instructions}[1]{}
	\newcommand{\ea}[1]{}
	\newcommand{\dl}[1]{}
	\newcommand{\mr}[1]{}
	\newcommand{\reviewer}[2]{}
	\newcommand{\todo}[1]{}
\fi

\ifdefined\VersionWithComments%
 	\definecolor{colorok}{RGB}{80,80,150}
\else
	\definecolor{colorok}{RGB}{0,0,0}
\fi

\newcommand{\eg}{\textcolor{colorok}{\textit{e.\,g.,}}\xspace}
\newcommand{\ie}{\textcolor{colorok}{\textit{i.\,e.,}}\xspace}
\newcommand{\viz}{\textcolor{colorok}{viz.,}\xspace}

\newcommand{\wrt}{\textcolor{colorok}{w.r.t.}\xspace}
\newcommand{\st}{\textcolor{colorok}{s.t.}\xspace}

\theoremstyle{plain} %

\def\eg{\emph{e.g.}}

\begin{document}

\title[Instructions]{Parametric updates in parametric timed automata}

\author[\'E.~Andr\'e]{\'Etienne Andr\'e\rsuper{a}}	%
\address{\lsuper{a}Université de Lorraine, CNRS, Inria, LORIA, F-54000 Nancy, France; JFLI, CNRS, Tokyo, Japan; and National Institute of Informatics, Tokyo, Japan}	%
\author[D.~Lime]{Didier Lime\rsuper{b}}	%
\address{\lsuper{b}\'Ecole Centrale de Nantes, LS2N, CNRS, UMR 6004, Nantes, France}	%
\author[M.~Ramparison]{Mathias Ramparison\rsuper{c}}	%
\address{\lsuper{c}Université Sorbonne Paris Nord, LIPN, CNRS, UMR 7030, F-93430, Villetaneuse, France; University of Luxembourg, Luxembourg; and Univ.\ Grenoble Alpes, CNRS, UMR 5104, Grenoble INP, VERIMAG, Grenoble, France}	%

\thanks{This work is partially supported by the ANR national research program PACS (ANR-14-CE28-0002) and by the ANR-NRF French-Singaporean research program \href{https://www.loria.science/ProMiS}{ProMiS} (ANR-19-CE25-0015). 
    \'{E}.~Andr\'{e} was partially supported by ERATO HASUO Metamathematics for Systems Design Project (No.\ JPMJER1603), JST}

\begin{abstract}
  \noindent 	We introduce a new class of Parametric Timed Automata (PTAs) where we allow
	clocks to be compared to parameters in guards, as in classic PTAs, but also to be updated to parameters.
	We focus here on the EF-emptiness problem: ``is the set of parameter valuations for which some given location is reachable in the instantiated timed automaton empty?''.
    This problem is well-known to be undecidable for PTAs, and so it is for our extension.
	Nonetheless, if we update all clocks each time we compare a clock
	with a parameter and each time we update a clock to a parameter, we obtain
    a syntactic subclass for which we can decide the EF-emptiness problem
    and even perform the exact synthesis of the set of rational
    valuations such that a given location is reachable. To the best of our knowledge, this is the first non-trivial subclass of PTAs, actually even extended with parametric updates, for which this is possible.
\end{abstract}

\maketitle

\setcounter{footnote}{0}

\ifdefined\VersionWithComments%
	\textcolor{red}{\textbf{This is the version with comments. To disable comments, comment out line~3 in the \LaTeX{} source.}}
\fi

\ifdefined\VersionWithComments%
	\setcounter{tocdepth}{3}
	\tableofcontents
\fi

\newcommand{\figpowpta}{
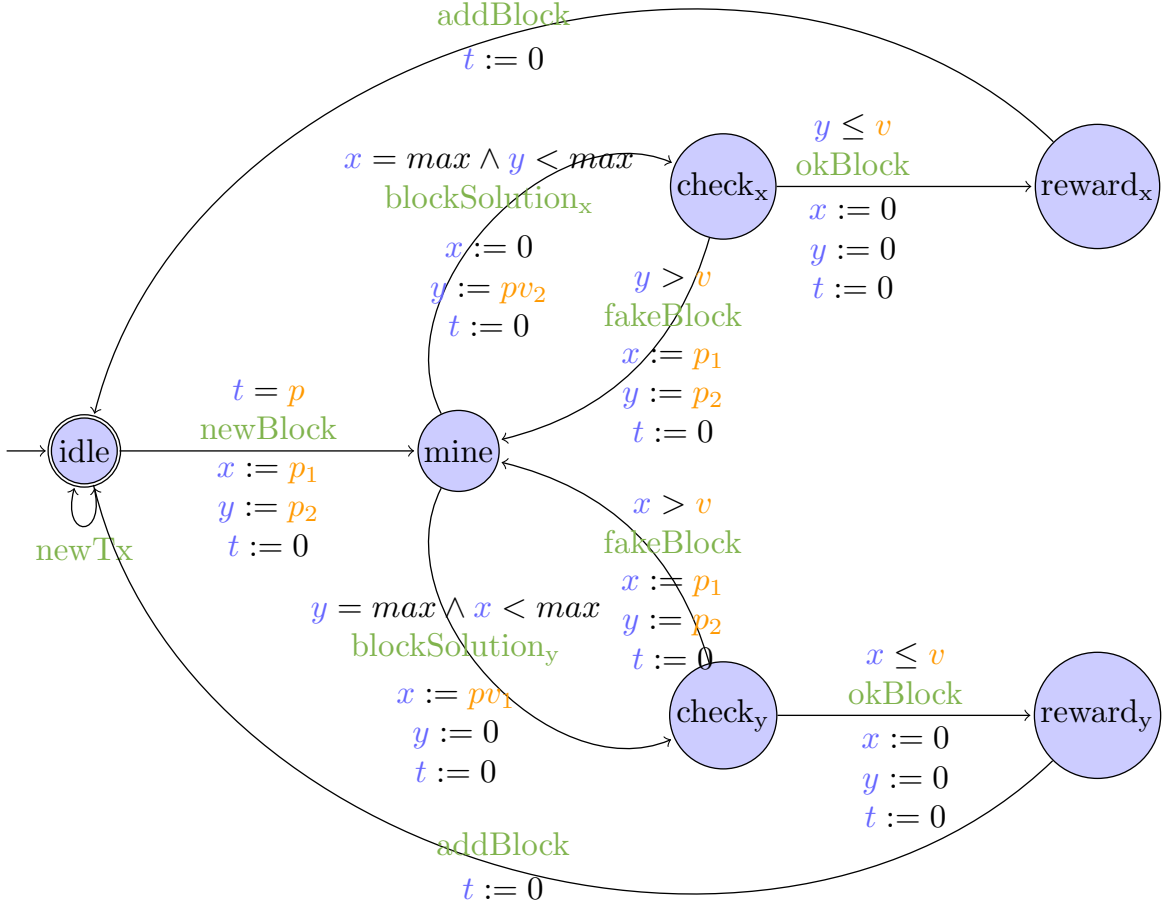
\begin{figure*}[tb!]
\vspace{-4em}
\resizebox{\linewidth}{!}{
{\centering
\small
\hspace{-2em}\begin{tikzpicture}[shorten >=1pt, node distance=4cm, on grid, auto]
\node[location, initial, final]       (A)   {$\styleloc{idle}$};
\node[location, inner sep=0pt,inner sep=1pt]       (B)  [right=of A]     {$\styleloc{mine}$};
\node[location, inner sep=0pt,inner sep=1pt]       (C)  [above right =of B]     {$\styleloc{check_x}$};
\node[location, inner sep=0pt,inner sep=1pt]       (D)  [below right=of B]     {$\styleloc{check_y}$};
\node[location, inner sep=0pt,inner sep=1pt]       (E)  [right=of C]     {$\styleloc{reward_x}$};
\node[location, inner sep=0pt,inner sep=1pt]       (F)  [right=of D]     {$\styleloc{reward_y}$};

\path[->]
	(A)  edge[loop below] node  [swap,align=center]   {$\styleact{newTx}$}   (A)
	(A)  edge node  [swap,align=center, above]   {$\styleclock{t}=\styleparam{\param}$\\$\styleact{newBlock}$} node [swap,align=center,below]{$\styleclock{x} := \styleparam{\param_1}$\\$\styleclock{y} := \styleparam{\param_2}$\\ $\styleclock{t} :=0$}  (B)
	(B)  edge[bend left=70] node  [swap,align=center, above]   {$\styleclock{x} = max\wedge\styleclock{y} < max$\\$\styleact{blockSolution_x}$} node [swap,align=center,below] {$\styleclock{x}\r 0$\\$\styleclock{y}\r\styleparam{\param v_2}$\\$\styleclock{t}\r 0$}   (C)
	(B)  edge[bend right=70] node  [swap,align=center, above, xshift=-10]   {$\styleclock{y} = max\wedge\styleclock{x} < max$\\$\styleact{blockSolution_y}$} node  [swap,align=center, below, xshift=-10] {$\styleclock{x}\r \styleparam{\param v_1}$\\$\styleclock{y}\r 0$\\$\styleclock{t}\r 0$}  (D)
	(C)  edge[below] node  [swap,align=center, above, xshift=-16]   {$\styleclock{y} \leq \styleparam{v}$\\$\styleact{okBlock}$} node  [swap,align=center, below, xshift=-16]   {$\styleclock{x}\r 0$\\$\styleclock{y}\r 0$\\$\styleclock{t}\r 0$} (E)
	(C)  edge[bend left] node  [swap,align=center, above, xshift=12, yshift=10]   {$\styleclock{y} > \styleparam{v}$\\$\styleact{fakeBlock}$} node  [swap,align=center, below, xshift=12, yshift=10]   {$\styleclock{x}\r \styleparam{\param_1}$\\$\styleclock{y}\r \styleparam{\param_2}$\\ $\styleclock{t}\r 0$} (B)

	(D) edge node  [swap,align=center, above]   {$\styleclock{x} \leq \styleparam{v}$\\$\styleact{okBlock}$} node  [swap,align=center, below]   {$\styleclock{x}\r 0$\\$\styleclock{y}\r 0$\\$\styleclock{t}\r 0$} (F)
	(D)  edge[bend right] node  [swap,align=center, above, xshift=12, yshift=-10]   {$\styleclock{x} > \styleparam{v}$\\$\styleact{fakeBlock}$} node  [swap,align=center, below, xshift=12, yshift=-10]   {$\styleclock{x}\r \styleparam{\param_1}$\\$\styleclock{y}\r \styleparam{\param_2}$\\ $\styleclock{t}\r 0$} (B)

	(E) edge[bend right=60] node  [swap,align=center, above]   {$\styleact{addBlock}$} node [swap,align=center,below]{$\styleclock{t} :=0$} (A)
	(F) edge[bend left=60] node  [swap,align=center, above]   {$\styleact{addBlock}$} node [swap,align=center,below]{$\styleclock{t} :=0$} (A);
\end{tikzpicture}
}
}
\vspace{-3em}
\caption{A blockchain proof-of-work modeled with a bounded \RPGRtoPPTA{}.}%
\label{figure:peer}
\hspace{-10em}
\end{figure*}
}
\section{Introduction}

Timed automata (TAs) are a powerful formalism to model and verify timed concurrent systems, both expressive enough to model many interesting systems and enjoying several decidability properties.
In particular, the reachability of a discrete state is \journalVersion{decidable and }PSPACE-complete~\cite{AD94}.
In TAs, clocks can be compared with constants in guards, and can be updated to~0\journalVersion{ (``reset'')} along edges.

Timed automata may turn insufficient to verify systems where the timing constants themselves are subject to some uncertainty, %
or when they are simply not known at the early design stage.
Parametric timed automata (PTAs)~\cite{AHV93} address this drawback by allowing parameters (unknown constants) in the timing constraints; this high expressive power comes at the cost of the undecidability of most interesting problems\journalVersion{ (see \eg{}~\cite{Andre19STTT})}.
In particular, the basic problem of EF-emptiness (``is the set of valuations for which a given location is reachable in the instantiated timed automaton empty?'')\ is ``robustly'' undecidable: even for a single rational-valued~\cite{Miller00} or integer-valued parameter~\cite{AHV93,BBLS15}, or when only strict constraints are used~\cite{Doyen07}.
A famous syntactic subclass of PTAs that enjoys limited decidability is L/U-PTAs~\cite{HRSV02}, where the parameters set is partitioned into lower-bound and upper-bound parameters, \ie{} parameters that can only be compared to a clock as a lower-bound (resp.\ upper-bound).
The EF-emptiness problem is decidable for L/U-PTAs~\cite{HRSV02,BlT09} and for PTAs under several restrictions~\cite{BO14}; however, most other problems
are undecidable (\eg{}~\cite{BlT09,Quaas14,JLR15,ALR16ICFEM,ALime17,ALM20})\journalVersion{ (see~\cite{Andre19STTT} for a survey)}.

Recall that the EF-emptiness problem is decidable for L/U-PTAs~\cite{HRSV02,BlT09} and for PTAs under several restrictions~\cite{BO14}; however, most other problems
are undecidable (\eg{}~\cite{BlT09,Quaas14,JLR15,ALR16ICFEM,ALime17})\journalVersion{ (see~\cite{Andre19STTT} for a survey)}.

\subsection{Contribution}

We investigate parametric updates, which can model an unknown timing configuration in a system where processes need to synchronize together on common events, as in \eg{} programmable controller logic programs with concurrent tasks execution.
We show that the EF-emptiness problem is decidable for PTAs augmented with parametric updates\longVersion{ (\ie{} \RtoPPTA)}, with the additional condition that whenever a clock is compared to a parameter in a guard or updated to a parameter%
, all clocks must be updated (possibly to parameters)---this gives \RPGRtoPPTA{}.
This result holds when the parameters are \emph{bounded rationals in guards}, and possibly \emph{unbounded rationals in updates}.
Non-trivial decidable subclasses of PTAs are a rarity (to the best of our knowledge, only L/U-PTAs~\cite{HRSV02} and integer-points (IP-)PTAs~\cite{ALR16ICFEM}); 
this makes our positive result very welcome.
In addition, not only the emptiness is decidable, but \emph{exact synthesis} for bounded rational-valued parameters can be performed---which contrasts with L/U-PTAs and IP-PTAs for which synthesis was shown to be intractable~\cite{JLR15,ALR16ICFEM}.

\paragraph{About this manuscript}
This is the extended version of~\cite{ALR19FORTE}.
In addition to additional explanations and all proofs of our results, we added the whole new \cref{section:stopwatch} adding stopwatches to our formalism.

\subsection{Related work}

Our construction is reminiscent of the parametric difference bound matrices (PDBMs) defined in~\cite[section III.C]{QSW17} where the authors in this paper revisit the result of the binary reachability relation over both locations and clock valuations in TAs;
however, parameters of~\cite{QSW17} are used to bound in time a run that reaches a given location, while we use parameters directly in guards and resets along the run, which make them active components of the run specifically for intersection with parametric guards, a key point not tackled in~\cite{QSW17}.\ea{il faudrait aussi dire ce que Ã§a fait~\cite{QSW17}!}
Related DBMs with an additional parameter have been studied, such as shrunk DBMs~\cite{SBM14,BGMRS19} and infinitesimally enlarged DBMs~\cite{Sankur15}.

Allowing parameters in clock updates is inspired
by the updatable TA \longVersion{formalism }defined in~\cite{BDFP04} where clocks can be updated not only to~$0$ (``reset'') but also to rational constants (``update'').
In~\cite{ALR18ACSD}, we extended the result of~\cite{BDFP04} by allowing parametric updates (and no parameter elsewhere, \eg{} in guards):
the EF-emptiness is undecidable even in the restricted setting of bounded rational-valued parameters, but becomes decidable when parameters are restricted to (unbounded) integers.

Synthesis is obviously harder than EF-emptiness: only three results have been proposed to synthesize the exact set of valuations for subclasses of PTAs, but they are all concerned with \emph{integer}-valued parameters~\cite{BlT09,JLR15,ALR18ACSD}.
\journalVersion{%
	More precisely, it is possible to synthesize unbounded integers for L- or U-PTAs (L/U-PTAs with only lower-bound, or only upper-bound, parameters)~\cite{BlT09};
	bounded integers for PTAs~\cite{JLR15}
	unbounded integers for timed automata with parametric updates~\cite{ALR18ACSD}.

}%
In contrast, we deal here with (bounded) rational-valued parameters---which makes this result the first of its kind.
The idea of updating all clocks when compared to parameters comes from our class of \emph{reset-PTAs} briefly mentioned in~\cite{ALR16ICFEM}, but not thoroughly studied.
\journalVersion{

}%
Finally, updating clocks on each transition in which a parameter appears is reminiscent of \longVersion{the }initialized rectangular hybrid automata\journalVersion{ formalism defined in}~\cite{HKPV98}, which remains one of the few decidable subclasses of hybrid automata.
\journalVersion{Indeed, timed automata can be defined as a subclass of initialized rectangular hybrid automata where clocks evolve at the same fixed rate, in which diagonal constraints are allowed but not systematically used in practice.
}
However, besides the fact that in PTAs variables (clocks) evolve at the same rate, in initialized rectangular hybrid automata variables are reset whenever one of the derivatives of those variables changes, which is not at all the condition we use for global updates in our \RPGRtoPPTA{}.

\subsection{Outline}
\cref{section:preliminaries} recalls preliminaries. \cref{section:decidablesubclass} presents \RPGRtoPPTA{} along with our decidability result. %
\cref{section:operations,section:region} introduce operations on our \mPDBMs{} and our extended region automaton.
\cref{section:decidability} proves the main decidability result.
\cref{section:stopwatch} extends our results to stopwatches.
\cref{section:casestudy} gives a concrete application of our result.
\cref{section:conclusionRtoPPTA} concludes the paper.

\section{Preliminaries}\label{section:preliminaries}

\journalVersion{
	Let $\grandn$, $\grandz$, $\grandqplus$ and $\grandrplus$ denote the sets of non-negative integers, integers, non-negative rational numbers and non-negative real numbers respectively.
}

Throughout this paper, we assume a set~$\Clock = \{ \clock_1, \dots, \clock_\ClockCard \} $ of \emph{clocks}, \ie{} real-valued variables evolving at the same rate.
A clock valuation is \journalVersion{a function }$\clockval : \Clock \rightarrow \grandrplus$.
We write $\ClocksZero$ for the clock valuation that assigns $0$ to all clocks.
Given $d \in \grandrplus$, $\clockval + d$ (resp.\ $\clockval - d$) denotes the valuation such that $(\clockval + d)(\clock) = \clockval(\clock) + d$ (resp.\ $(\clockval - d)(\clock) = \clockval(\clock) - d$ if~$\clockval(\clock) - d>0$, $0$ otherwise), for all $\clock \in \Clock$.
We assume a set~$\Param = \{ \param_1, \dots, \param_\ParamCard \} $ of \emph{parameters}, \ie{} unknown constants.
A parameter \emph{valuation} $\pval$ is a function $\pval : \Param \rightarrow \grandqplus$.
We identify a valuation~$\pval$ with the point $(\pval(\param_1), \dots, \pval(\param_{\ParamCard}))$ of $\grandqplus^\ParamCard$.
Given $d \in \grandn$, $\pval + d$ (resp.\ $\pval - d$) denotes the valuation such that $(\pval + d)(\param) = \pval(\param) + d$ (resp.\ $(\pval - d)(\param) = \pval(\param) - d$ if~$\pval(\param) - d>0$, $0$ otherwise), for all $\param \in \Param$.\ea{warning: si ca devient pas n\'egatif}

In the following, we assume ${\compOpLeq} \in \{<, \leq\}$ %
and ${\compOp} \in \{<, \leq, \geq, >\}$.

A \emph{parametric guard}~$\guard$ is a constraint over $\Clock \cup \Param$ defined as the conjunction of inequalities of the form
	$\clock \compOp z$, where $\clock$ is a clock and $z$ is either a parameter or a constant in~$\grandz$.
A \emph{non-parametric guard} is a parametric guard without parameters (\ie{} over~$\Clock$).

Given a parameter valuation~$\pval$, $\valuate{\guard}{\pval}$ denotes the constraint over~$\Clock$ obtained by replacing in~$\guard$ each parameter~$\param$ with~$\pval(\param)$.
We extend this notation to an \emph{expression}: a sum or difference of parameters and constants.
Likewise, given a clock valuation~$\clockval$, $\valuate{\valuate{\guard}{\pval}}{\clockval}$ denotes the expression obtained by replacing in~$\valuate{\guard}{\pval}$ each clock~$\clock$ with~$\clockval(\clock)$.
A clock valuation~$\clockval$ \emph{satisfies} constraint~$\valuate{\guard}{\pval}$ (denoted by $\clockval \models \valuate{\guard}{\pval}$) if~$\valuate{\valuate{\guard}{\pval}}{\clockval}$ evaluates to true.
We say that %
$\pval$ \emph{satisfies}~$\guard$,
denoted by $\pval \models \guard$,
if the set of clock valuations satisfying~$\valuate{\guard}{\pval}$ is nonempty.
We say that $\guard$ is \emph{satisfiable} if $\exists \clockval, \pval \text{ s.t.\ } \clockval \models \valuate{\guard}{\pval}$.

A \emph{parametric update} is a partial function $\resetfun : \Clock \rightharpoonup \grandn \cup \Param$ which assigns to some of the clocks an integer constant or a parameter.
For $\pval$ a parameter valuation, we define a partial function $\pval(\resetfun) : \Clock \rightharpoonup \grandqplus$ as follows:
for each clock $\clock\in\Clock$, $\valuate{\clock}{\valuate{\resetfun}{\pval}}=k\in\grandn$ if $\valuate{\clock}{\resetfun}=k$
and $\valuate{\clock}{\valuate{\resetfun}{\pval}}=\valuate{\param}{\pval}\in\grandqplus$ if $\valuate{\clock}{\resetfun}=\param$ a parameter.
A non-parametric update is $\resetfunnp : \Clock \rightharpoonup \grandn$.
\journalVersion{The term \emph{reset} has been used for clock updates to values different from~$0$ in~\cite{BY03}.}%
For a clock valuation~$\clockval$ and a parameter valuation~$\pval$, we denote by~$\reset{\clockval}{\valuate{\resetfun}{\pval}}$
the clock valuation obtained after applying~$\valuate{\resetfun}{\pval}$.
\mr{definir le reset dans la phrase du dessus}%
We first define a new class of parametric timed automata and then define
plain parametric timed automata and timed automata as special cases.

\begin{defi}\label{def:PTA}
	An update-to-parameter PTA (\RtoPPTA{})
	$\A$ is a tuple \[\A = (\Actions, \Loc, \locinit, \Clock, \Param, \Edges),\] where: %
  \begin{ienumeration}
		\item $\Actions$ is a finite set of actions,
		\item $\Loc$ is a finite set of locations,
		\item $\locinit \in \Loc$ is the initial location,
		\item $\Clock$ is a finite set of clocks,
		\item $\Param$ is a finite set of parameters,
		\item $\Edges$ is a finite set of edges  $\edge = \langle\loc,\guard,\action,\resetfun,\loc'\rangle$
		where
		$\loc,\loc'\in \Loc$ are the source and target locations, $\guard$ is a parametric guard, $\action \in \Actions$ and
		$\resetfun : \Clock \rightharpoonup \grandn \cup \Param$ is a parametric update function.
  \end{ienumeration}
\end{defi}
\figpowpta{}

\noindent
An \RtoPPTA{} is depicted in \cref{figure:peer}. Note that all clocks are updated whenever there is a comparison with a parameter (as in \styleact{newBlock}) or a clock is updated to a parameter (as in \styleact{blockSolution_\clock}).

Given a parameter valuation $\pval$, we denote by $\valuate{\A}{\pval}$ the structure where all occurrences of a parameter~$\param_i$ have been replaced by~$\pval(\param_i)$.
If $\valuate{\A}{\pval}$ is such that all constants in guards and updates are integers, then $\valuate{\A}{\pval}$ is a \emph{updatable timed automaton}~\cite{BDFP04}
but will be called \emph{timed automaton} (TA) for the sake of simplicity in this paper.
\journalVersion{%
	In the following, we may denote as a timed automaton any structure $\valuate{\A}{\pval}$, by assuming a rescaling of the constants:
	by multiplying all constants in $\valuate{\A}{\pval}$ by their least common denominator,
		we obtain an equivalent timed automaton (with integer constants).\ea{ben en même temps, on dit déjà que si c'est rationnel, c'est déjà un TA, alors pourquoi rescaler…?}
}

A \emph{bounded} \RtoPPTA{} is a \RtoPPTA{} with a bounded parameter domain that assigns to each parameter a minimum integer bound and a maximum integer bound.
That is, each parameter~$\param_i$ ranges in an interval $[a_i, b_i]$, with $a_i,b_i \in \grandn$.
Hence, a bounded parameter domain is a hyperrectangle of dimension~$\ParamCard$.

A parametric timed automaton (PTA)~\cite{AHV93} is a \RtoPPTA{} where, for any edge $\edge = \langle\loc,\guard,\action,\resetfun,\loc'\rangle \in \Edges$, $\resetfun : \Clock \rightharpoonup \{0\}$.

\begin{defi}[Concrete semantics of a TA]\label{definition:concretesemantics}
	Given a \RtoPPTA{} $\A = (\Actions, \Loc, \locinit, \Clock, \Param, \Edges)$, %
	and a parameter valuation~\(\pval\),
	the concrete semantics of $\valuate{\A}{\pval}$ is given by the timed transition system $(\States, \sinit, \flecheRel)$, with
	\begin{itemize}
		\item $\States = \{ (\loc, \clockval) \in \Loc \times \grandrplus^\ClockCard \}$,
            $\sinit = (\locinit, \ClocksZero)$;
		\item  $\flecheRel$ consists of the discrete and (continuous) delay transition relations:
				\begin{itemize}
			\item discrete transitions: $(\loc,\clockval) \longueflecheRel{\edge} (\loc',\clockval')$, %
                if $(\loc, \clockval) , (\loc',\clockval') \in \States$, there exists
                \begin{mathpar}
                \edge = \langle\loc,\guard,\action,\resetfun,\loc'\rangle \in \Edges, \and
                \clockval'= \reset{\clockval}{\pval(\resetfun)},\; \text{and} \and
                \clockval\models \pval(\guard)
                \end{mathpar}
			\item delay transitions: $(\loc,\clockval) \longueflecheRel{d} (\loc, \clockval+d)$, with $d \in \grandrplus$. %
		\end{itemize}
	\end{itemize}
\end{defi}

\noindent
Moreover, we write $(\loc, \clockval)\longuefleche{\edge} (\loc',\clockval')$ for a combination of a delay and discrete transitions where
	$((\loc, \clockval), \edge, (\loc', \clockval')) \in \flecheRel$ if
		$\exists d, \clockval'' :  (\loc,\clockval) \longueflecheRel{d} (\loc,\clockval'') \longueflecheRel{\edge} (\loc',\clockval')$.

Given a TA~$\valuate{\A}{\pval}$ with concrete semantics $(\States, \sinit, \flecheRel)$,
we refer to the states of~$\States$ as the \emph{concrete states} of~$\valuate{\A}{\pval}$.
A (concrete) \emph{run} of~$\valuate{\A}{\pval}$ is a possibly infinite alternating sequence of concrete states of $\valuate{\A}{\pval}$ and edges starting from\journalVersion{ the initial concrete state}~$\sinit$ of the form
$\sinit \longuefleche{\edge_0} \state_1\longuefleche {\edge_1} \cdots \longuefleche{\edge_{m-1}} \state_m \longuefleche{\edge_{m}} \cdots$, such that for all
$i = 0, 1, \dots$, $\edge_i \in \Edges$, and $(\state_i , \edge_i , \state_{i+1}) \in \flecheRel$.
Given a state~$\state=(\loc, \clockval)$, we say that $\state$ is reachable (or that $\valuate{\A}{\pval}$ reaches~$\state$) if $\state$ belongs to a run of $\valuate{\A}{\pval}$.
By extension, we say that $\loc$ is reachable in~$\valuate{\A}{\pval}$, if there exists a state $(\loc, \clockval)$ that is reachable.

Throughout this paper, let~$\CONSTMAX$ denotes the largest constant in a given \RtoPPTA{}, \ie{} the maximum of the largest constant compared to a clock in a guard and the largest upper bound of a parameter (if the \RtoPPTA{} is bounded).

Let us recall the notion of clock region~\cite{AD94}.
Given a clock~$\clock$ and a clock valuation~$\clockval$, recall that $\lfloor \clockval(\clock)\rfloor$ denotes the integer part of $\clockval(\clock)$ while $\partieFrac(\clockval(\clock))$ denotes its fractional part.
We define the same notation for parameter valuations.

\begin{defi}[clock region]\label{regionAD}
	For two clock valuations $\clockval$ and $\clockval'$, $\sim$ is an equivalence relation defined by: $\clockval\sim \clockval'$ iff
	\begin{ienumeration}
		\item for all clocks~$\clock$, either $\lfloor \clockval(\clock)\rfloor = \lfloor \clockval'(\clock)\rfloor$ or $\clockval(\clock), \clockval'(\clock)>\CONSTMAX$;
		\item for all clocks~$\clock, \clockk$ with $\clockval(\clock),\clockval(\clockk)\leq \CONSTMAX$,
		$\partieFrac(\clockval(\clock))\leq \partieFrac(\clockval(\clockk))$ iff $\partieFrac(\clockval'(\clock))\leq \partieFrac(\clockval'(\clockk))$;
		\item for all clocks~$\clock$ with $\clockval(\clock)\leq \CONSTMAX$, $\partieFrac(\clockval(\clock))=0$ iff $\partieFrac(\clockval'(\clock))=0$.
	\end{ienumeration}

    \noindent
	A \emph{clock region} $\clockR$ is an equivalence class of $\sim$.
\end{defi}

Two clock valuations in the same clock region (cf. \cref{regionAD}) reach the same regions by time elapsing, satisfy the same guards and can take the same transitions~\cite{AD94}.

In this paper, we address the \styleTCTL{EF}-emptiness problem:
given a \RtoPPTA~$\A$ and a location~$\loc$, is set of valuations $\pval$ such that there is a run in $\valuate{\A}{\pval}$ reaching~$\loc$ is empty?
More formally, the problem can be written as:
\smallskip

\defProblem%
	{\styleTCTL{EF}-emptiness}
	{a \RtoPPTA~$\A$ and a location~$\loc$}
	{$\{\pval \mid \exists \sinit \longuefleche{\edge_0} (\loc_1,\clockval_1)\longuefleche {\edge_1} \cdots \longuefleche{\edge_{m-1}} (\loc, \clockval)$ a run of $\valuate{\A}{\pval}\}=\emptyset$?
}

\smallskip

\section{A decidable subclass of \texorpdfstring{\RtoPPTA{}}{U2P-PTA}s}\label{section:decidablesubclass}

We now impose that, whenever a guard or an update along an edge contains parameters, then all clocks must be updated (to constants or parameters).
Our main contribution is to prove that this restriction makes EF-emptiness decidable.

\begin{defi}\label{def:RtoPPTA}
	An \emph{\RPGRtoPPTA{}} is a \RtoPPTA{} where for any \longVersion{edge }$\langle\loc,\guard,\action,\resetfun,\loc'\rangle \in \Edges$,
 $\resetfun$ is a total function whenever:
	\begin{ienumeration}%
		\item $\guard$ is a parametric guard, or
		\item $\resetfun(\clock) \in \Param$ for some~$\clock \in \Clock$.
	\end{ienumeration}%
	\end{defi}

\noindent
Both conditions of \cref{def:RtoPPTA} are necessary. If we allow parametric guards to be passed without a full update of clocks, then we obtain a larger class of PTAs for which the \styleTCTL{EF}-emptiness problem is undecidable as it is for regular PTAs~\cite{AHV93}. If we allow partial parametric updates of clocks, then we obtain a larger class of Reset-to-Parameter Timed Automaton defined in~\cite{ALR18ACSD} for which we proved the \styleTCTL{EF}-emptiness problem is undecidable.

  In the following we only consider either non-parametric, or (necessarily total) fully parametric update functions.
A total update function which is not fully parametric (\ie{} an update of some clocks to parameters and all others to constants)
can be encoded as a total fully parametric update immediately followed by a (partial) non-parametric update function.

The main idea for proving decidability is the following: given an \RPGRtoPPTA{} $\A$ we will construct a finite region automaton that bisimulates~$\A$, as in TA~\cite{AD94}.
Our regions will contain both clocks and parameters and will be a finite number, due to the finite number of parameter and their construction similar to clock regions~\cite{AD94}.
Since parameters are allowed in guards,
we need to construct parameter regions and more restricted clock regions.

We will define a form of Parametric Difference Bound Matrices (\viz \mPDBMs{} for precise PDBMs, inspired by~\cite{HRSV02}) in which, once valuated by a parameter valuation,
two clock valuations have the same discrete behavior and satisfy the same non-parametric guards. A \mPDBM{} will define the \emph{set of clocks and parameter valuations} that satisfies it, while once valuated by a parameter valuation, a valuated \mPDBM{} will define the \emph{set of clock valuations} that satisfies it.
A key point is that in our \mPDBMs{} the parametric constraints used in the matrix will be defined from a \emph{finite} set of predefined expressions involving parameters and constants, and we will prove that this defines a finite number of \mPDBMs{}. Decidability will come from this fact: the region automaton will evolve in this finite and stable set of \mPDBMs{} under time elapsing and update operators.

We define this set of parametric constraints ($\PLT$ for parametric linear term) as follows:
	$\PLT=\{\partieFrac(\param_{i}), 1-\partieFrac(\param_{i}), \partieFrac(\param_{i})-\partieFrac(\param_{j}), \partieFrac(\param_{j})+1-\partieFrac(\param_{i}), 1, 0,\partieFrac(\param_{i})-1-\partieFrac(\param_{j}), -\partieFrac(\param_{i}), \partieFrac(\param_{i})-1\}$,
for all $1\leq i, j \leq\ParamCard$.
Given a parameter valuation~$\pval$ and~$d\in\PLT$, we denote by~$\pval(d)$ the term obtained by replacing in~$d$ each parameter~$\param$ by~$\pval(\param)$.
 Let us now define an equivalence relation between parameter valuations~$\pval$ and~$\pval'$.

\mr{note: $\frown$ n'est utilisé que 3 fois}
\begin{defi}[regions of parameters]\label{parameterRegions}
	We write that $\pval\frown\pval'$ if
	\begin{ienumeration}%
	\item for all parameters $\param$, $\lfloor \pval(\param)\rfloor = \lfloor \pval'(\param)\rfloor$;
	\item{for all~$d_1, d_2, d_3\in\PLT$, $\pval(d_1)\leq \pval(d_{2})+\pval(d_{3})$
	iff $\pval'(d_1)\leq \pval'(d_{2})+\pval'(d_{3})$.}
	\end{ienumeration}%
\end{defi}

\noindent
\emph{Parameter regions} are defined as the equivalence classes of $\frown$%
\longVersion{, and we will use the notation $\paramR$ for parameter regions.
The set of all \emph{parameter regions} is denoted by~$\paramregions$}.
The definition is in a way similar to \crefDef{regionAD} but also involves comparisons of sums of elements of~$\PLT$. In fact, we will need this kind of comparisons to define our \mPDBMs{}. Nonetheless we do not need more complicated comparisons as in \RPGRtoPPTA{} whenever a parametric guard or updated is met the update is a total function: this preserves us from the parameter accumulation, \eg{} obtaining expressions of the form~$5\partieFrac(\param_{i})-1-3\partieFrac(\param_{j})$ (that may occur in usual PTAs).

In the following, our \mPDBMs{} will be matrices of
projections on parameters of parametric clock constraints,
written as matrices of pairs of the form~$\D = (d, \compOpLeq)$ where~$d\in\PLT$.
We therefore need to define comparisons on these pairs.

We define an associative and commutative operator $\oplus$ as $\compOpLeq_1 \oplus \compOpLeq_2 ={<}$ if $\compOpLeq_1\neq\compOpLeq_2$, or $\compOpLeq_1$ if $\compOpLeq_1=\compOpLeq_2$.
We define $\D_1+\D_2=(d_1+d_2,\compOpLeq_{1} \oplus \compOpLeq_{2})$.
\ea{rajouter une explication intuitive en une phrase}%
Following the idea of parameter regions, we define the \emph{validity} of a comparison
between pairs of the form~$(d_i,\compOpLeq_i)$ within a given parameter region,
\ie{} whether the comparison is true for all parameter valuations~$\pval$ in the parameter region~$\paramR$.

\begin{defi}[validity of comparison]\label{def:validity}
	Let $\paramR$ be a parameter region.
		Given any two linear terms $d_1,d_2$ over~$\Param$ (\ie{} of the form $\sum_i \alpha_i \param_i + d$ with $\alpha_i,d \in \grandz$),
		the comparison $(d_{1},\compOpLeq_1) \compOpLeq (d_{2},\compOpLeq_2)$ is \emph{valid for~$\paramR$} if:
	\begin{enumerate}
		\item $\compOpLeq = {<}$, and either
		\begin{ienumeration}
			\item\label{valideia}
			for all ${\pval}\in{\paramR}$,  $\valuate{d_{1}}{\pval}<\valuate{d_{2}}{\pval}$ evaluates to true regardless of $\compOpLeq_1,\compOpLeq_2$,
			or
			\item\label{valideib}
			for all ${\pval}\in{\paramR}$,  $\valuate{d_{1}}{\pval} \leq \valuate{d_{2}}{\pval}$ evaluates to true,
			$\compOpLeq_1 = {<}$ and $\compOpLeq_2 = {\leq}$;
		\end{ienumeration}
 		\item $\compOpLeq = {\leq}$, and either
 		\begin{ienumeration}
		 \item\label{valideiiie}for all ${\pval}\in{\paramR}$,  $\valuate{d_{1}}{\pval}<\valuate{d_{2}}{\pval}$ evaluates to true regardless of $\compOpLeq_1,\compOpLeq_2$, or
 		 \item\label{valideiiid}for all ${\pval}\in{\paramR}$,  $\valuate{d_{1}}{\pval}\leq\valuate{d_{2}}{\pval}$ evaluates to true, and
 		$\compOpLeq_1=\compOpLeq_2$, or $\compOpLeq_1 = {<}$.
 		\end{ienumeration}
	\end{enumerate}
		\ea{On a donc dit non : vois si tu as besoin de l'\'egalit\'e, et si oui, syntaxique ou s\'emantique ``The comparison $(d_{1},\compOpLeq_1) = (d_{2},\compOpLeq_2)$ is \emph{valid for~$\paramR$} if 
		for all ${\pval}\in{\paramR}$,  $\valuate{d_{1}}{\pval}=\valuate{d_{2}}{\pval}$ evaluates to true,
		$\compOpLeq_1=\compOpLeq_2$, and \emph{not valid} otherwise. ''}

\end{defi}

\noindent
Transitivity is immediate from the definition: if $\D_{1} \compOpLeq_1 \D_{2}$ and $\D_{2}\compOpLeq_2 \D_3$ are valid for~$\paramR$, $\D_{1} (\compOpLeq_1 \oplus \compOpLeq_2) \D_{3}$ is valid for~$\paramR$.

The following lemma derives from \crefDef{def:validity}:
\newcommand{\lemmasumvalid}{%
Let $d_1,d_2,d_3,d_4\in\PLT$. Let $\paramR$ be a parameter region.
If $(d_1,\compOpLeq_1)\leq (d_2,\compOpLeq_2)$ and $(d_3,\compOpLeq_3)\leq (d_4,\compOpLeq_4)$ are valid for $\paramR$
then $(d_1,\compOpLeq_1)+(d_3,\compOpLeq_3)\leq(d_2,\compOpLeq_2)+(d_4,\compOpLeq_4)$ is valid for $\paramR$.
}
\begin{lem}[validity of addition]\label{lemma:sumvalid}
\lemmasumvalid{}
\end{lem}
\newcommand{\prooflemmasumvalid}{%
\label{proof:lemmasumvalid}
Four cases show up: for all~$\pval\in\paramR$,
\begin{itemize}
	\item $\pval(d_1)<\pval(d_2)$ and $\pval(d_3)<\pval(d_4)$, then clearly $\pval(d_1)+\pval(d_3)<\pval(d_2)+\pval(d_4)$ and we have our result from \crefDef{def:validity}~(\refc{valideiiie}).

	\item $\pval(d_1)<\pval(d_2)$ and $\pval(d_3)\leq\pval(d_4)$, then $\pval(d_1)+\pval(d_3)<\pval(d_2)+\pval(d_4)$ and we have our result from \crefDef{def:validity}~(\refc{valideiiie}).
	\item $\pval(d_1)\leq\pval(d_2)$ and $\pval(d_3)<\pval(d_4)$, then $\pval(d_1)+\pval(d_3)<\pval(d_2)+\pval(d_4)$ and we have our result from \crefDef{def:validity}~(\refc{valideiiie}).

	\item $\pval(d_1)\leq\pval(d_2)$ and $\pval(d_3)\leq\pval(d_4)$, then $\pval(d_1)+\pval(d_3)\leq\pval(d_2)+\pval(d_4)$ and
		\begin{enumerate}
			\item if $\compOpLeq_1=\compOpLeq_2$ and $\compOpLeq_3=\compOpLeq_4$ then $\compOpLeq_1 \oplus \compOpLeq_3=\compOpLeq_2\oplus\compOpLeq_4$ and we have our result from \crefDef{def:validity}~(\refc{valideiiid}).
			\item if $\compOpLeq_1=\compOpLeq_2$ and $\compOpLeq_3= {<}$, $\compOpLeq_4= {\leq}$ then $\compOpLeq_1 \oplus \compOpLeq_3= {<}$ and $\compOpLeq_2\oplus\compOpLeq_4$ is either $<$ or~$\leq$ and we have our result from \crefDef{def:validity}~(\refc{valideiiid}).
			\item if $\compOpLeq_1= {<}$, $\compOpLeq_2= {\leq}$ and $\compOpLeq_3=\compOpLeq_4$ then $\compOpLeq_1 \oplus \compOpLeq_3= {<}$ and $\compOpLeq_2\oplus\compOpLeq_4$ is either $<$ or~$\leq$ and we have our result from \crefDef{def:validity}~(\refc{valideiiid}).
			\item if $\compOpLeq_1=\compOpLeq_3=<$ and $\compOpLeq_2=\compOpLeq_4=\leq$ then $\compOpLeq_1 \oplus \compOpLeq_3=<$ and $\compOpLeq_2\oplus\compOpLeq_4=\leq$ and we have our result from \crefDef{def:validity}~(\refc{valideiiid}).
		\end{enumerate}
\end{itemize}
From \crefDef{def:validity}~(\refc{valideiiie},~\ref{valideiiid}) we have that $(d_1,\compOpLeq_1)+(d_3,\compOpLeq_3)\leq(d_2,\compOpLeq_2)+(d_4,\compOpLeq_4)$ is valid for $\paramR$.
}

\begin{proof}
\versionProofIn{
\prooflemmasumvalid{}
}
\versionProofOut{
See \cref{appendix:proof:lemmasumvalid}.
}
\end{proof}

\journalVersion{
}

We can now define our data structure, namely \mPDBMs{}\longVersion{~(for \textit{precise Parametric Difference Bound Matrices})},
inspired by the PDBMs of~\cite{HRSV02}\journalVersion{; PDBMs were} themselves inspired by \journalVersion{the }DBMs\journalVersion{~of}~\cite{Dill89}. %
However, our \mPDBM{} compare differences of \emph{fractional parts} of clocks, instead of clocks as in classical DBMs; therefore, our \mPDBMs{} are closer to
clock regions than to DBMs and \emph{fully contained} into clock regions of~\cite{AD94}.
A \mPDBM{} is a pair made of an integer vector (encoding the clocks integer part), and a matrix (encoding the parametric differences between any two clock fractional parts).
Their interpretation also follows that of PDBMs and DBMs:
for $i\neq 0$, the matrix cell $\Dio=(\dio,\compOpLeq_{i0})$ is interpreted as the constraint $\partieFrac(\clock_{i})\compOpLeq_{i0} \dio$,
and $\Doi=(\dzeroi,\compOpLeq_{0i})$ as the constraint $-\partieFrac(\clock_i) \compOpLeq_{0i} \dzeroi$.
For $i\neq 0$ and $j\neq 0$,
the matrix cell $\Dij=(\dij,\compOpLeq_{ij})$ is interpreted as $\partieFrac(\clock_{i})-\partieFrac(\clock_{j}) \compOpLeq_{ij} \dij$.
Finally for all~$i$, $\Dii=(0,\leq)$.

Our \mPDBMs{} are partitioned into two types: \oPDBMs{} and \pointPDBMs{}.
A \pointPDBM{} is a clock region defined by only parameters which contains only one clock valuation;
that is, it corresponds to a set of inequalities of the form $\clock_i \leq \param_j \wedge \param_j \leq \clock_i$.
In contrast, an \oPDBM{} is a clock region which can contain several clock valuations satisfying some possibly parametric constraints,
or contain at least one clock valuation satisfying non-parametric constraints (as the corner-point of~\cite{AD94}).
In particular, the initial clock region $\{0^{\ClockCard}\}$ and any clock region $\{\E_{i}^{\ClockCard}\}$ where $\E_i$ is an integer for all clock $\clock_i$, is an \oPDBM{}.

Basically, only the first \mPDBM{} after a (necessarily total) parametric clock update will be a \pointPDBM{}; any following \mPDBM{} will be an \oPDBM{} until the next (total) parametric update.
The following two definitions impose several conditions to \mPDBMs{} that ensure we build satisfiable ones.

\begin{defi}[\oPDBM{}]\label{def:med-PDBM} %
	Let $\paramR$ be a parameter region.
	An \oPDBM{} for $\paramR$ is a pair $(\E,\D)$ with $\E = (\E_{1}, \dots, \E_{\ClockCard})$ a vector of $\ClockCard$ integers (or $\infty$) which is the integer part of each clock,
		and $\D$ is an ${(\ClockCard+1)}^{2}$ matrix where each element $\Dij$ is a pair $(\dij, \compOpLeq_{ij})$ for all $0\leq i,j\leq\ClockCard$,
        where $\dij \in \PLT$. Moreover, for all $0\leq i \leq\ClockCard$, $\Dii=(0,\leq)$.
    In addition:
	\begin{enumerate}
		\item\label{PDBMi}For all $i$, $(-1,<)\leq\Doi\leq (0,\leq)$ and $(0,\leq)\leq\Dio\leq (1,<)$ are valid for $\paramR$,
		\item\label{PDBMv} For all $i\neq 0,j\neq 0$, either $(0,\leq)\leq\Dij\leq (1,<)$ is valid for $\paramR$ and $(-1,<)\leq\Dji\leq (0,\leq)$ is valid for $\paramR$ or $(0,\leq)\leq\Dji\leq (1,<)$ is valid for $\paramR$ and $(-1,<)\leq\Dij\leq (0,\leq)$ is valid for $\paramR$.
		\item\label{PDBMvi} For all $i,j$, if~$\dij=-\dji$ and is different from~$1$ then $\compOpLeq_{ij}=\compOpLeq_{ji}={\leq}$, else $\compOpLeq_{ij}=\compOpLeq_{ji}={<}$,
		\item\label{PDBMii} For all $i,j,k$, $\Dij\leq\Dik+\Dkj$ is valid for $\paramR$ (canonical form), and
		\item
			\begin{enumerate}
				\item\label{PDBMiii} There is at least one $i$ \st{} $\Dio=\Doi=(0,\leq)$, or
				\item\label{PDBMiv} there is at least one $i$ \st{} $\Dio=(1,<)$ and for all $j$ \st{} $\Doj=(0,\compOpLeq_{0j})$, then we have ${\compOpLeq_{0j}} = {<}$.
			\end{enumerate}

	\end{enumerate}
\end{defi}

\noindent
Condition~\refc{PDBMi} ensures fractional parts of clocks valuations have only non negative values. Condition~\refc{PDBMv} ensures the consistency of differences of clocks \ie{} $\partieFrac(\clock)-\partieFrac(\clockk)\leq 0$ iff $0\leq\partieFrac(\clockk)-\partieFrac(\clock)$. Condition~\refc{PDBMvi} ensures the only possible closed sets of clock valuations are parametric singleton of clock valuations. Condition~\refc{PDBMii} is the canonical form which ensures, as described in~\cite{HRSV02,BY03}, that the \oPDBM{} has the tightest possible bounds \ie{} no constraint $\partieFrac(\clock)-\partieFrac(\clockk) \compOpLeq_{\clock\clockk} \dxy$ can be strengthened without losing solutions.

An \oPDBM{} satisfying condition~\ref{PDBMiii} can be seen as a subregion of an open line segment or a corner point region of~\cite[fig.~9 example 4.4]{AD94} (it can be seen as a \emph{border} region)
and one satisfying condition~\ref{PDBMiv} can be seen as a subregion of an open region of~\cite[fig.~9 example 4.4]{AD94} (it can be seen as a \emph{center} region).
Remark that sets of the form~$\{\partieFrac(\clockval(\clock))\mid 0\leq \partieFrac(\clockval(\clock))\leq 1\}$ are forbidden by \crefDef{def:med-PDBM}~(\ref{PDBMvi}), as in the regions of~\cite{AD94}.

Let $\paramR$ be a parameter region.
In the following, \OPDBMs{} is the set of all possible \oPDBMs{} $(\E,\D)$ for $\paramR$.
\journalVersion{This definition is similar to that of~\cite[def. 3.1]{HRSV02}.}

The second type is the~\pointPDBM{}. It represents the unique clock valuation (for a given parameter valuation) obtained after a total parametric update in an~\RtoPPTA{}.

\begin{defi}[\pointPDBM{}]\label{definition:pointPDBM}
    Let $\paramR$ be a parameter region.
    A \pointPDBM{} for $\paramR$ is a pair $(\E,\D)$ where $\D$ is an  ${(\ClockCard+1)}^{2}$ matrix
    where each element $\Dij$ is a pair $(\dij, \leq)$ and for all $0\leq i,j\leq\ClockCard$,
    $\dio=\partieFrac(\param_{1})=-\dzeroi$,
    and $\dij=\partieFrac(\param_{1})-\partieFrac(\param_{2})=-\dji$,
    for any $\param_{1}, \param_{2}\in\Param$. %
    and for all $1 \leq i \leq \ClockCard$, $\E_i=\lfloor\param_k\rfloor$ if $\dio=\partieFrac(\param_k)$, for $1 \leq k \leq \ParamCard$.
    In addition:
	\begin{enumerate}
		\item\label{PDBM2i}For all $i$, $(-1,<)\leq\Doi\leq (0,\leq)$ and $(0,\leq)\leq\Dio\leq (1,<)$ are valid for $\paramR$,%
		\item\label{PDBM2ii} For all $i,j,k$, $\Dij\leq\Dik+\Dkj$ is valid for $\paramR$ (canonical form).
	\end{enumerate}
 \end{defi}

\noindent
The fact that $\D$ is antisymmetric \ie{} for all $i, j$, $\Dij=-\Dji$,
means that each clock is valuated to a parameter and each difference of clocks is valuated to a difference of parameters. Conditions \refc{PDBM2i} and \refc{PDBM2ii} are the same as for \oPDBMs{}.

The set of all \pointPDBM{} for $\paramR$ is denoted by \CPr, and the set of all \mPDBMs{} for $\paramR$ \journalVersion{is denoted }by $\PDBMRp{}$ (hence $\PDBMRp = \OPDBMs \cup \CPr$).

The use of \emph{validity} ensures the consistency of the \mPDBM{}. We denote the set of all \mPDBMs{} that are \emph{valid for} $\paramR$ by $\PDBMRp{}$.
Given a \mPDBM~$(\E,\D)$, it defines the subset of~$\grandr^\ClockCard\cup\grandq^\ParamCard$
satisfying the constraints $\bigwedge_{i, j \in [0, \ClockCard]}
	\partieFrac(\clock_i)-\partieFrac(\clock_j)\compOpLeq_{i,j}\dij	\land
	\bigwedge_{i \in [1, \ClockCard]} \lfloor\clock_i\rfloor = \E_i\text{.}$
Given a \mPDBM~$(\E,\D)$ and a parameter valuation~$\pval$, we denote by~$(\E,\valuate{\D}{\pval})$ the \emph{valuated \mPDBM{}}, \ie{} the set of clock valuations defined by\journalVersion{~the inequalities}:
\[\bigwedge_{i, j \in [0, \ClockCard]}
	\partieFrac(\clock_i)-\partieFrac(\clock_j)\compOpLeq_{i,j}\pval(\dij)
	\land
	\bigwedge_{i \in [1, \ClockCard]} \lfloor\clock_i\rfloor = \E_i\text{.}
\]
For a clock valuation~$\clockval$, we write $\clockval\in(\E,\valuate{\D}{\pval})$ if it satisfies all constraints of~$(\E,\valuate{\D}{\pval})$%
\journalVersion{\footnote{%
If $\pval$ is a valuation assigning an \emph{integer} to each parameter, then $(\E,\valuate{\D}{\pval})$ is DBM as defined in~\cite{BY03}.
}}.

The following two lemmas derive from the above definitions of \pointPDBM{} and \mPDBMs{}:

\newcommand{\lemmaDijDji}{
    Let $\paramR$ be a parameter region and $(\E,\D)$ be a \mPDBM{} for $\paramR$.
  For all clocks $i,j$, $(0,\leq)\leq\Dij+\Dji$ is valid for $\paramR$.
  }
\begin{lem}[positivity of reflexivity]\label{lemmaDijDjiGeq}
  \lemmaDijDji{}
\end{lem}

\newcommand{\prooflemmaDijDjiGeq}{
	By condition~(\refc{PDBMii}) in \crefDef{def:med-PDBM} and \crefDef{definition:pointPDBM}~(\refc{PDBM2ii}), we have that $\Dii\leq\Dij+\Dji$ is valid for $\paramR$;
	the result follows from the fact that $\Dii=(0,\leq)$ (again from \crefDef{def:med-PDBM} and \crefDef{definition:pointPDBM}).
  }

  \begin{proof}
  \versionProofIn{
  \prooflemmaDijDjiGeq{}
  }
  \versionProofOut{
  See \cref{appendix:proof:lemmaDijDjiGeq}.
  }
\end{proof}

\newcommand{\lemmaDiiDij}{
    Let $\paramR$ be a parameter region and $(\E,\D)$ be a \mPDBM{} for $\paramR$.
    For all clocks $i,j$, $\Dij\leq\Dij+\Djj$ and  $\Dij\leq\Dii+\Dij$ are valid for $\paramR$.
  }
\begin{lem}[neutral element of the set of cells]\label{lemmaDiiDijGeq}
  \lemmaDiiDij%
\end{lem}

\newcommand{\prooflemmaDiiDijGeq}{%
\label{proof:lemmaDiiDijGeq}
	Let~$\paramR$ be a parameter region and~$(\E,\D)$ be a \mPDBM{} for~$\paramR$.
	Let~$\Dij=(\dij,\compOpLeq_{ij})$~with $\dij\in\PLT$.
	By \crefDef{def:med-PDBM} and \crefDef{definition:pointPDBM} for all clock~$i$, $\Dii=(0,\leq)$.
	We have $\Dji+\Dii=(\dji + 0, \compOpLeq_{ij} \oplus \leq)=\Dji$.
	Moreover from \crefDef{def:validity}~(\refc{valideiiid}) $\Dij\leq\Dij$ is valid for~$\paramR$.
	Hence~$\Dij\leq\Dii+\Dij$ is valid for~$\paramR$.
	The same way we prove~$\Dij\leq\Dij+\Djj$ is valid for~$\paramR$.
  }

  \begin{proof}
  \versionProofIn{
  \prooflemmaDiiDijGeq{}
  }
  \versionProofOut{
  See \cref{appendix:proof:lemmaDiiDijGeq}.
  }
\end{proof}

But let us first clarify our needs graphically.
Intuitively, our \mPDBMs{} are partitioned into three types.

(1) The \pointPDBM{} is a clock region defined by \emph{only parameters} which contains only one clock valuation;
it represents the unique clock valuation (for a given parameter valuation) obtained after a total parametric update in an~\RtoPPTA{}.
Each clock is valuated to a parameter and each difference of clocks is valuated to a difference of parameters (it corresponds to constraints of the form $\clock = \param$ and~$\clock-\clockk=\param_i-\param_j$).

Let $\pval$ be a parameter valuation. We assume $\lfloor\pval(\param_2)\rfloor=\lfloor\pval(\param_1)\rfloor=k\in\grandn$ and $\partieFrac(\pval(\param_1))>\partieFrac(\pval(\param_2))$.
The \mPDBM{} obtained after an update $\resetfun(\clock)=\pval(\param_2)$
and $\resetfun(\clockk)=\pval(\param_1)$ is represented using the following pair (where the indices $\mathbf{0}, \mathbf{\clock}, \mathbf{\clockk}$ are shown for the sake of comprehension)
{\scriptsize
\[\hspace{-1.5em}(\E,\D)=\Big(	\begin{pmatrix}
k \\
k
\end{pmatrix},
\begin{pmatrix}
                    & \mathbf{0}                         & \mathbf{\clock}                                     & \mathbf{\clockk}\\
\mathbf{0}          & (0,\leq)                           & (-\partieFrac(\param_2),\leq)                       & (-\partieFrac(\param_1),\leq)\\
\mathbf{\clock}     & (\partieFrac(\param_2),\leq)       & (0,\leq)                                            & (\partieFrac(\param_2)-\partieFrac(\param_1),\leq) \\
\mathbf{\clockk}    & (\partieFrac(\param_1),\leq)       & (\partieFrac(\param_1)-\partieFrac(\param_2),\leq)  & (0,\leq)
\end{pmatrix}
\Big)\] }%

 \begin{wrapfigure}{r}{0.43\textwidth}
 	\vspace{-1em}

 	\scalebox{0.75}{
    \begin{tikzpicture}
    \filldraw
    (-3,4) circle (0pt) node[align=left,   left] {$\clockk$} --
    (-3,3) circle (0pt) node[align=left,   left] {$(k,k+1)$} --
    (-3,2.3) circle (0pt) node[align=left,   left] {$\partieFrac(\pval(\param_1))$} --
    (-3,0) circle (0pt) node[align=left,   left] {$(k,k)$} --
    (-2,0) circle (0pt) node[align=left,   below] {$\partieFrac(\pval(\param_2))$} --
    (0,0) circle (0pt) node[align=left,   below] {$(k+1,k)$} --
    (1,0) circle (0pt) node[align=right,   below] {$\clock$};
    \draw (-3,3) -- (-3,0) -- (0,0)-- (0,3) -- cycle;
    \draw (-3,0) -- (0,3);

    \draw (-2.3,3) circle (0pt) node[align=right,   above right] {$1-\partieFrac(\pval(\param_1))$} ;
    \draw[red, thick] (-3,2.3) -- (-2.3,3) ;
    \filldraw[blue]  (-3,2.3) circle (2pt) node {};

    \draw [dashed] (-3,2.3)-- (-2,2.3) -- (-2,0);
    \filldraw  (-2,2.3) circle (2pt) node {};

    \end{tikzpicture}
    }
		\caption{Graphical representations of \mPDBMs{} and~\cite{AD94} regions}%
		\label{fig:exAD}
\end{wrapfigure}
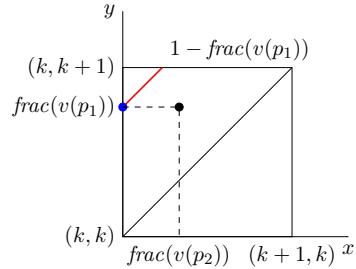

\noindent Once valuated with~$\pval$, it contains a unique clock valuation. We represent it as the black dot in \cref{fig:exAD}.

(2) In contrast, an \oPDBM{} satisfying condition~(\refc{PDBMiii}) is a clock region which can contain several clock valuations satisfying some possibly parametric constraints,
or contain at least one clock valuation satisfying non-parametric constraints (as the corner-point region of~\cite{AD94}).
In particular, the initial clock region $\{0^{\ClockCard}\}$ and any clock region
that is a single integer clock valuation
is a \mPDBM{}.
An \oPDBM{} satisfying condition~\ref{PDBMiii} is characterized by at least one clock~$\clock$ \st{} $\Dxo=\Dox=(0,\leq)$ and can be seen as a subregion of an open line segment or a corner point region of~\cite[fig.~9 example 4.4]{AD94}.
After an immediate update of~$\clock$ to~$k$, the above \mPDBM{}~$(E,D)$ becomes
{\scriptsize
\[\hspace{-1.5em}(\E,\D)=\Big(	\begin{pmatrix}
k \\
k
\end{pmatrix},
\begin{pmatrix}
                    & \mathbf{0}                         & \mathbf{\clock}                                     & \mathbf{\clockk}\\
\mathbf{0}          & (0,\leq)                           & (0,\leq)                                            & (-\partieFrac(\param_1),\leq)\\
\mathbf{\clock}     & (0,\leq)                           & (0,\leq)                                            & (-\partieFrac(\param_1),\leq) \\
\mathbf{\clockk}    & (\partieFrac(\param_1),\leq)       & (\partieFrac(\param_1),\leq)                           & (0,\leq)
\end{pmatrix}
\Big)\] }%

\noindent We represent it once valuated with~$\pval$ as the blue dot in \cref{fig:exAD}. The open line segment of~\cite[fig.~9 example 4.4]{AD94} can be represented as
{\scriptsize
\[\hspace{-1.5em}\Big(	\begin{pmatrix}
k \\
k
\end{pmatrix},
\begin{pmatrix}
                    & \mathbf{0}                         & \mathbf{\clock}                                     & \mathbf{\clockk}\\
\mathbf{0}          & (0,\leq)                           & (0,\leq)                                            & (0,<)\\
\mathbf{\clock}     & (0,\leq)                           & (0,\leq)                                            & (0,<) \\
\mathbf{\clockk}    & (1,<)                              & (1,<)                                               & (0,\leq)
\end{pmatrix}
\Big)\] }%

\noindent and is depicted as the vertical left black line in \cref{fig:exAD}.

(3) An \oPDBM{} satisfying condition~(\refc{PDBMiv}) is a clock region which can contain several clock valuations satisfying some possibly parametric constraints
(as the open region of~\cite{AD94}).
An \oPDBM{} satisfying condition~(\refc{PDBMiv}) is characterized by at least one clock~$\clockk$ \st{} $\Dyo=(1,<)$ and for all~$\clock$ \st{} $\Dox=(0,\compOpLeq_{ox})$, then we have ${\compOpLeq_{ox}} = {<}$ and can be seen as a subregion of an open region of~\cite[fig.~9 example 4.4]{AD94}.
After some time elapsing, and \emph{before} any clock valuation reaches the next integer~$k+1$---therefore the next \oPDBM{} satisfying condition~\ref{PDBMiii}---, the above \mPDBM{}~$(E,D)$ becomes
{\scriptsize
\[\hspace{-1.5em}(\E,\D)=\Big(	\begin{pmatrix}
k \\
k
\end{pmatrix},
\begin{pmatrix}
                    & \mathbf{0}                         & \mathbf{\clock}                                     & \mathbf{\clockk}\\
\mathbf{0}          & (0,\leq)                           & (0,<)                                               & (-\partieFrac(\param_1),<)\\
\mathbf{\clock}     & (1-\partieFrac(\param_1),<)        & (0,\leq)                                            & (-\partieFrac(\param_1),\leq) \\
\mathbf{\clockk}    & (1,<)                              & (\partieFrac(\param_1),\leq)                        & (0,\leq)
\end{pmatrix}
\Big)\] }%

\noindent We represent it once valuated with~$\pval$ as the red line in \cref{fig:exAD}. The open region of~\cite[fig.~9 example 4.4]{AD94} can be represented as
{\scriptsize
\[\hspace{-1.5em}\Big(	\begin{pmatrix}
k \\
k
\end{pmatrix},
\begin{pmatrix}
                    & \mathbf{0}                         & \mathbf{\clock}                                     & \mathbf{\clockk}\\
\mathbf{0}          & (0,\leq)                           & (0,<)                                               & (0,<)\\
\mathbf{\clock}     & (1,<)                              & (0,\leq)                                            & (0,<) \\
\mathbf{\clockk}    & (1,<)                              & (1,<)                                               & (0,\leq)
\end{pmatrix}
\Big)\] }%

\noindent and is depicted as the top left black triangle in \cref{fig:exAD}.\ea{c'est quoi ce top left triangle ? celui délimité par la ligne rouge?} 

Remark that sets of the form~$\{\partieFrac(\clockval(\clock))\mid 0\leq \partieFrac(\clockval(\clock))\leq 1\}$ are in contradiction with \crefDef{def:med-PDBM}~(\ref{PDBMvi}) and therefore cannot be part of a \mPDBM{}, as in the regions of~\cite{AD94}.
Basically, only the first \mPDBM{} after a (necessarily total) parametric clock update will be a \pointPDBM{}; any following \mPDBM{} will be a \oPDBM{} satisfying condition~\ref{PDBMiii} or~\ref{PDBMiv} until the next (total) parametric update.

The differentiation made in the previous paragraph between \oPDBMs{} satisfying condition~\ref{PDBMiii} and~\ref{PDBMiv} is intended to give an intuition to the reader about the inclusion of \mPDBMs{} into~\cite{AD94} clock regions.
Technical details are given in the following \cref{section:operations}.
In the following subsections \cref{section:reset,section:preset,section:te,section:npg,section:pg}, we are going to define operations on~\mPDBMs{} (\ie{} update of clocks, time elapsing and guards satisfaction), and will show that the set of~\mPDBMs{} is stable under these operations.

\section{Operations on \texorpdfstring{\mPDBMs{}}{p-PDBMs}}\label{section:operations}
\subsection{Non-parametric update}\label{section:reset}

To apply a non-parametric update on a \mPDBM{}, %
following classical algorithms for DBMs~\cite{BY03}, we define an update operator, given in \cref{algorithm:reset}\ea{réf dynamique manquante}.
Given a \mPDBM{} $(\E,\D)$ and $\resetfunnp$ a non-parametric update function that updates a clock~$\clock$ to~$k\in\grandn$,
$\resetF((\E, \D), \resetfunnp)$ defines a new \mPDBM{} by
\begin{ienumeration}
\item updating~$\E_{\clock}$ to~$k$;
\item setting the fractional part of~$\clock$ to~$0$: $\Dxo\r\Dox\r(0,\leq)$;
\item updating the new difference between fractional parts with all other clocks~$i$, which is the range of values~$i$ can currently take: $\Dxi\r\Doi$ and $\Dix\r\Dio$.
\end{ienumeration}

\begin{exa}
Here is an \oPDBM{} satisfying condition~\ref{PDBMiv} on the left of the figure below. Formally, it is written:
{\scriptsize
\[\hspace{-1.5em}(\E,\D)=\Big(	\begin{pmatrix}
k \\
k
\end{pmatrix},
\begin{pmatrix}
            & \mathbf{0} 									& \mathbf{\clock} 								& \mathbf{\clockk}\\
\mathbf{0} 				& (0,\leq) 										& (-\partieFrac(\param_2),<) 						& (-\partieFrac(\param_1),<)\\
\mathbf{\clock}				& (\partieFrac(\param_2)+1-\partieFrac(\param_1),<) 	& (0,\leq) 										& (-\partieFrac(\param_1)+\partieFrac(\param_2),\leq) \\
\mathbf{\clockk} 			& (1,<) 										& (\partieFrac(\param_1)-\partieFrac(\param_2),\leq) 	& (0,\leq)
\end{pmatrix}
\Big)\] }%

\noindent  After an update of~$\clockk$ to~$k$ prior to reaching~$k+1$, here is the \oPDBM{} satisfying condition~\ref{PDBMiii} obtained, on the right of the figure below. Formally, it is written:
{\scriptsize
\[\hspace{-1.5em}(\E,\D)=\Big(	\begin{pmatrix}
k \\
k
\end{pmatrix},
\begin{pmatrix}
            & \mathbf{0} 									& \mathbf{\clock} 								& \mathbf{\clockk}\\
\mathbf{0} 				& (0,\leq) 										& (-\partieFrac(\param_2),<) 						& (0,\leq)\\
\mathbf{\clock}				& (\partieFrac(\param_2)+1-\partieFrac(\param_1),<) 	& (0,\leq) 										& (\partieFrac(\param_2)+1-\partieFrac(\param_1),<) \\
\mathbf{\clockk} 			& (0,\leq) 										& (-\partieFrac(\param_2),<)						& (0,\leq)
\end{pmatrix}
\Big)\] }%

\begin{figure*}[h!]
\includegraphics[width=1\linewidth]{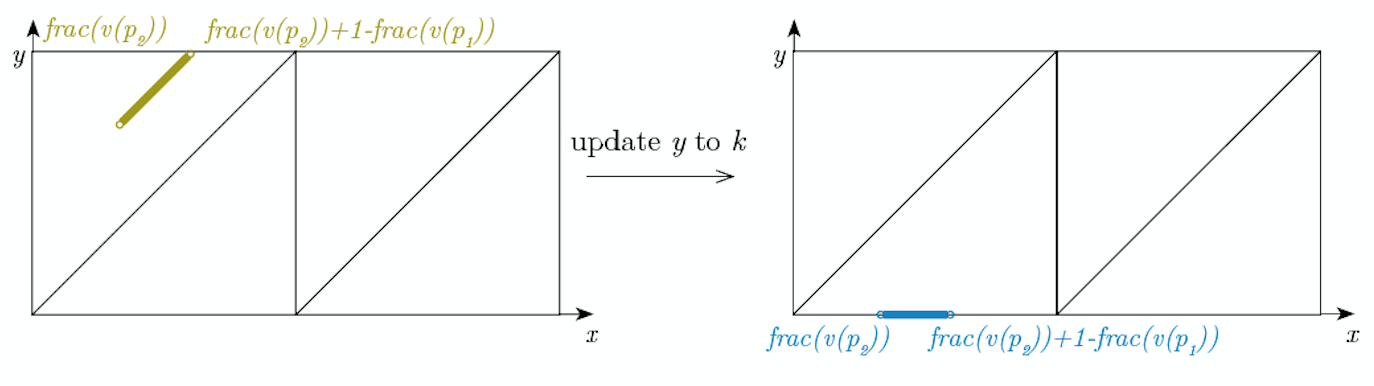}%
\label{fig:pPDBM1}
\vspace{-2em}
\end{figure*}

\end{exa}

\medskip
\begin{algorithm}[H]
  \DontPrintSemicolon%
  \SetAlgoLined%
      \ForEach{$\clock$ where~$\resetfunnp(\clock)$ is defined}{
        $\Dxo\r\Dox=(0,\leq)$\;
        \For{$i$ from $1$ to $\ClockCard$}{
        $\Dxi=\Doi$\;
        $\Dix=\Dio$\;
        }
      }
    \caption{$\resetF(\D,\resetfunnp)$: for all clock~$\clock$ where~$\resetfunnp$ is defined, update $\partieFrac(\clock):=0$}%
  \label{algorithm:reset}
\end{algorithm}

\medskip

\begin{defi}[update of a \mPDBM]\label{defReset}
	Let $\resetfunnp$ be a non-parametric update function.
	Given $(\E, \D) \in \PDBMRp$,
	we define the \emph{update} of~$(\E,\D)$,
	denoted by $(\E',\D')=\resetF((\E, \D), \resetfunnp)$ as: $\D'$ is the result of \cref{algorithm:reset}\ea{réf dynamique manquante; test : \cref{algorithm:reset}}\mr{ok} 
	and for each clock $\clock$ if $\resetfunnp(\clock)$ is defined
	$\E_{\clock}'\r\valuate{\clock}{\resetfunnp}$, $\E_{\clock}'\r\E_{\clock}$ otherwise.
\end{defi}

\newcommand{\lemmaStableReset}{
 	Let $\paramR$ be a parameter region and \[(\E,\D)\in\PDBMRp.\]
	Let $\resetfunnp$ be a non-parametric update.
	Then $\resetF((\E, \D), \resetfunnp)\in\OPDBMs$.
}

\begin{lem}[stability under update]\label{lemmaStableReset}
\lemmaStableReset{}
\end{lem}
\newcommand{\prooflemmaStableReset}{%
\label{proof:lemmaStableReset}

The case of a trivial non-parametric update \ie{} that updates no clock, is straightforward.

We split this proof in two parts: the first one treats the case of~\pointPDBMs{} and the second one of~\oPDBMs{}.

\bigskip

First we show that applying an~$\resetF$ on any~\pointPDBM{} transforms it into an~\oPDBM{}.

\newcommand{\lemmaResetTypei}{
	Let $\paramR$
	be a parameter region and $(\E,\D)\in\CPr$.
	Let $\resetfunnp$ be a non-parametric update.
	Then $\resetF((\E, \D), \resetfunnp)\in\OPDBMs$.
}
\begin{clm}[$\CPr$ becomes \OPDBMs{} after~$\resetF$]\label{T1toT2reset}
  \lemmaResetTypei{}
\end{clm}
\begin{preuve}%
	Let $\paramR$ be a parameter region and $(\E,\D)\in\CPr$. Consider $(\E',\D') = \resetF((\E,\D), \resetfunnp)$.
	After applying \cref{algorithm:reset},\ea{doublon, et référence dynamique manquante !} 
	for all clock~$\clock_i$ of~$(\E,\D)$ where~$\resetfunnp$ is defined,
	$\E_{i}'=\valuate{\clock_{i}}{\resetfunnp}$; moreover for all clock~$j$, $\Dij'=\Doj$ and~$\Dji'=\Djo$.\ea{la seconde partie de la phrase, c'est pas par d\'efinition, c'est apr\`es application de Algorithm 1}
	First note that if $\clock_i, \clock_j$ have been updated, $\Dij'=\Dji'=\Doj'=\Djo'=\Doi'=\Dio'=(0,\leq)=\Doo$.
	\ea{ok, mais tu dis ca pourquoi ?}\mr{je l'utilise dans f, g, h} 
	For all clocks $i,j,k$, the following inequalities are valid for $\paramR$:
	\begin{enumerate}
	\item \begin{enumerate}
		      \item if $\clock_i$ is updated: $\Dio'=(0,\leq)=\Doi'$ and therefore trivially it holds that $-1\leq\Doi'\leq 0$ and $0\leq\Dio'\leq 1$ are valid for~$\paramR$;
		      \item if $\clock_i$ is not updated: $\Dio'=\Dio$ and therefore $-1\leq\Doi'\leq 0$ and $0\leq\Dio'\leq 1$  are valid for~$\paramR$ because these constraints were already satisfied in~$(\E,\D)$.
	      \end{enumerate}
	\item For all~$\clock_i,\clock_j$, if neither~$\clock_i$ nor~$\clock_j$ is updated, $\Dij$ and~$\Dji$ are not modified so condition \crefDef{def:med-PDBM}~(\refc{PDBMv}) still holds.
	If either~$\clock_i$ is updated, as~$\Dij'=\Doj$ and~$\Dji'=\Djo$ condition \crefDef{def:med-PDBM}~(\refc{PDBMv}) still holds as it holds for~$\Doj$ and~$\Djo$ and we apply the same reasoning if~$\clock_j$ is updated.
	If both~$\clock_i,\clock_j$ are updated, condition \crefDef{def:med-PDBM}~(\refc{PDBMv}) trivially holds.
	\item For all~$\clock_i$, if it is updated then~$\Doi'=\Dio'=(0,\leq)$, hence~$\dzeroi=-\dio=0$ and~$\compOpLeq_{0i}=\compOpLeq_{i0}=\leq$; condition \crefDef{def:med-PDBM}~(\refc{PDBMvi}) holds.
	For all~$\clock_i,\clock_j$, if neither~$\clock_i$ nor~$\clock_j$ is updated, $\Dij'=\Dij$ and~$\Dji'=\Dji$ so condition \crefDef{def:med-PDBM}~(\refc{PDBMvi}) holds as it holds for~$\Dij$ and~$\Dji$.
	If either~$\clock_i$ is updated, as~$\Dij'=\Doj$ and~$\Dji'=\Djo$, condition \crefDef{def:med-PDBM}~(\refc{PDBMvi}) holds as it holds for~$\Doj$ and~$\Djo$.
	We treat the case where~$\clock_j$ is updated similarly.
	If both~$\clock_i,\clock_j$ are updated, condition \crefDef{def:med-PDBM}~(\refc{PDBMvi}) trivially holds.
	\item Canonical form is preserved:
	      	      \begin{enumerate}
		      \item if~$\clock_i,\clock_j,\clock_k$ are not updated:
					since no clock is updated we have~$\Dij'=\Dij$, $\Djk'=\Djk$ and $\Dik'=\Dik$
					since~$(\E,\D)\in\CPr$ from \crefDef{definition:pointPDBM} (\refc{PDBM2ii}), we know that~$\Dik \leq \Dij+\Djk$ is valid for~$\paramR$;
					therefore it remains valid.
		      \item if~$\clock_k$ is updated and~$\clock_i, \clock_j$ are not updated: $\Dij'=\Dij$
					and~$\Djk'=\Djo$, $\Dik'=\Dio$ because~$\clock_k$ is updated.
					Since $(\E,\D)\in\CPr$ from \crefDef{definition:pointPDBM} (\refc{PDBM2ii}), we know that~$\Dio \leq \Dij+\Djo$ is valid for~$\paramR$;
					therefore, $\Dik'\leq\Dij'+\Djk'$ is valid for~$\paramR$.%

		      \item if $\clock_j$ is updated and $\clock_i, \clock_k$ are not updated:
					then~$\Dik'=\Dik$ because neither~$\clock_i$ nor~$\clock_k$ are updated;
					since~$\clock_k$ is updated we have~$\Djk'=\Dok$ and $\Dij'=\Dio$;
					since~$(\E,\D)\in\CPr$ from \crefDef{definition:pointPDBM} (\refc{PDBM2ii}), we know that~$\Dik \leq \Dio+\Dok$ is valid for~$\paramR$;
					therefore, $\Dik'\leq\Dij'+\Djk'$ is valid for~$\paramR$.%

		      \item if $\clock_j, \clock_k$ are updated and $\clock_i$ is not updated:
					then~$\Dik'=\Dio$ because~$\clock_k$ is updated;
					since~$\clock_j$ is updated we have~$\Dij'=\Dio$ and $\Djk'=\Doo$;
					since $(\E,\D)\in\CPr$ from \crefDef{definition:pointPDBM} (\refc{PDBM2ii}) and \crefLemma{lemmaDiiDijGeq}, we know that~$\Dio \leq \Dio+\Doo$ is valid for~$\paramR$;
					therefore, $\Dik'\leq\Dij'+\Djk'$ is valid for~$\paramR$.%

		      \item if $\clock_i$ is updated and $\clock_j, \clock_k$ are not updated:
		      			then~$\Dik'=\Dok$, $\Dij'=\Doj$ because~$\clock_i$ is updated;
		      			since~$\clock_j, \clock_k$ are not updated, we have $\Djk'=\Djk$;
		      			since~$(\E,\D)\in\CPr$ from \crefDef{definition:pointPDBM} (\refc{PDBM2ii}), we know that~$\Dok \leq \Doj+\Djk$ is valid for~$\paramR$;
					therefore $\Dik'\leq\Dij'+\Djk'$ is valid for~$\paramR$.%

		      \item if $\clock_i, \clock_k$ are updated and $\clock_j$ is not updated: %
		      			we have~$\Dik'=(0,\leq)=\Doo$, $\Dij'=\Doj$ and $\Djk'=\Djo$ because~$\clock_i, \clock_k$ are updated.
					Since $(\E,\D)\in\CPr$ from \crefDef{definition:pointPDBM} (\refc{PDBM2ii}), we know that~$\Doo \leq \Doj+\Djo$ is valid for~$\paramR$;
					therefore, $\Dik'\leq\Dij'+\Djk'$ is valid for~$\paramR$.%

		      \item if $\clock_i, \clock_j$ are updated and $\clock_k$ is not updated: %
					we have~$\Dik'=\Dok$, $\Dij'=(0<\leq)=\Doo$ and~$\Djk'=\Dok$ because~$\clock_i, \clock_j$ are updated.
					Since $(\E,\D)\in\CPr$ from \crefDef{definition:pointPDBM} (\refc{PDBM2ii}) and \crefLemma{lemmaDiiDijGeq}, we know that~$\Dok \leq \Doo+\Dok$ is valid for~$\paramR$;
					therefore, $\Dik'\leq\Dij'+\Djk'$ is valid for~$\paramR$.%

		      \item if $\clock_i,\clock_j, \clock_k$ are updated: %
					we have~$\Dik'=\Doo$, $\Dij'=\Doo$ and~$\Djk'=\Doo$ because~$\clock_i,\clock_j,\clock_k$ are updated.
					Since $(\E,\D)\in\CPr$ from \crefDef{definition:pointPDBM} (\refc{PDBM2ii}) and \crefLemma{lemmaDiiDijGeq}, we know that~$\Doo \leq \Doo+\Doo$ is valid for~$\paramR$;
					therefore, $\Dik'\leq\Dij'+\Djk'$ is valid for~$\paramR$.%
	      \end{enumerate}
	\item there is at least one clock~$\clock$ \st{} $\Dxo'=\Dox'=(0,\leq)$.\ea{pourquoi $\clock_i$? Dans les cas ci-dessus elle n'est pas forc\'ement r\'einitialis\'ee?}
	\end{enumerate}
	Therefore, $(\E',\D')\in\OPDBMs$.
\end{preuve}
\ea{ok, j'ai relu rapidement et ca semble mieux }

\bigskip

Now we show that applying an~$\resetF$ on any~\oPDBM{} transforms it into an~\oPDBM{} respecting \crefDef{def:med-PDBM}~(\refc{PDBMv}).

\newcommand{\lemmaResetStableTypeii}{
	Let $\paramR$ be a parameter region and $(\E,\D)\in\OPDBMs$.
	Let $\resetfunnp$ be a non-parametric update.
	Then $\resetF((\E, \D), \resetfunnp)\in\OPDBMs$.
}
\begin{clm}[stability of~$\OPDBMs$ under~$\resetF$]%
\label{T2toT2reset}
  \lemmaResetStableTypeii{}
\end{clm}

\begin{preuve}%
Most cases are similar to the proof of \refClaim{T1toT2reset}.

The remaining cases to treat are the cases of \crefDef{def:med-PDBM}~(\refc{PDBMv}).
If $i,j$ are different from~$0$, and
		\begin{enumerate}
			\item if $i,j$ are not updated then $\Dij'=\Dij$ and since it is the case in $(\E,\D)$, condition \crefDef{def:med-PDBM}~(\refc{PDBMv}) holds.
			\item if $j$ is updated and $i$ is not updated then $\Dij'=\Dio$ and $\Dji'=\Doi$
			and as condition \crefDef{definition:pointPDBM}~(\refc{PDBM2i}) holds for~$\Dio$ and~$\Doi$ in~$(\E,\D)$,
			condition \crefDef{def:med-PDBM}~(\refc{PDBMv}) holds in~$(\E',\D')$.
			\item if $i$ is updated and $j$ is not updated then $\Dij'=\Doj$ and $\Dji'=\Djo$
			and as condition \crefDef{definition:pointPDBM}~(\refc{PDBM2i}) holds for~$\Djo$ and~$\Doj$ in~$(\E,\D)$,
			condition \crefDef{def:med-PDBM}~(\refc{PDBMv}) holds in~$(\E',\D')$.
			\item if $i,j$ are updated then trivially $\Dij'=\Dji'=(0,\leq)$ and condition \crefDef{def:med-PDBM}~(\refc{PDBMv}) holds. \qedhere
		\end{enumerate}
\end{preuve}
}

\begin{preuve}
\versionProofIn{
\prooflemmaStableReset{}
}
\versionProofOut{
Intuitively, we update in~$(\E,\D)$ the lower and upper bounds of some clocks to~$(0,\leq)$ and the difference between two clocks~$\Dij$ to~$\Doj$ if~$\clock_i$ is updated: that is, the new difference between two clocks if one has been updated is just the lower/upper bound of the one that is not updated. This allows us to conserve the canonical form as we only ``moved'' some cells in~$\D$ that already verified the canonical form. Therefore $\resetF((\E, \D), \resetfunnp)$ is a \mPDBM{}. \longVersion{See \cref{appendix:proof:lemmaStableReset} for details.}%
}
\end{preuve}

Applying a non-parametric~$\resetF$ on any~\pointPDBM{} transforms it into an~\oPDBM{}, and \oPDBMs{} are stable under~$\resetF$.
It can seem a paradox that the (non-parametric) update of a \pointPDBM{} becomes an \oPDBM{};
in fact, it remains geometrically speaking a point, \ie{} a singleton containing one clock valuation.
Recall that our \oPDBMs{} include \mPDBMs{} geometrically corresponding to a point for each valuation.
In contrast, \pointPDBMs{} are also punctual (for each valuation), but are fully parametric.

The following lemma states that the update operator behaves as expected.

\newcommand{\lemmaResetLong}{%
	Let $\paramR$ be a parameter region and $(\E,\D) \in \PDBMRp$. %
	Let $\pval\in\paramR$.
	Let $\resetfunnp$ be a non-parametric update.
	For all~$\clockval$, $\reset{\clockval}{\resetfunnp}\in\resetF((\E, \valuate{\D}{\pval}), \resetfunnp)$ iff $\clockval\in(\E, \valuate{\D}{\pval})$.
}
\begin{lem}[semantics of $\resetF$ on~$\PDBMRp$]%
\label{ssiReset}
	\lemmaResetLong{}
\end{lem}

\newcommand{\proofssiReset}{%
\label{proofssiReset}

We first treat the case of \oPDBMs{}, the case of \pointPDBMs{} will be handled similarly at the end.
We also prove this lemma for a singleton update (only one clock, say~$\clock_i$) since updating several clocks can be done by applying several singleton updates in a~$0$ delay.

\medskip

\noindent \textbf{$\Longrightarrow$ for \oPDBMs{}}

\medskip

Let $\paramR$ be a parameter region and $(\E,\D)\in\OPDBMs$.
Let $\pval\in\paramR$.
Let $\resetfunnp$ be a non-parametric update which updates $\clock_i$ to an integer $n$ and lets the value of other clocks unchanged.
Consider $(\E',\D')=\resetF((\E, \valuate{\D}{\pval}), \resetfunnp)$
and suppose $\clockval'\in(\E',\D')$.
We want to construct a valuation
$
\clockval\in(\E,\pval(\D))
$ \st{} $
\clockval'=\resetfunnp(\clockval)
$.

Let $\clockval$ be a clock valuation \st{}
for all clock $\clock_j$ where $i\neq j$, $\clockval(\clock_j)=\clockval'(\clock_j)$.
That means that for all $j\neq i$,
\[
\partieFrac(\clockval(\clock_j))\compOpLeq_{j0}\pval(\djo),
\quad
-\partieFrac(\clockval(\clock_j))\compOpLeq_{0j}\pval(\doj)
\quad
\mbox{and}
\quad
\lfloor \clockval(\clock_j)\rfloor=\E_j
\]
hold from \crefDef{defReset} since it is the case in $(\E',\D')$ and these values are left untouched by the update.
Moreover for all $j\neq i$, $k\neq i$,
\[
\partieFrac(\clockval(\clock_j))-\partieFrac(\clockval(\clock_k))\compOpLeq_{jk}\pval(\djk)
\quad
\mbox{and}
\quad
\partieFrac(\clockval(\clock_k))-\partieFrac(\clockval(\clock_j))\compOpLeq_{kj}\pval(\dkj)
\]
again hold from \crefDef{defReset} since it is the case in $(\E',\D')$ and these values are left untouched by the update.

We want a valuation for $\clockval(\clock_i)$ \st{}
\[
\partieFrac(\clockval(\clock_i))\compOpLeq_{i0}\pval(\dio)
\quad
-\partieFrac(\clockval(\clock_i))\compOpLeq_{0i}\pval(\dzeroi)
\quad
\mbox{and}
\quad
\lfloor \clockval(\clock_i)\rfloor=\E_i
\]
hold, and for all $j\neq i$, $k\neq i$,
 \begin{equation}\label{eq:un}
\partieFrac(\clockval(\clock_i))-\partieFrac(\clockval(\clock_j))\compOpLeq_{ij}\pval(\dij)
\quad
\mbox{and}
\quad
\partieFrac(\clockval(\clock_k))-\partieFrac(\clockval(\clock_i))\compOpLeq_{ki}\pval(\dki)
\quad
\end{equation}
hold.
Let us prove that such a valuation~$\clockval$ exists.
We set $\lfloor \clockval(\clock_i)\rfloor=\E_i$.

\bigskip

The following lemma proves transitivity of constraints on clocks with respect to constraints in a \mPDBM{}.
\begin{lem}\label{DjkleqDjiDik}
  Let $\paramR$ be a parameter region and $(\E,\D)\in\PDBMRp$. Let $\pval\in\paramR$.
  Let $\clockval \in (\E, \pval(\D))$.
    For all clocks $i,j,k$,
    $\partieFrac(\clockval(\clock_j))-\partieFrac(\clockval(\clock_k))(\compOpLeq_{ji} \oplus \compOpLeq_{ik})\pval(\dji)+\pval(\dik)$.
\end{lem}
\begin{preuve}
      Let $\paramR$ be a parameter region and $(\E,\D)\in\PDBMRp$. Let $\pval\in\paramR$.
    Let $\clockval \in (\E, \pval(\D))$.

      Since $(\E,\D)\in\PDBMRp$, for all~$i,j,k$ we have from \crefDef{def:med-PDBM}~(\refc{PDBMii}),
      \[
       \Djk\leq\Dji+\Dik
      \]
      is valid for $\paramR$ hence since $\pval\in \paramR$, we have $\pval(\Djk)\leq\pval(\Dji)+\pval(\Dik)$.
      Precisely
      that is $(\pval(\djk),\compOpLeq_{jk})\leq(\pval(\dji),\compOpLeq_{ji})+(\pval(\dik),\compOpLeq_{ik})$ \ie{}
      \[
       (\pval(\djk),\compOpLeq_{jk})\leq(\pval(\dji)+\pval(\dik),\compOpLeq_{ji} \oplus \compOpLeq_{ik})\text{.\ea{rappelle le num\'ero de la d\'ef ?}} 
      \]\mr{en fait c'est juste une addition, il n'y a pas de d\'ef}\ea{c'est vrai}
      For all clocks $j,k$ satisfying constraints of $(\E,\D)$,
      \[
      \partieFrac(\clockval(\clock_j))-\partieFrac(\clockval(\clock_k))\compOpLeq_{jk}\pval(\djk)\text{.\ea{ca vient d'o\`u ca ? et pourquoi y'a pas de $i$ dans l'expression alors que tu le quantifies au-dessus ?}} 
      \]
      \mr{c'est l'interpr\'etation des contraintes de la PDBM}\ea{ok (mets-le), mais ca n'explique toujours pas la pr\'esence du $i$ (ni l'absence du $k$) dans les quantificateurs}
      Then for all~$i,j,k$, either:
      \begin{itemize}
    \item from \crefDef{def:validity}~(\refc{valideiiie}):
          $\pval(\djk)<\pval(\dji)+\pval(\dik)$ and then, regardless of $\compOpLeq_{jk}$ and~$\compOpLeq_{ji} \oplus \compOpLeq_{ik}$
          we have $\partieFrac(\clockval(\clock_j))-\partieFrac(\clockval(\clock_k))(\compOpLeq_{ji} \oplus \compOpLeq_{ik})\pval(\dji)+\pval(\dik)$,\ea{et pourquoi pas l'autre cas du~\ref{valideiiie}??}
          or
    \item from \crefDef{def:validity}~(\refc{valideiiid}):
      \begin{itemize}
      \item $\pval(\djk)\leq\pval(\dji)+\pval(\dik)$ and $\compOpLeq_{jk}={<}$, $ \compOpLeq_{ji} \oplus \compOpLeq_{ik}={\leq}$ and then
      we have \[\partieFrac(\clockval(\clock_j))-\partieFrac(\clockval(\clock_k))(\compOpLeq_{ji} \oplus \compOpLeq_{ik})\pval(\dji)+\pval(\dik),\] or
      \item $\pval(\djk)\leq\pval(\dji)+\pval(\dik)$ and $\compOpLeq_{jk}=\compOpLeq_{ji} \oplus \compOpLeq_{ik}$ and then
      we have \[\partieFrac(\clockval(\clock_j))-\partieFrac(\clockval(\clock_k))(\compOpLeq_{ji} \oplus \compOpLeq_{ik})\pval(\dji)+\pval(\dik)\]
      which completes the proof.
      \end{itemize}
      \end{itemize}
  This completes the proof of \crefLemma{DjkleqDjiDik}.
\end{preuve}

\bigskip

For all $j \neq i$ and $k\neq i$, since $\pval(\Djk)\leq\pval(\Dji)+\pval(\Dik)$ from \crefDef{def:med-PDBM}~(\refc{PDBMii}),
we have $\partieFrac(\clockval(\clock_j))-\partieFrac(\clockval(\clock_k))\compOpLeq_{jk}\pval(\djk)$ and
\[
\partieFrac(\clockval(\clock_j))-\partieFrac(\clockval(\clock_k))(\compOpLeq_{ji} \oplus \compOpLeq_{ik})\pval(\dji)+\pval(\dik)
\]
holds from \crefLemma{DjkleqDjiDik}. Hence
\begin{equation}\label{eq:deux}
\partieFrac(\clockval(\clock_j))-\pval(\dji)(\compOpLeq_{ji} \oplus \compOpLeq_{ik})\partieFrac(\clockval(\clock_k))+\pval(\dik)
\quad
\end{equation}
holds. Note that $\compOpLeq_{ji} \oplus \compOpLeq_{ik}$ is either $\leq$ or~$<$.
Note the following trick is inspired by~\cite[Proof of Lemma 3.5]{HRSV02} and~\cite[Proof of Lemma 3.13]{HRSV02}.
Hence \ea{par ici il faudrait pr\'eciser que ce raisonnement est inspir\'e de~\cite{HRSV02}, non ?} 
\[
I=\{t \in \grandrplus \mid {\partieFrac(\clockval(\clock_j))-\pval(\dji)} \leq {t} \leq {\partieFrac(\clockval(\clock_k))+\pval(\dik)} \mbox{ for all clocks } j,k\}
\]%
is a non empty set.
That means that choosing a $\partieFrac(\clockval(\clock_i))$ with respect to constraints~(\refc{eq:un}), recall that they are
\[
\partieFrac(\clockval(\clock_j))-\partieFrac(\clockval(\clock_i))\compOpLeq_{ji}\pval(\dji)
\quad
\mbox{and}
\quad
\partieFrac(\clockval(\clock_i))-\partieFrac(\clockval(\clock_k))\compOpLeq_{ik}\pval(\dik)
\]
is equivalent to choose a $\partieFrac(\clockval(\clock_i))$ \st{}
\[
\partieFrac(\clockval(\clock_j))-\pval(\dji)\compOpLeq_{ji}\partieFrac(\clockval(\clock_i))
\quad
\mbox{and}
\quad
\partieFrac(\clockval(\clock_i))\compOpLeq_{ik}\partieFrac(\clockval(\clock_k))+\pval(\dik)
\]
which is a nonempty set from formula~(\refc{eq:deux}).
Finally we choose a $\partieFrac(\clockval(\clock_i))\in I$, then $\clockval\in(\E,\pval(\D))$
and it completes the proof.

\medskip

\noindent \textbf{$\Longleftarrow$ for \oPDBMs{}}

\medskip

Let $\paramR$ be a parameter region and $(\E,\D)\in \OPDBMs{}$.
Let $\pval\in\paramR$.
Let $\resetfunnp$ be a non-parametric update which updates $\clock_i$ to an integer $n$ and lets the value of other clocks unchanged.
Consider $(\E',\D')=\resetF((\E, \valuate{\D}{\pval}), \resetfunnp)$.
Now suppose $\clockval\in(\E, \valuate{\D}{\pval})$ and let $\clockval'=\reset{\clockval}{\resetfunnp}$.
\begin{itemize}
\item for $\clock_i$, since $\resetfunnp$ is defined,
$\clockval'(\clock_i)=\resetfunnp(\clock_i)=\E'_{\clock_i}$ (\ie{} $\partieFrac(\clockval'(\clock_i))=0$) by applying~$\resetF$ as defined in \crefDef{defReset}.
By applying~$\resetF$ as defined in \crefDef{defReset}, $\Dio'=\Doi'=(0,\leq)$, hence
\[
-\partieFrac(\clockval'(\clock_i))\compOpLeq_{0i}\valuate{\dzeroi'}{\pval}
\quad
\mbox{and}
\quad
\partieFrac(\clockval'(\clock_i))\compOpLeq_{i0}\valuate{\dio'}{\pval}
\]
hold from \crefDef{defReset} and \refClaim{T1toT2reset}.
Moreover we know that for all $j\neq i$
\begin{equation}\label{eq:trois}
-\valuate{\Dij'}{\pval}=-\valuate{\Doj'}{\pval}
\quad
\mbox{and}
\quad
\valuate{\Dji'}{\pval}=\valuate{\Djo'}{\pval}
\quad
\end{equation}
holds from \crefDef{defReset}, and we also know that
\begin{equation}\label{eq:quatre}
 \partieFrac(\clockval'(\clock_{j}))-\partieFrac(\clockval'(\clock_i))=\partieFrac(\clockval'(\clock_{j}))
 \quad
\end{equation}
since $\partieFrac(\clockval'(\clock_i))=0$.
Hence, combining (\refc{eq:trois}) and (\refc{eq:quatre}), clearly
since
\[
 -\partieFrac(\clockval'(\clock_{j}))\compOpLeq_{0j}\valuate{\doj'}{\pval}
\quad
\mbox{and}
\quad
\partieFrac(\clockval'(\clock_{j}))\compOpLeq_{j0}\valuate{\djo'}{\pval}
\]
hold in~$(\E',\D')$,
\[
 \partieFrac(\clockval'(\clock_{j}))-\partieFrac(\clockval'(\clock_i))\compOpLeq_{ji}\valuate{\dji'}{\pval}
\quad
\mbox{and}
\quad
 \partieFrac(\clockval'(\clock_{i}))-\partieFrac(\clockval'(\clock_j))\compOpLeq_{ij}\valuate{\dij'}{\pval}
\]
hold.

\item for any two clocks $\clock_j, \clock_k$ where $\resetfunnp$ is not defined, $\clockval(\clock_j)=\clockval'(\clock_j)$ and $\clockval(\clock_k)=\clockval'(\clock_k)$. Hence
\[-\valuate{\Doj'}{\pval}\compOpLeq_{0j}\partieFrac(\clockval'(\clock_j))\compOpLeq_{j0}\valuate{\Djo'}{\pval}
 \]
and
\[
-\valuate{\Dkj'}{\pval}\compOpLeq_{kj}\partieFrac(\clockval'(\clock_{j}))-\partieFrac(\clockval'(\clock_k))\compOpLeq_{jk}\valuate{\Djk'}{\pval}
\]
hold from \crefDef{defReset} and \refClaim{T1toT2reset} since bounds remain unchanged.
\end{itemize}
Then $\clockval'\in\resetF((\E, \valuate{\D}{\pval}), \resetfunnp)$.

This concludes the case $(\E,\D) \in \OPDBMs$.\ea{j'ai rajout\'e ca}

\bigskip

Let us now treat the case $(\E,\D)\in\CPr$.

\ea{note pour moi : j'ai relu la preuve de ce \crefLemma{ssiReset} jusque-l\`a} 

\medskip

\noindent \textbf{$\Longrightarrow$ for \pointPDBMs{}}

\medskip

Let $\paramR$ be a parameter region and $(\E,\D)\in\CPr$.
Let $\pval\in\paramR$.
Let $\resetfunnp$ be a non-parametric update which updates $\clock_i$ to an integer $n$ and lets the value of other clocks unchanged.
Consider $(\E',\D')=\resetF((\E, \valuate{\D}{\pval}), \resetfunnp)$
and suppose $\clockval'\in(\E',\D')$.
We want to construct a valuation
\[
\clockval\in(\E,\pval(\D))
\quad
\mbox{\st{}}
\quad
\clockval'=\resetfunnp(\clockval)
\]
Let $\clockval$ be a clock valuation \st{}
for all clock $\clock_j$ where $j\neq i$, $\clockval(\clock_j)=\clockval'(\clock_j)$.
That means for all $j\neq i$,
\[
\partieFrac(\clockval(\clock_j))\compOpLeq_{j0}\pval(\djo),
\quad
-\partieFrac(\clockval(\clock_j))\compOpLeq_{0j}\pval(\doj)
\quad
\mbox{and}
\quad
\lfloor \clockval(\clock_j)\rfloor=\E_j
\]
hold from \crefDef{defReset} since it is the case in $(\E',\D')$ and bounds remain unchanged \ie{} $\Doj=\Doj'$ and $\Djo=\Djo'$.
Moreover for all $k\neq i$ and $k\neq j$,
\[
\partieFrac(\clockval(\clock_j))-\partieFrac(\clockval(\clock_k))\compOpLeq_{jk}\pval(\djk)
\quad
\mbox{and}
\quad
\partieFrac(\clockval(\clock_k))-\partieFrac(\clockval(\clock_j))\compOpLeq_{kj}\pval(\dkj)
\]
also hold from \crefDef{defReset} since it is the case in $(\E',\D')$ and bounds remain unchanged \ie{} $\Dkj=\Dkj'$ and $\Djk=\Djk'$.

Recall that $(\E,\D)$ contains only one clock valuation for each parameter valuation $\pval\in\paramR$.

Let $\partieFrac(\clockval(\clock_i))=\pval(\dio)$ (or equivalently $\partieFrac(\clockval(\clock_i))=-\pval(\dzeroi)$
since by \crefDef{definition:pointPDBM} we have $(\dio,\compOpLeq_{i0})=(-\dzeroi,\compOpLeq_{0i})$).
Then, as it is the case in~$(\E,\D)$,\ea{prime}
\[
\partieFrac(\clockval(\clock_i))\compOpLeq_{i0}\pval(\dio),
\quad
-\partieFrac(\clockval(\clock_i))\compOpLeq_{0i}\pval(\dzeroi)
\quad
\mbox{and}
\quad
\lfloor \clockval(\clock_i)\rfloor=\E_i
\]
hold, and for all $j\neq i$, $k\neq i$,
\[
\partieFrac(\clockval(\clock_i))-\partieFrac(\clockval(\clock_j))\compOpLeq_{ij}\pval(\dij)
\quad
\mbox{and}
\quad
\partieFrac(\clockval(\clock_k))-\partieFrac(\clockval(\clock_i))\compOpLeq_{ki}\pval(\dki)
\]
hold, which completes the proof, as~$\clockval\in(\E,\pval(\D))$ and~$\clockval'=\resetfunnp(\clockval)$.

\medskip

\noindent \textbf{$\Longleftarrow$ for \pointPDBMs{}}

\medskip

This case is straightforward and similar to the case $(\Leftarrow)$ above of \oPDBMs{}.
\ea{relu assez en d\'etails}
}

\begin{preuve}
\versionProofIn{
\proofssiReset{}
}
\versionProofOut{
	The technical part is~$(\Rightarrow)$. The idea is to prove that, given~$\clockval'\in\resetF((\E, \valuate{\D}{\pval}), \resetfunnp)$ there is a non-empty set of clock valuations~$\clockval$ \st{} $\clockval'=\reset{\clockval}{\resetfunnp}$ that is precisely defined by the constraints in~$(\E, \valuate{\D}{\pval})$.
	\longVersion{See \cref{appendix:proofssiReset} for details.}%
	}
\end{preuve}

\subsection{Parametric update}\label{section:preset}

Given~$(\E,\D)\in\PDBMRp$ we write $\resetpF((\E, \D), \resetfun)$ to denote the update of $(E,D)$ by $\resetfun$, when $\resetfun$ is a total parametric update function, \ie{} updating the set of clocks exclusively to parameters.
We therefore obtain a \pointPDBM{}, containing the parametric constraints defining a unique clock valuation. The semantics is straightforward.
	Recall that a total update function which is not fully parametric (\ie{} an update of some clocks to parameters and some others to constants) can be encoded as a total parametric update immediately followed by a partial non-parametric update function.

\subsection{Time elapsing}\label{section:te}

Given a parameter region~$\paramR$, recall that constraints satisfied by parameters are known, and we can order elements of~$\PLT$.
Thanks to this order, within a~\mPDBM{} $(\E,\D)$
the clocks with the (possibly parametric) largest fractional part
  \ie{} the clocks that have a larger fractional part than any other clock,
  can always be identified by their bounds in~$\D$.
For a \mPDBM{} $(\E,\D)$, we define the set of clocks with the largest fractional part~($\LFP$) as $\LFPd{\D}=\{\clock \in [1,\ClockCard] \mid (0,\leq)\leq\Dxi$ is valid for $\paramR$, for all $0\leq i\leq\ClockCard \}$.
Clocks belonging to $\LFP$ are the first to reach the upper bound~$1$ by letting time elapse.

\begin{defi}[clocks with the largest fractional part in a \mPDBM{}]%
\label{def:largestclock}
	Let $\paramR$ be a parameter region and $(\E,\D)\in\PDBMRp$.
	A \emph{clock with the (possibly parametric) largest fractional part} is a clock $\clock$
    \st{} for all $0 \leq i\leq\ClockCard$, $(0,\leq) \leq\Dxi$ is valid for $\paramR$.\dl{set of clocks?}\mr{en effet}
\end{defi}

There is at least one clock with the (possibly parametric) largest fractional part:

\newcommand{\lemmaLargestClock}{
	Let $\paramR$ be a parameter region and $(\E,\D)\in\PDBMRp$.
	There is at least one clock~$\clock$ \st{} for all $0\leq i\leq\ClockCard$, $(0,\leq) \leq\Dxi$ is valid for $\paramR$.
}
\begin{lem}[existence of a clock with the largest fractional part]\label{lemmaLargestFracClock}
  \lemmaLargestClock{}
\end{lem}

\newcommand{\prooflemmaLargestFracClock}{%
\label{proof:lemmaLargestFracClock}
Reductio ad absurdum: %
    Let $\paramR$ be a parameter region and $(\E,\D)\in\OPDBMs$ with at least 2 clocks $i,j$.
	Suppose for all clock $\clock_i$ there is another clock $\clock_j$ \st{} $\Dij<0$ is valid for $\paramR$.
	Let $\pval\in\paramR$. Then $\pval(\Dij)<0$.
	\begin{itemize}
		\item Suppose for $\clock_j$, $\clock_i$ is the clock \st{} $\Dji<0$ is valid for $\paramR$.
		Then $\pval(\Dji)<0$.
		We have $\pval(\Dij)+\pval(\Dji)<0$ holds, therefore $0\leq \pval(\Dij) + \pval(\Dji)$ does not hold, and hence $0\leq\Dij+\Dji$ is not valid for $\paramR$.
	Then~$(\E,\D)$ does not respect \crefLemma{lemmaDijDjiGeq} and violates condition~(\refc{PDBMii}) of \crefDef{def:med-PDBM}. So~$(\E,\D)\not\in\OPDBMs$.
		\item Suppose for $\clock_j$, a third clock $\clock_k$ is the clock \st{} $\Djk<0$ is valid for $\paramR$. Then $\pval(\Djk)<0$.
	Suppose we have only three clocks. Then for $\clock_k$, either $\clock_i$ or $\clock_j$ is the clock \st{} $\Dki<0$ is valid for $\paramR$.
		\begin{itemize}
			\item Assume this is $\clock_i$.
				Then $\pval(\Dki)<0$.
				We have $\pval(\Dki)+\pval(\Dij)<0$ and $\pval(\Dkj)\leq\pval(\Dki)+\pval(\Dij)$ by \crefDef{def:med-PDBM}~(\refc{PDBMii}).
				Follows that $\pval(\Dkj)+\pval(\Djk)<0$ and $0\leq\Dkj+\Djk$ is not valid for $\paramR$.
				Then~$(\E,\D)$ does not respect \crefLemma{lemmaDijDjiGeq} and violates condition~(\refc{PDBMii}) of \crefDef{def:med-PDBM}. So~$(\E,\D)\not\in\OPDBMs$.
			\item Assume this is $\clock_j$.
				This case is similar (and simpler).
		\end{itemize}
	\end{itemize}
    We apply the same reasoning for more than 3 clocks.
	\mr{en fait ce lemme n'\'etait que pour les $\OPDBMs$, je le d\'eplace ici et ajoute l'argument ci-dessous pour les $\CPr$}%
	Now suppose $(\E,\D)\in\CPr$. We apply the same reasoning, replacing the argument of condition~(\refc{PDBMii}) of \crefDef{def:med-PDBM}
	by the fact from \crefDef{definition:pointPDBM} that~$\D$ is antisymmetric.
  }

  \begin{proof}
  \versionProofIn{
  \prooflemmaLargestFracClock{}
  }
  \versionProofOut{
  See \cref{appendix:proof:lemmaLargestFracClock}.
  }
\end{proof}

Note that several clocks may have the largest fractional parts (up to some syntactic replacements
\footnote{%
			Let~$\pval\in\paramR$ and suppose, we have two different syntactic expressions, such as~$\param, 1-\param$ that are equal once valuated \ie{} $\pval(\param)=1-\pval(\param)$. From \crefDef{parameterRegions} remark that if it is for~$\pval$, it is for any~$\pval'\in\paramR$. We choose one \eg{} $1-\pval(\param)$ and replace the second, $\pval(\param)$, everywhere it appears.
		}\ea{je me suis permis de supprimer la footnote car je ne savais pas bien où la déplacer, et on manque de place}, in that case they satisfy the same constraints in $(\E,\D$)\dl{L\`a encore ca demande des pr\'ecisisions: on pourrait avoir deux expressions param\'etr\'ees \'egales pour toutes les valeurs de param\`etres, non ? p.\ ex.\ $1-p$ et $p$ si $p$ est restreint \`a la valeur $0.5$}). 

\ea{on avait dit que, dans ce cas, on sait qui est \'egal \`a qui (pour Rp) et donc on fait un remplacement syntaxique (par exemple si $p = 1 - p$ alors on remplace statiquement $1 - p$ par $p$ partout)}\ea{\`a faire quelque part (ici ?)}
\journalVersion{Suppose, given a parameter valuation~$\pval\in\paramR$, we have two different syntactic expressions that are equal once valuated
(such as, given~$\param$, $\pval(\param)=1-\pval(\param)$ and by \crefDef{parameterRegions} if it is for~$\pval$, it is for any~$\pval'\in\paramR$). Then we choose one and replace the second in every constraint where it appears (\eg{} replace~$1-\pval(\param)$ by~$\pval(\param)$ everywhere).}\dl{OK, je crois que je suis convaincu mais du coup, c'est pas obvious mais plutot up to some syntactic replacements o\`u quelque chose du genre. Il faut aussi surement explicitement invoquer la def des precises regions pour justifier que les expressions sont \'egales pour tout $v$}
For a \mPDBM{} $(\E,\D)$, we define the set of clocks with the largest fractional part~($\LFP$) as $\LFPd{\D}=\{\clock \in \Clock \mid 0\leq\Dxi$ is valid for $\paramR$, for all $0\leq i\leq\ClockCard \}$.

As we are able, thanks to the parameter regions, to order our parameter valuations (\ie{} whether one is greater or less than another one),
we can define~$\LFP$ from the constraints defined in the \pointPDBM{}.
We will define and apply successively two time-elapsing algorithms:
the first one starts from a \pointPDBM{} or an \oPDBM{} respecting condition \crefDef{def:med-PDBM}~(\refc{PDBMiii}).
We will prove that we obtain an \oPDBM{} respecting condition \crefDef{def:med-PDBM}~(\refc{PDBMiv}).
The second one starts from an \oPDBM{} respecting condition \crefDef{def:med-PDBM}~(\refc{PDBMiv}) and will define constraints
defining the possible clocks valuations exactly when any clock of~$\LFP$ has reached its upper bound~$1$.
We will prove that we obtain an \oPDBM{} respecting condition \crefDef{def:med-PDBM}~(\refc{PDBMiii}).
As we will obtain at each iteration of the algorithm an \oPDBM{} respecting either condition \crefDef{def:med-PDBM}~(\refc{PDBMiii}) or~(\refc{PDBMiv}),
this will prove that we have a stable set of \oPDBM{}s. Now we explain our algorithms more precisely.

Clocks belonging to $\LFP$ are the first to reach the upper bound~$1$ by letting time elapse.
Since $\LFP$ can contain multiple clocks and they have the same fractional part, we can consider any~$\clock\in\LFP$.

Let~$(\E,\D)\in\PDBMRp$ and~$\clock\in\LFPd{\D}$. To formalize time elapsing until the largest fractional part~$\partieFrac(\clock)$ reaches~$1$,
we define a time elapsing operator that will have two variants depending on the input:
\oPDBM{} (\crefDef{def:med-PDBM}) satisfying condition~(\refc{PDBMiii}) and \pointPDBM{} (\crefDef{definition:pointPDBM})
or \oPDBM{} (\crefDef{def:med-PDBM} satisfying condition~(\refc{PDBMiv})).

Given an \oPDBM{} satisfying condition~\ref{PDBMiii} or a \pointPDBM{} $(\E,\D)$ with~$\E_{\clock}=k$,
$\TEF((\E, \D))$ described in \cref{algorithm:TEup}\ea{réf dynamique manquante}
and named~$\TEF_{<}$, defines a new \oPDBM{} satisfying condition~\ref{PDBMiv} by
\begin{ienumeration}
\item setting~$\Dxo\r(1,<)$ as~$\clock$ is the first one that will reach~$k+1$;
\item updating the upper bound of all other clocks~$i$, which has increased: $\Dio\r\Dix+(1,<)$;
\item updating all lower bounds as they have to leave the \emph{border}: $\Doi\r\Doi+(0,<)$ ($\clock$ included).
\end{ienumeration}
This gives the range of possible clock valuations \emph{before} $\partieFrac(\clock)$ reaches~$1$.
Intuitively it represents the transformation from an open line segment or the corner-point region of~\cite{AD94} into an open region of~\cite{AD94}.

\begin{exa}
Here is an \oPDBM{} satisfying condition~\ref{PDBMiii}, on the left of the figure below. Formally, it is written:
 	{\scriptsize
\[\hspace{-1.5em}(\E,\D)=\Big(	\begin{pmatrix}
	k \\
 	k
	\end{pmatrix},
	\begin{pmatrix}
							& \mathbf{0} 									& \mathbf{\clock} 								& \mathbf{\clockk}\\
	\mathbf{0} 				& (0,\leq) 										& (-\partieFrac(\param_2),<) 						& (0,\leq)\\
	\mathbf{\clock}				& (\partieFrac(\param_2)+1-\partieFrac(\param_1),<) 	& (0,\leq) 										& (\partieFrac(\param_2)+1-\partieFrac(\param_1),<) \\
	\mathbf{\clockk} 			& (0,\leq) 										& (-\partieFrac(\param_2),<)						& (0,\leq)
	\end{pmatrix}
\Big)\] }%

 \noindent Time elapsing before~$\clock\in\LFP$ reaches the next integer gives the following \oPDBM{} satisfying condition~\ref{PDBMiv}, on the right of the figure below. Formally, it is written:
	{\scriptsize
\[\hspace{-1.5em}(\E,\D)=\Big(	\begin{pmatrix}
	k \\
 	k
	\end{pmatrix},
	\begin{pmatrix}
							& \mathbf{0} 						& \mathbf{\clock} 						& \mathbf{\clockk}\\
	\mathbf{0} 				& (0,\leq) 							& (-\partieFrac(\param_2),<) 				& (0,<)\\
	\mathbf{\clock}				& (1,<) 							& (0,\leq) 								& (\partieFrac(\param_2)+1-\partieFrac(\param_1),<) \\
	\mathbf{\clockk} 			& (1-\partieFrac(\param_2),<) 			& (-\partieFrac(\param_2),<)				& (0,\leq)
	\end{pmatrix}
\Big)\] }%

\begin{figure*}[h!]
\includegraphics[width=1\linewidth]{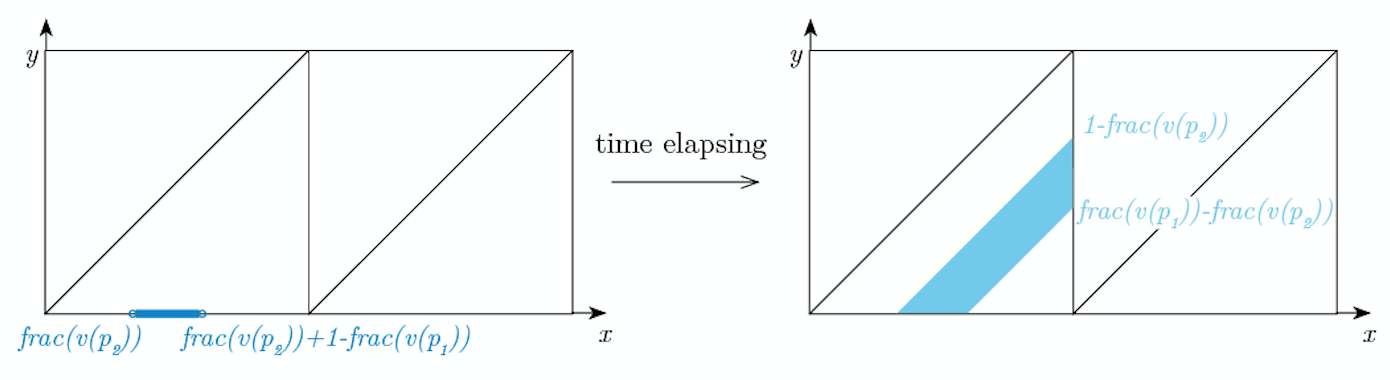}%
\label{fig:pPDBM2}
\vspace{-2em}
\end{figure*}

\end{exa}

\medskip
  \begin{algorithm}[H]
    \DontPrintSemicolon%
    \SetAlgoLined%
      pick $\clock\in\LFPd{\D}$\;
      \For{$i$ from $1$ to $\ClockCard$}{
        \eIf{$i\in\LFPd{\D}$}{
          $\Dio\r(1,<)$\;
        }
        {
          $\Dio\r\Dix+(1,<)$\;
        }
        $\Doi\r\Doi+(0,<)$\;
      }
  \caption{$\TEF_{<}((\E,\D))$: set upper bound of all~$\partieFrac(\clock)\in\LFPd{\D}$ to~$1$}%
  \label{algorithm:TEup}
  \end{algorithm}

\medskip

The result of \cref{algorithm:TEup}\ea{réf dynamique manquante} is denoted by $\TEF_{<}((\E,\D))$ and leaves $\E$ unchanged.

The time elapsing operator also operates the transformation from an open region of~\cite{AD94} to the upper open line segment or the corner-point region of~\cite{AD94}, given in \cref{algorithm:TEeq}\ea{réf dynamique manquante}
as~$\TEF_{=}$.
Given an \oPDBM{} $(\E,\D)$ satisfying condition~\ref{PDBMiv} where $\E_{\clock}=k$,
$\TEF((\E, \D))$ defines a new \oPDBM{} satisfying condition~\ref{PDBMiii} by
\begin{ienumeration}
\item setting~$\Dxo\r\Dox\r(0,\leq)$ (intuitively both became~$(1,\leq)$) and $\E_{\clock}=k+1$ (if  $\E_{\clock}\leq\CONSTMAX+1$), as~$\clock$ is now in the upper \emph{border};
\item updating the upper/lower bounds of all other clocks~$i$: $\Dio\r\Dix+(1,\leq)$ and $\Doi\r\Dxi+(-1,\leq)$;
\item updating the new difference between fractional parts with all other clocks~$i$, which is the range of values~$i$ can currently take (as in the update operator): $\Dxi\r\Doi$ and $\Dix\r\Dio$.
\end{ienumeration}

\begin{exa}
 Here is an \oPDBM{} satisfying condition~\ref{PDBMiv}, on the left of the figure below. Formally, it is written:
 {\scriptsize
\[\hspace{-1.5em}(\E,\D)=\Big(	\begin{pmatrix}
 k \\
 k
 \end{pmatrix},
 \begin{pmatrix}
             & \mathbf{0} 						& \mathbf{\clock} 						& \mathbf{\clockk}\\
 \mathbf{0} 				& (0,\leq) 							& (-\partieFrac(\param_2),<) 				& (0,<)\\
 \mathbf{\clock}				& (1,<) 							& (0,\leq) 								& (\partieFrac(\param_2)+1-\partieFrac(\param_1),<) \\
 \mathbf{\clockk} 			& (1-\partieFrac(\param_2),<) 			& (-\partieFrac(\param_2),<)				& (0,\leq)
 \end{pmatrix}
\Big)\] }%

\noindent When~$\clock\in\LFP$ reaches~$k+1$, the \oPDBM{} satisfying condition~\ref{PDBMiii} obtained is given on the right of the figure below. Formally, it is written:
 {\scriptsize
\[\hspace{-1.5em}(\E,\D)=\Big(	\begin{pmatrix}
 k+1\\
 k
 \end{pmatrix},
 \begin{pmatrix}
             & \mathbf{0} 						& \mathbf{\clock} 					& \mathbf{\clockk}\\
 \mathbf{0} 				& (0,\leq) 							& (0,\leq) 							& (-\partieFrac(\param_1)+\partieFrac(\param_2),<)\\
 \mathbf{\clock}				& (0,\leq) 							& (0,\leq) 							& (-\partieFrac(\param_1)+\partieFrac(\param_2),<) \\
 \mathbf{\clockk} 			& (1-\partieFrac(\param_2),<) 			& (1-\partieFrac(\param_2),<)			& (0,\leq)
 \end{pmatrix}
\Big)\] }%

\begin{figure*}[h!]
\includegraphics[width=1\linewidth]{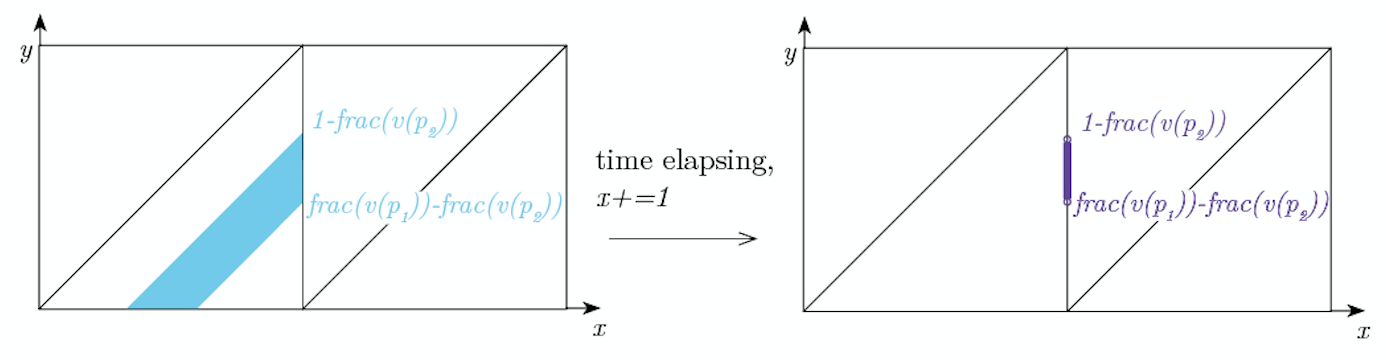}%
\label{fig:pPDBM3}
\vspace{-2em}
\end{figure*}

\end{exa}

\medskip
\begin{algorithm}[H]
\DontPrintSemicolon%
\SetAlgoLined%
  pick $\clock\in\LFPd{\D}$\;
  \For{$i$ from $1$ to $\ClockCard$}{
    \eIf{$i\in\LFPd{\D}$}{
      $\Dio\r(0,\leq)$\;
      $\Doi\r(0,\leq)$\;
      $E_i\r\E_i+1$\;
      }
      {
      $\Dio\r\Dix+(1,\leq)$\;
      $\Doi\r\Dxi+(-1,\leq)$\;
      }
  }
  \For{$i$ from $1$ to $\ClockCard$}{
    $\Dix\r\Dio$\;
    $\Dxi\r\Doi$\;
  }
\caption{$\TEF_{=}((\E,\D))$: set upper and lower bound of all~$\partieFrac(\clock)\in\LFPd{\D}$ to~$1$}%
\label{algorithm:TEeq}
\end{algorithm}
\medskip

The result of \cref{algorithm:TEeq}\ea{réf dynamique manquante} is denoted by $\TEF_{=}((\E,\D))$.

\begin{defi}[time elapsing in a \mPDBM]\label{defTE}
Let $\paramR$ be a parameter region and $(\E,\D)\in\CPr\cup\OPDBMs$.
    We define $(\E',\D') = \TEF((\E, \D))$ as applying
    either $\TEF_{<}$ if $(\E, \D)$ respects condition~\ref{PDBMiii}
    or $(\E,\D)\in\CPr$,
    or $\TEF_{=}$ if $(\E,\D)$ respects condition~\ref{PDBMiv}.

\end{defi}

\newcommand{\lemmaTEstable}{%
	Let $\paramR$ be a parameter region.
	Let~$(\E,\D)\in\PDBMRp$.
	Then $\TEF((\E,\D)) \in \PDBMRp$.
}

\begin{lem}[stability under time elapsing]\label{lemmaTEstable}
\lemmaTEstable{}
\end{lem}

\newcommand{\prooflemmaTEstable}{
We prove our lemma for the two types of \oPDBMs{} and for~\pointPDBMs{}.

\subsubsection{$\rightarrow$ \crefDef{def:med-PDBM} type (\refc{PDBMiii}) to (\refc{PDBMiv})}

\newcommand{\lemmaStablePDBMiv}{
Let $\paramR$ be a parameter region and $(\E,\D)\in\OPDBMs$ respecting condition~\ref{PDBMiii},
then $\TEF_{<}((\E,\D))\in\OPDBMs$ respecting condition~\ref{PDBMiv}.
}
\begin{clm}[modification of an \oPDBM{} respecting condition~(\ref{PDBMiii}) under~$\TEF_<$]%
\label{TEmPDBM}
  \lemmaStablePDBMiv{}
\end{clm}

\begin{preuve}%
 \label{proofTEmPDBM}

 Suppose $(\E,\D)\in\OPDBMs$ respects condition (\refc{PDBMiii}) of \crefDef{def:med-PDBM},
 \ie{} we have at least an~$\clock$ \st{} $\Dxo=\Dox=(0,\leq)$.
 Since, in $\paramR$, we know which parameters have the largest fractional part,
 we can determine~$\LFPd{\D}$ from \crefLemma{lemmaLargestFracClock}.
 If more than one clock belong to $\LFPd{\D}$ then their valuations have the same fractional part.
 Indeed, from \crefDef{def:largestclock} if~$\clock_i, \clock_j\in\LFPd{\D}$ then both~$(0,\leq)\leq\Dij$
 and~$(0,\leq)\leq\Dji$ are valid for~$\paramR$, and from \crefDef{def:med-PDBM}~(\refc{PDBMv})
 we must have~$\Dij=\Dji=(0,\leq)$($\star$).\ea{macro} 

 Let~$\pval\in\paramR$.
 Assume $\clock_i\in\LFPd{\D}$ and $\clockval\in(\E,\pval(\D))$,
 by letting time elapse, $\partieFrac(\clockval(\clock_i))$ is the first that might reach $1$.
 Moreover, for all $\clock_j\in\Clock\setminus\LFPd{\D}$, $\partieFrac(\clockval(\clock_j))$ cannot reach $1$ before $\partieFrac(\clockval(\clock_i))$.
 We are going to construct a new~$(\E',\D')=\TEF_<((\E,\D))$, which will be an~\oPDBM{} respecting condition~\ref{PDBMiv} of \crefDef{def:med-PDBM}.
 While detailing the procedure of $\TEF_<$, we are going to prove that \crefDef{def:med-PDBM}~(\refc{PDBMi}) and~(\refc{PDBMv}) hold for $(\E', \D')$.
 Further we will prove that~(\refc{PDBMii}) and~(\refc{PDBMiv}) also hold.

 \medskip

 \noindent \textbf{proof that \crefDef{def:med-PDBM}~(\refc{PDBMi}) holds.}

 \medskip

According to the definition of~$\TEF_<$ (\cref{algorithm:TEup}\ea{réf dynamique manquante})
the first step is to set a new upper bound
 \[
  \Dio'=(1,<)
  \quad
  \mbox{for all } \clock_i\in\LFPd{\D}
 \]
 and obviously $(0,\leq)\leq\Dio'\leq (1,\leq)$ is valid for~$\paramR$.
 Then we set new upper bounds for all other clock $\clock_j\in\Clock\setminus\LFPd{\D}$ by setting
 \[
  \Djo'=\Dji+(1,<)\text{.}
 \]

 Indeed, $\Dji$ is the constraint on the lower bound of $\partieFrac(\clockval(\clock_j))-\partieFrac(\clockval(\clock_i))$
 and since the upper bound of $\clock_i$ has increased, this gives the new upper bound of $\clock_j$.
 Note that since $\clock_i\in\LFPd{\D}$, from \crefDef{def:largestclock} and \crefDef{def:med-PDBM}~(\refc{PDBMv})
 we have that $-1\leq\Dji\leq 0$ is valid for~$\paramR$ for all clock~$\clock_j$.
 Precisely, $\dji \in \{0,-\param_1,\param_2-\param_1, \param_1-1-\param_2,\param_1-1\}$ for some $\param_1,\param_2\in\Param$
 where $\param_2\leq\param_1$ \valid{}.
 Hence as $\dji+1 \in \{1,1-\param_1,\param_2+1-\param_1, \param_1-\param_2,\param_1\}$,
 we have that $\djo'\in\PLT$, $\compOpLeq_{ji'}=\compOpLeq_{ji}\oplus{<}={<}$ so $(0,\leq)\leq\Djo'\leq (1,<)$ is valid for~$\paramR$.

 \longVersion{Note that we cannot have $(\dji,\compOpLeq_{ji})=(-1,<)$ because even if~$(\dij,\compOpLeq_{ij})=(1,<)$, since~$(\E,\D)\in\OPDBMs$ we do not have have $0\leq\Dji+\Dij$ is valid for~$\paramR$ from \crefDef{def:med-PDBM}~(\refc{PDBMii}) and \crefLemma{lemmaDijDjiGeq}.}

 Secondary we set for all clock $\clock$ regardless of whether they are in $\LFPd{\D}$ %
 \[
  \Dox'=\Dox+(0,<)\text{.}
 \]
  Since some time elapsed, lower bounds of all clocks are increased.
  Moreover, as~$(-1,<)\leq\Dox\leq (0,\leq)$ is valid for~$\paramR$ from \crefDef{def:med-PDBM}~(\refc{PDBMi}),
  $(-1,\leq)\leq\Dox'\leq (0,\leq)$ is also valid for~$\paramR$.

 Therefore, \crefDef{def:med-PDBM}~(\refc{PDBMi}) holds.

\medskip

 \noindent \textbf{proof that \crefDef{def:med-PDBM}~(\refc{PDBMv}) holds}

\medskip

  Third we set for all clocks $\clock, \clockk$ regardless of whether they are in $\LFPd{\D}$
  \[
  \Dxy'=\Dxy
  \]
  so as \crefDef{def:med-PDBM}~(\refc{PDBMv}) holds in~$(\E,\D)$, it still does.
  More intuitively since no fractional part has reached~$1$, constraints on differences of clocks and integer parts remain unchanged.

\medskip

 \noindent \textbf{proof that \crefDef{def:med-PDBM}~(\refc{PDBMvi}) holds}

\medskip

   For all~$\clock_i$:
\begin{itemize}
 \item if~$\clock_i\in\LFPd{\D}$, $\Dio'=(1,<)$, $\Doi'=\Doi+(0,<)$ hence~$\dio'\neq\dzeroi'$ and~$\compOpLeq_{i0'}=\compOpLeq_{0i'}={<}$, condition \crefDef{def:med-PDBM}~(\refc{PDBMvi}) holds;
 \item if~$\clock_i\in\Clock\setminus\LFPd{\D}$, $\clock\in\LFPd{\D}$, $\Dio'=\Dix+(1,<)$, $\Doi'=\Doi+(0,<)$ hence as~$(0,\leq)\leq\Dio'$ \valid{} and~$\Doi'\leq(0,\leq)$ \valid{},
 we have~$\dio'\neq\dzeroi'$ and~$\compOpLeq_{i0'}=\compOpLeq_{0i'}={<}$ and condition \crefDef{def:med-PDBM}~(\refc{PDBMvi}) holds.
\end{itemize}
  For all~$\clock_i, \clock_j$:
  \begin{itemize}
   \item if~$\clock_i, \clock_j\in\Clock\setminus\LFPd{\D}$, $\Dij'=\Dij$ and~$\Dji'=\Dji$, condition \crefDef{def:med-PDBM}~(\refc{PDBMvi}) holds as it holds for~$\Dij$ and~$\Dji$.
   \item if~$\clock_i\in\Clock\setminus\LFPd{\D}$, $\clock_j\in\LFPd{\D}$, $\Dio'=\Dij+(1,<)$, $\Doi'=\Doi+(0,<)$ hence as~$(0,\leq)\leq\Dio'$ \valid{} and~$\Doi'\leq(0,\leq)$ \valid{},
    we have~$\dio'\neq\dzeroi'$ and~$\compOpLeq_{i0'}=\compOpLeq_{0i'}={<}$, condition \crefDef{def:med-PDBM}~(\refc{PDBMvi}) holds. The case~$\clock_j\in\Clock\setminus\LFPd{\D}$, $\clock_i\in\LFPd{\D}$ is treated similarly.
    \item if~$\clock_i, \clock_j\in\LFPd{\D}$, $\Dij'=\Dji'=(0,\leq)$, hence~$\dij'=-\dji'=0$ and~$\compOpLeq_{ij'}=\compOpLeq_{ji'}=\leq$ and condition \crefDef{def:med-PDBM}~(\refc{PDBMvi}) holds.
  \end{itemize}

\medskip

 \noindent \textbf{proof that \crefDef{def:med-PDBM}~(\refc{PDBMii}) holds}

\medskip

 Now we prove that \crefDef{def:med-PDBM}~(\refc{PDBMii}) holds, \ie{} for all clocks~$\clock_i, \clock_j,\clock_k$,
 valid conditions such as~$\Dij'\leq\Dik'+\Dkj'$ remain valid in~$\paramR$.
 Indeed, when time elapses, all clocks have the same behavior, hence the difference between two clocks does not change without an update.
 Precisely, for all clocks~$\clock_i,\clock_j,\clock_k$, are valid for~$\paramR$:
	\begin{enumerate}
		      \item if~$\clock_i,\clock_j,\clock_k\in\Clock\setminus\LFPd{\D}$: let $\clock\in\LFPd{\D}$ and
					\begin{itemize}
						  \item if~$i,j,k$ are different from $0$, we have~$\Dik'=\Dik$, $\Dij'=\Dij$ and~$\Djk'=\Djk$;
							since~$(\E,\D)\in\OPDBMs$ from \crefDef{def:med-PDBM}~(\refc{PDBMii}), we know that~$\Dik \leq \Dij+\Djk$ is valid for~$\paramR$;
							therefore, $\Dik'\leq\Dij'+\Djk'$ is valid for~$\paramR$.
						  \item if~$i,j$ are different from $0$, $k=0$, we have~$\Dio'=\Dix+(1,<)$, $\Dij'=\Dij$ and~$\Djo'=\Djx+(1,<)$;
							since~$(\E,\D)\in\OPDBMs$ from \crefDef{def:med-PDBM}~(\refc{PDBMii}), we know that~$\Dix \leq \Dij+\Djx$ is valid for~$\paramR$;
							then $\Dix+(1,<)\leq\Dij+\Djx+(1,<)$ is valid for~$\paramR$ from \crefLemma{lemma:sumvalid} and therefore, $\Dio'\leq\Dij'+\Djo'$ is valid for~$\paramR$.
						  \item if~$i,k$ are different from~$0$,~$j=0$, we have~$\Dik'=\Dik$, $\Dio'=\Dix+(1,<)$ and~$\Dok'=\Dok+(0,<)$;
							we claim that
							\begin{equation}\label{eq:cinq}
							 \Dik\leq\Dix+(1,<)+\Dok+(0,<)
							 \quad
							\end{equation}
							is valid for~$\paramR$, which is equivalent to $\Dik' \leq \Dio'+\Dok'$ is valid for~$\paramR$.
							Since~$(\E,\D)\in\OPDBMs$ from \crefDef{def:med-PDBM}~(\refc{PDBMi}), we know that \begin{equation}\label{eq:six}\Dxo \leq (1,<)\text{;}\end{equation}
							moreover we have
							\begin{equation}\label{eq:sept}
							(1,<)+(0,<)=(1+0,< \oplus <)=(1,<)
							\quad
							\end{equation}
							Since~$(\E,\D)\in\OPDBMs$ from \crefDef{def:med-PDBM}~(\refc{PDBMii}), we know that~$\Dxk \leq \Dxo+\Dok$ is valid for~$\paramR$;
							combining with (\refc{eq:six}) and (\refc{eq:sept}) %
							we obtain
							\begin{equation}\label{eq:huit}
							\Dxk\leq (1,<)+\Dok+(0,<).
							\quad
							\end{equation}
 							Now, since~$(\E,\D)\in\OPDBMs$ from \crefDef{def:med-PDBM}~(\refc{PDBMii}), we know that~$\Dik \leq \Dix+\Dxk$ is valid for~$\paramR$
 							and combining with (\refc{eq:huit}) we obtain (\refc{eq:cinq}) and therefore our result.
 							\ea{j'ai regard\'e jusqu'ici}
 						\item if~$i$ is different from~$0$,~$j=k=0$, we have~$\Dio'=\Dix+(1,<)$;
							from \crefDef{def:validity}~(\refc{valideiiid}) we have that
							\[
							\Dix+(1,<)\leq\Dix+(1,<)
							\]
							is valid for~$\paramR$.
							Hence from \crefLemma{lemmaDiiDijGeq}
							\[
							\Dio'\leq\Dio'+\Doo'
							\]
							is valid for~$\paramR$.
						\item if~$j,k$ are different from~$0$,~$i=0$, we have~$\Dok'=\Dok+(0,<)$, $\Doj'=\Doj+(0,<)$ and $\Djk'=\Djk$;
							since~$(\E,\D)\in\OPDBMs$, from \crefDef{def:med-PDBM}~(\refc{PDBMii})
							we know that $\Dok \leq \Doj+\Djk$ is valid for~$\paramR$.
							Moreover we have that
							\[
							\Dok+(0,<) = (\dok,<)
							\quad
							\mbox{and}
							\quad
							\Doj+(0,<)+\Djk	= (\doj+\djk,<)
							\]
							so we have from \crefDef{def:validity}~(\refc{valideiiid})
							\[
							\Dok+(0,<)\leq\Doj+(0,<)+\Djk
							\]
							is valid for~$\paramR$.
							Hence $\Dok' \leq \Doj'+\Djk'$ is valid for $\paramR$.
						\item if~$j$ is different from~$0$,~$i=k=0$, we have~$\Doo'=(0,\leq)$, $\Doj'=\Doj+(0,<)$ and $\Djo'=\Djx+(1,<)$;
							since $(\E,\D)\in\OPDBMs$, from \crefDef{def:med-PDBM}~(\refc{PDBMii})
							we know that~$\Dox \leq \Doj+\Djx$ is valid for~$\paramR$;
							moreover, from\longVersion{ \crefDef{def:validity}~(\refc{valideiiid}) and} \crefLemma{lemma:sumvalid},
							\[
							\Dox+(0,<) \leq \Doj+(0,<)+\Djx
							\]
							is valid for~$\paramR$.
							Recall that from \crefLemma{lemmaDijDjiGeq} $(0,\leq)\leq\Dox+\Dxo$ is valid for~$\paramR$
							and since $\Dxo\leq (1,<)$ from \crefDef{def:med-PDBM}~(\refc{PDBMi}),
							we have
							\[
							(0,\leq)\leq\Dox+(1,<)
							\]
							is valid for~$\paramR$.
							As we have~$(1,<)+(0,<)=(1+0,< \oplus <)=(1,<)$,
							we obtain that
							\[
							\Dox+(1,<)\leq\Doj+\Djx+(1,<)
							\]
							is valid for~$\paramR$ and therefore~$\Doo' \leq \Doj'+\Djo'$ is valid for $\paramR$.
						\item if~$k$ is different from~$0$,~$i=j=0$, we have~$\Dok'=\Dok+(0,<)$;
							From \crefDef{def:validity}~(\refc{valideiiid}) and \crefLemma{lemma:sumvalid} we have that
							\[
							\Dok+(0,<)\leq\Dok+(0,<)
							\]
							is valid for~$\paramR$.
							Hence from \crefLemma{lemmaDiiDijGeq}
							\[
							\Dok'\leq\Doo'+\Dok'
							\]
							is valid for~$\paramR$.
						\item if~$i=j=k=0$, from \crefDef{def:med-PDBM}~(\refc{PDBMii}) and \crefLemma{lemmaDiiDijGeq}
							we trivially have
							\[
							\Doo'\leq\Doo'+\Doo'
							\]
							is valid for~$\paramR$.

					\end{itemize}
		      \item if~$\clock_k\in\LFPd{\D}$ and~$\clock_i, \clock_j\in\Clock\setminus\LFPd{\D}$: $k\neq 0$ and
		      		\begin{itemize}
						  \item if~$i,j$ are different from $0$, we have~$\Dik'=\Dik$, $\Dij'=\Dij$ and~$\Djk'=\Djk$;
							since~$(\E,\D)\in\OPDBMs$ from \crefDef{def:med-PDBM}~(\refc{PDBMii}), we know that~$\Dik \leq \Dij+\Djk$;
							therefore, $\Dik'\leq\Dij'+\Djk'$.
						  \item if~$i\neq 0$,~$j=0$, we have~$\Dik'=\Dik$, $\Dio'=\Dik+(1,<)$ and~$\Dok'=\Dok+(0,<)$;
						  	we claim that $\Dik\leq\Dik+(1,<)+\Dok+(0,<)$ is valid for~$\paramR$, \ie{}
							\begin{equation}\label{eq:neuf}
							 (0,\leq)\leq (1,<)+\Dok+(0,<)
							 \quad
							\end{equation}
							is valid for~$\paramR$, which is equivalent to $\Dik' \leq \Dio'+\Dok'$ is valid for~$\paramR$.
							We have
							\begin{equation}\label{eq:dix}
							(1,<)+(0,<)=(1+0,< \oplus <)=(1,<)\text{.}
							\quad
							\end{equation}
							Since~$(\E,\D)\in\OPDBMs$, from \crefDef{def:med-PDBM}~(\refc{PDBMii})
							we know that~$(0,\leq)\leq\Dok+\Dko$ \valid{}
							and from \crefDef{def:med-PDBM}~(\refc{PDBMi}) that~$\Dko \leq (1,<)$ is valid for~$\paramR$;
							combining with (\refc{eq:neuf}) and (\refc{eq:dix}) we obtain our result.
						\item if~$i=0$,~$j\neq 0$, we have~$\Dok'=\Dok+(0,<)$, $\Doj'=\Doj+(0,<)$ and $\Djk'=\Djk$;
							since~$(\E,\D)\in\OPDBMs$, from \crefDef{def:med-PDBM}~(\refc{PDBMii}) we know that~$\Dok \leq \Doj+\Djk$.
							Moreover we have that
							\[
							\Dok+(0,<) = (\dok,<)
							\quad
							\mbox{and}
							\quad
							\Doj+(0,<)+\Djk	= (\doj+\djk,<)
							\]
							so we have from \crefDef{def:validity}~(\refc{valideiiid})
							\[
							\Dok+(0,<)\leq\Doj+(0,<)+\Djk
							\]
							is valid for~$\paramR$.
							Hence $\Dok' \leq \Doj'+\Djk'$ is valid for $\paramR$.
						\item if~$i=j=0$, from \crefDef{def:med-PDBM}~(\refc{PDBMii}) and \crefLemma{lemmaDiiDijGeq} we trivially have
							\[
							\Dok'\leq\Doo'+\Dok'
							\]
							is valid for~$\paramR$.
					\end{itemize}

		      \item if $\clock_j\in\LFPd{\D}$ and $\clock_i, \clock_k\in\Clock\setminus\LFPd{\D}$: $j\neq 0$ and
					\begin{itemize}
						  \item if~$i,k$ are different from $0$, we have~$\Dik'=\Dik$, $\Dij'=\Dij$ and~$\Djk'=\Djk$;
							since~$(\E,\D)\in\OPDBMs$, from \crefDef{def:med-PDBM}~(\refc{PDBMii})
							we know that~$\Dik \leq \Dij+\Djk$ is valid for~$\paramR$;
							therefore, $\Dik'\leq\Dij'+\Djk'$ is valid for~$\paramR$.
						  \item if~$i\neq 0$, $k=0$, we have~$\Dio'=\Dij+(1,<)$, $\Dij'=\Dij$ and~$\Djo'=(1,<)$;
						   	From \crefDef{def:validity}~(\refc{valideiiid}) we trivially have that
							$\Dij+(1,<)\leq\Dij+(1,<)$ is valid for~$\paramR$ and therefore, $\Dio'\leq\Dij'+\Djo'$ is valid for~$\paramR$.
						\item if~$i=0$, $k\neq 0$, we have~$\Dok'=\Dok+(0,<)$, $\Doj'=\Doj+(0,<)$ and $\Djk'=\Djk$;
							since~$(\E,\D)\in\OPDBMs$, from \crefDef{def:med-PDBM}~(\refc{PDBMii}) we know that~$\Dok \leq \Doj+\Djk$ is valid for~$\paramR$.
							Moreover we have that
							\[
							\Dok+(0,<) = (\dok,<)
							\quad
							\mbox{and}
							\quad
							\Doj+(0,<)+\Djk	= (\doj+\djk,<)
							\]
							so we have from\longVersion{ \crefDef{def:validity}~(\refc{valideiiid}) and} \crefLemma{lemma:sumvalid}
							\[
							\Dok+(0,<)\leq\Doj+(0,<)+\Djk
							\]
							is valid for~$\paramR$.
							Hence $\Dok' \leq \Doj'+\Djk'$ is valid for~$\paramR$.
						\item if~$i=k=0$, we have~$\Doo'=(0,\leq)$, $\Doj'=\Doj+(0,<)$ and $\Djo'=(1,<)$;
							since~$(\E,\D)\in\OPDBMs$, from \crefLemma{lemmaDijDjiGeq} we know that~$(0,\leq) \leq \Doj+\Djo$ is valid for $\paramR$,
							and since from \crefDef{def:med-PDBM}~(\refc{PDBMi}) $\Djo\leq (1,\leq)$ is valid for~$\paramR$, that means $(0,\leq)\leq\Doj+(1,<)$ is valid for $\paramR$.
							As we have
							\[
							(1,<)+(0,<)=(1+0,< \oplus <)=(1,<)
							\]
							we obtain that
							\[
							(0,\leq)\leq\Doj+(0,<)+(1,<)
							\]
							is valid for~$\paramR$ and therefore~$\Doo' \leq \Doj'+\Djo'$ is valid for~$\paramR$.
					\end{itemize}

		      \item if $\clock_j, \clock_k\in\LFPd{\D}$ and $\clock_i\in\Clock\setminus\LFPd{\D}$: $j\neq0, k\neq 0$ and
					\begin{itemize}
						  \item if~$i$ is different from $0$, we have~$\Dik'=\Dik$, $\Dij'=\Dij$ and~$\Djk'=\Djk$;
							since~$(\E,\D)\in\OPDBMs$, from \crefDef{def:med-PDBM}~(\refc{PDBMii}) we know that~$\Dik \leq \Dij+\Djk$;
							therefore, $\Dik'\leq\Dij'+\Djk'$.
					\item if~$i=0$, we have~$\Dok'=\Dok+(0,<)$, $\Doj'=\Doj+(0,<)$ and $\Djk'=\Djk$;
							since~$(\E,\D)\in\OPDBMs$, from \crefDef{def:med-PDBM}~(\refc{PDBMii}) we know that~$\Dok \leq \Doj+\Djk$.
							Moreover we have that
							\[
							\Dok+(0,<) = (\dok,<)
							\quad
							\mbox{and}
							\quad
							\Doj+(0,<)+\Djk	= (\doj+\djk,<)
							\]
							so we have from \crefDef{def:validity}~(\refc{valideiiid}) and \crefLemma{lemma:sumvalid}
							\[
							\Dok+(0,<)\leq\Doj+(0,<)+\Djk
							\]
							is valid for~$\paramR$.
							Hence $\Dok' \leq \Doj'+\Djk'$ is valid for~$\paramR$.
					\end{itemize}
		      \item if $\clock_i\in\LFPd{\D}$ and $\clock_j,\clock_k\in\Clock\setminus\LFPd{\D}$: $i\neq 0$ and
					\begin{itemize}
						  \item if~$j,k$ are different from $0$, we have~$\Dik'=\Dik$, $\Dij'=\Dij$ and~$\Djk'=\Djk$;
							since~$(\E,\D)\in\OPDBMs$, from \crefDef{def:med-PDBM}~(\refc{PDBMii}) we know that~$\Dik \leq \Dij+\Djk$;
							therefore, $\Dik'\leq\Dij'+\Djk'$.
						  \item if~$j\neq 0$, $k=0$, we have~$\Dio'=(1,<)$, $\Dij'=\Dij$ and~$\Djo'=\Dji+(1,<)$;
						  	from \crefDef{def:med-PDBM}~(\refc{PDBMii}) and \crefLemma{lemmaDijDjiGeq} we know that $(0,\leq)\leq \Dij+\Dji$ is valid for~$\paramR$.
						  	Since, from \crefDef{def:validity}~(\refc{valideiiid}) $(1,<)\leq(1,<)$ is valid for~$\paramR$,
							then from \crefLemma{lemma:sumvalid}
							\[
							(1,<)\leq\Dij+\Dji+(1,<)
							\]
							is valid for~$\paramR$ and therefore, $\Dio'\leq\Dij'+\Djo'$ is valid for~$\paramR$.
						  \item if~$j=0$, $k\neq 0$, we have~$\Dik'=\Dik$, $\Dio'=(1,<)$ and~$\Dok'=\Dok+(0,<)$;
							we claim that
							\[
							 \Dik\leq (1,<)+\Dok+(0,<)
							\]
							is valid for~$\paramR$, which is equivalent to $\Dik' \leq \Dio'+\Dok'$ is valid for~$\paramR$.
							Since~$(\E,\D)\in\OPDBMs$ from \crefDef{def:med-PDBM}~(\refc{PDBMii}), we know that~$\Dik \leq \Dio+\Dok$ is valid for~$\paramR$;
							moreover, from \crefDef{def:med-PDBM}~(\refc{PDBMi}), we know that~$\Dio \leq (1,<)$ is valid for~$\paramR$.
							We have
							\[
							(1,<)+(0,<)=(1+0,< \oplus <)=(1,<)
							\]
							so we obtain that
							\[
							\Dik\leq \Dio+\Dok\leq (1,<)+\Dok=(1,<)+\Dok+(0,<)
							\]
 							is valid for~$\paramR$ and therefore our result.
 						\item if~$i$ is different from~$0$,~$j=k=0$, we have~$\Dio'=(1,<)$, $\Doo'=(0,\leq)$;
							from \crefDef{def:validity}~(\refc{valideiiid}) we have that
							\[
							(1,<)\leq (1,<)
							\]
							is valid for~$\paramR$.
							Hence from \crefLemma{lemmaDiiDijGeq}
							\[
							\Dio'\leq\Dio'+\Doo'
							\]
							is valid for~$\paramR$.
					\end{itemize}

		      \item if $\clock_i, \clock_k\in\LFPd{\D}$ and $\clock_j\in\Clock\setminus\LFPd{\D}$: $i\neq 0, k\neq 0$ and
		      			\begin{itemize}
						  \item if~$j\neq 0$, we have~$\Dik'=\Dik$, $\Dij'=\Dij$ and~$\Djk'=\Djk$;
							since~$(\E,\D)\in\OPDBMs$, from \crefDef{def:med-PDBM}~(\refc{PDBMii}) we know that~$\Dik \leq \Dij+\Djk$;
							therefore, $\Dik'\leq\Dij'+\Djk'$ is valid for~$\paramR$.
						  \item if~$j=0$, we have~$\Dik'=\Dik$, $\Dio'=(1,<)$ and~$\Dok'=\Dok+(0,<)$;
							we claim that
							\[
							 \Dik\leq (1,<)+\Dok+(0,<)
							\]
							is valid for~$\paramR$, which is equivalent to $\Dik' \leq \Dio'+\Dok'$ is valid for~$\paramR$.
							Since~$(\E,\D)\in\OPDBMs$ from \crefDef{def:med-PDBM}~(\refc{PDBMii}), we know that~$\Dik \leq \Dio+\Dok$ is valid for~$\paramR$;
							moreover, from \crefDef{def:med-PDBM}~(\refc{PDBMi}), we know that~$\Dio \leq (1,<)$ is valid for~$\paramR$.
							We have
							\[
							(1,<)+(0,<)=(1+0,< \oplus <)=(1,<)
							\]
							so we obtain that
							\[
							\Dik\leq \Dio+\Dok\leq (1,<)+\Dok=(1,<)+\Dok+(0,<)
							\]
 							is valid for~$\paramR$ and therefore our result.
		      			\end{itemize}

		      \item if $\clock_i, \clock_j\in\LFPd{\D}$ and $\clock_k\in\Clock\setminus\LFPd{\D}$: $i\neq 0, j\neq 0$ and
					\begin{itemize}
						  \item if~$k\neq 0$, we have~$\Dik'=\Dik$, $\Dij'=\Dij$ and~$\Djk'=\Djk$;
							since~$(\E,\D)\in\OPDBMs$, from \crefDef{def:med-PDBM}~(\refc{PDBMii}) we know that~$\Dik \leq \Dij+\Djk$ is valid for~$\paramR$;
							therefore, $\Dik'\leq\Dij'+\Djk'$ is valid for~$\paramR$.
						  \item if~$k=0$, we have~$\Dio'=(1,<)$, $\Dij'=\Dij=(0,\leq)$ since both $\clock_i, \clock_j\in\LFPd{\D}$ (cf.\ ($\star$)) and~$\Djo'=(1,<)$;
							then $(1,<)\leq(0,\leq)+(1,<)$ is valid for~$\paramR$ and therefore, $\Dio'\leq\Dij'+\Djo'$ is valid for~$\paramR$.
					\end{itemize}

		      \item if $\clock_i,\clock_j, \clock_k\in\LFPd{\D}$:
		     				$i,j,k$ are different from $0$, we have~$\Dik'=\Dik$, $\Dij'=\Dij$ and~$\Djk'=\Djk$;
						since~$(\E,\D)\in\OPDBMs$, from \crefDef{def:med-PDBM}~(\refc{PDBMii}) we know that~$\Dik \leq \Dij+\Djk$ is valid for~$\paramR$;
						therefore, $\Dik'\leq\Dij'+\Djk'$ is valid for~$\paramR$.
	\end{enumerate}

\medskip

 \noindent \textbf{proof that \crefDef{def:med-PDBM}~(\refc{PDBMiv}) holds}

\medskip

 Finally, for $\clock_i\in\LFPd{\D}$, $\Dio'=(1,<)$ and for all clock~$j$ \st{} $\Doj'=(0,\compOpLeq_{0j'})$, then we have$~{\compOpLeq_{0j'}}={<}$.
 Condition \crefDef{def:med-PDBM}~(\refc{PDBMiv}) is satisfied.

  We denote by $(\E,\D')$ the obtained \mPDBM{} and $(\E,\D')\in\OPDBMs$.
\end{preuve}

 \subsubsection{$\rightarrow$ \crefDef{def:med-PDBM} type (\refc{PDBMiv}) to (\refc{PDBMiii})}

\longVersion{
  \begin{lem}\label{lemma:dijstrictzero}
  Let~$(\E,\D)\in\OPDBMs$; let~$\clock_i\in\LFPd{\D}$, $\clock_j\in\Clock\setminus\LFPd{\D}$.
  If $(\dij,\compOpLeq_{ij})=(0,\compOpLeq)$, then $\compOpLeq= {<}$
 \end{lem}
 \begin{preuve}
 Let~$\clock_i\in\LFPd{\D}$, $\clock_j\in\Clock\setminus\LFPd{\D}$.
 Suppose $(\dij,\compOpLeq_{ij})=(0,\leq)$. From \crefDef{def:med-PDBM}~(\refc{PDBMv}) we should have that~$(\dji,\compOpLeq_{ji})=(0,\leq)$
  so \crefLemma{lemmaDijDjiGeq} is satisfied, and then~$\clock_j\in\LFPd{\D}$.
 \end{preuve}
 }

\bigskip

 \newcommand{\lemmaStablePDBMiii}{
 Let $\paramR$ be a parameter region and $(\E,\D)\in\OPDBMs$ respecting condition~\ref{PDBMiv},
 then $\TEF_{=}(\E,\D)\in\OPDBMs$ respecting condition~\ref{PDBMiii}.
 }
\begin{clm}[modification of an \oPDBM{} respecting condition~\ref{PDBMiv} under~$\TEF_=$]%
\label{TE2mPDBM}
  \lemmaStablePDBMiii{}
\end{clm}

\begin{preuve}%
\label{proofTE2mPDBM}
 Suppose $(\E,\D)\in\OPDBMs$ respects condition (\refc{PDBMiii}) of \crefDef{def:med-PDBM}
 \ie{} we have at least an~$\clock$ \st{} $\Dxo=(1,<)$ and for all other~$j$ \st{} $\Doj=(0,\compOpLeq_{0j})$, ${\compOpLeq_{0j}} = {<}$.
 First we can determine~$\LFPd{\D}$. Let $\clock\in\LFPd{\D}$.
  If more than one clock belong to~$\LFPd{\D}$ then their valuations have the same fractional part.
 Indeed, from \crefDef{def:largestclock} if~$\clock_i, \clock_j\in\LFPd{\D}$ then both~$(0,\leq)\leq\Dij$
 and~$(0,\leq)\leq\Dji$ are valid for~$\paramR$, and from \crefDef{def:med-PDBM}~(\refc{PDBMv})
 we must have~$\Dij=\Dji=(0,\leq)$.

 Let $\pval\in\paramR$. Let $\clock_i\in\LFPd{\D}$ and $\clockval\in(\E,\pval(\D))$.
 By letting time elapse, $\partieFrac(\clockval(\clock))$ is the first to actually reach $1$.
 Moreover, for all $\clock_j\in\Clock\setminus\LFPd{\D}$, $\partieFrac(\clockval(\clock_j))$ cannot reach $1$ before $\partieFrac(\clockval(\clock_i))$.
  We are going to construct a new~$(\E',\D')=\TEF_=((\E,\D))$ which is an~\oPDBM{} respecting condition~\ref{PDBMiv}.
 While detailing the procedure of~$\TEF_=$, we are going to prove that \crefDef{def:med-PDBM}~(\refc{PDBMi}) and~(\refc{PDBMv}) hold for $(\E', \D')$. Further we will prove that~(\refc{PDBMii}) and~(\refc{PDBMiii}) also hold.

\medskip

 \noindent \textbf{proof that \crefDef{def:med-PDBM}~(\refc{PDBMi}) holds}

\medskip

 According to the definition of~$\TEF_=$ \cref{algorithm:TEeq}\ea{réf dynamique manquante},
 the first step is to fix the value of~$\partieFrac(\clock_i)$ to~$0$ by setting
 \[
 \Dio'=(0,\leq)
 \quad
  \mbox{and}
 \quad
 \Doi'=(0,\leq)
 \quad
 \mbox{for all } \clock_i\in\LFPd{\D}\text{.}
 \]

 Indeed, when~$\partieFrac(\clock_i)$ reaches~$1$,
 in the constraints expressed by~$(\E,\pval(\D))$ we have to increase the integer part by~$1$ and set the new constraints on the fractional part to~$0$.

 Secondary we set new upper and lower bound for all other clock~$\clock_j\in\Clock\setminus\LFPd{\D}$
 \[
 \Doj'=\Dij+(-1,\leq)
 \quad
 \mbox{and}
 \quad
 \Djo'=\Dji+(1,\leq)\text{.}
 \]
 We have to force now upper and lower bounds for other clocks since we know the interval of time that elapsed when~$\clock_i$ reached $1$.

 Note that since~$\clock_i\in\LFPd{\D}$, $\clock_j\in\Clock\setminus\LFPd{\D}$
  from \crefDef{def:largestclock} we have that~$(0,\leq)\leq\Dij\leq (1,<)$ is valid for~$\paramR$ for all clock~$\clock_j$.
  Nonetheless, since~$\clock_j\in\Clock\setminus\LFPd{\D}$, we even have $\Dij\neq(0,\leq)$:
  suppose $(\dij,\compOpLeq_{ij})=(0,\leq)$:
   from \crefDef{def:med-PDBM}~(\refc{PDBMv}) we should have that~$(\dji,\compOpLeq_{ji})=(0,\leq)$
   so \crefLemma{lemmaDijDjiGeq} is satisfied, and then~$\clock_j\in\LFPd{\D}$.
   The same reasoning leads to~$\Dji\neq(0,\leq)$.

\longVersion{Obviously, we have $\Dij\neq(0,<)$: suppose~$\Dij=(0,<)$,
since~$\clock_i\in\LFPd{\D}$
then from \crefDef{def:largestclock} $(0,\leq)\leq\Dij$ should be valid for~$\paramR$,
which is not from \crefDef{def:validity}~(\refc{valideiiid}).
  }

 Precisely, $\dij{\in}\{1,1-\param_1,\param_2+1-\param_1, \param_1-\param_2,\param_1\}$ for any two $\param_1,\param_2\in\Param$
 where $\param_2\leq\param_1$ \valid{}.
 Hence as $-1+\dij{\in}\{0,-\param_1,\param_2-\param_1, \param_1-1-\param_2,\param_1-1\}$,
 we have that $\Doj'\in\PLT$ and $(-1,<)\leq\Doj'\leq (0,\leq)$ is valid for~$\paramR$\longVersion{from \crefLemma{lemma:dijstrictzero}}.

 Also note that since $\clock_i\in\LFPd{\D}$, from \crefDef{def:largestclock} and \crefDef{def:med-PDBM}~(\refc{PDBMv})
 we have that~$(-1,<)\leq\Dji\leq (0,\leq)$ is valid for~$\paramR$ for all clock~$\clock_j$.
 Precisely, $\dji \in \{0,-\param_1,\param_2-\param_1, \param_1-1-\param_2,\param_1-1\}$ for some $\param_1,\param_2\in\Param$
 where $\param_2\leq\param_1$ \valid{}.
 Hence as $\dji+1 \in \{1,1-\param_1,\param_2+1-\param_1, \param_1-\param_2,\param_1\}$,
 we have that $\djo'\in\PLT$ and $(0,\leq)\leq\Djo'\leq (1,<)$ is valid for~$\paramR$.

 Clearly \crefDef{def:med-PDBM}~(\refc{PDBMi}) holds.

\medskip

 \noindent \textbf{proof that \crefDef{def:med-PDBM}~(\refc{PDBMv}) holds}

\medskip

 Third we set for all two clocks $i,j$ where $\clock_i\in\LFPd{\D}$, $\clock_j\in\Clock\setminus\LFPd{\D}$
 \[
 \Dij'=\Doj'
 \quad
 \mbox{and}
 \quad
 \Dji'=\Djo'\text{,}
 \]
  for all two clocks $\clock_j,\clock_k\in\Clock\setminus\LFPd{\D}$
  \[
 \Djk'=\Djk
 \]
 and for all two clocks $\clock,\clockk\in\LFPd{\D}$
  \[
 \Dxy'=\Dyx'=(0,\leq)\text{.}
 \]
 Here as we have already proven above that~$(-1,<)\leq\Doj'\leq(0,\leq)$ and $(0,\leq)\leq\Doj'\leq(1,<)$ are valid for~$\paramR$,
 \crefDef{def:med-PDBM}~(\refc{PDBMv}) holds.

\medskip

 \noindent \textbf{proof that \crefDef{def:med-PDBM}~(\refc{PDBMvi}) holds}

\medskip

   For all~$\clock_i$:
\begin{itemize}
 \item if~$\clock_i\in\LFPd{\D}$, $\Dio'=(0,\leq)$, $\Doi'=(0,\leq)$ hence~$\dio'=-\dzeroi'$ and~$\compOpLeq_{i0'}\compOpLeq_{0i'}={\leq}$, condition \crefDef{def:med-PDBM}~(\refc{PDBMvi}) holds;
 \item if~$\clock_i\in\Clock\setminus\LFPd{\D}$, $\clock\in\LFPd{\D}$, $\Dio'=\Dix+(1,\leq)$, $\Doi'=\Dxi+(-1,\leq)$ as condition \crefDef{def:med-PDBM}~(\refc{PDBMvi}) holds
 for~$\Dix$ and~$\Dxi$ and~$\compOpLeq_{ij}{\oplus}\leq=\compOpLeq_{ij}$, $\compOpLeq_{ji}{\oplus}\leq=\compOpLeq_{ji}$, condition \crefDef{def:med-PDBM}~(\refc{PDBMvi}) holds for~$\Dio'$ and~$\Doi'$.
\end{itemize}
  For all~$\clock_i, \clock_j$:
  \begin{itemize}
   \item if~$\clock_i, \clock_j\in\Clock\setminus\LFPd{\D}$, $\Dij'=\Dij$ and~$\Dji'=\Dji$, condition \crefDef{def:med-PDBM}~(\refc{PDBMvi}) holds as it holds for~$\Dij$ and~$\Dji$.
   \item if~$\clock_i\in\Clock\setminus\LFPd{\D}$, $\clock_j\in\LFPd{\D}$, $\Dij'=\Dij+(1,\leq)$, $\Dji'=\Dji+(-1,\leq)$ condition \crefDef{def:med-PDBM}~(\refc{PDBMvi}) holds
    for~$\Dij$ and~$\Dji$ and~$\compOpLeq_{ij}{\oplus}\leq=\compOpLeq_{ij}$, $\compOpLeq_{ji}{\oplus}\leq=\compOpLeq_{ji}$, condition \crefDef{def:med-PDBM}~(\refc{PDBMvi}) holds for~$\Dij'$ and~$\Dji'$.
    The case~$\clock_j\in\Clock\setminus\LFPd{\D}$, $\clock_i\in\LFPd{\D}$ is treated similarly.
    \item if~$\clock_i, \clock_j\in\LFPd{\D}$, $\Dij'=\Dji'=(0,\leq)$, hence~$\dij'=-\dji'=0$ and~$\compOpLeq_{ij'}\compOpLeq_{ji'}=\leq$ and condition \crefDef{def:med-PDBM}~(\refc{PDBMvi}) holds.
  \end{itemize}

\medskip

 \noindent \textbf{proof that \crefDef{def:med-PDBM}~(\refc{PDBMii}) holds}

\medskip

  Now we prove that \crefDef{def:med-PDBM}~(\refc{PDBMii}) holds, \ie{} for all clocks~$\clock_i, \clock_j,\clock_k$,
  valid conditions such as $\Dij'\leq\Dik'+\Dkj'$ remain valid in $\paramR$.
 This is not trivial since, in this construction some clocks have been updated.
 Precisely, for all clocks $\clock_i,\clock_j,\clock_k$, are valid for $\paramR$:
	\begin{enumerate}
		      \item if~$\clock_i,\clock_j,\clock_k\in\Clock\setminus\LFPd{\D}$: let $\clock\in\LFPd{\D}$ and
					\begin{itemize}
						  \item if~$i,j,k$ are different from $0$, we have~$\Dik'=\Dik$, $\Dij'=\Dij$ and~$\Djk'=\Djk$;
							since~$(\E,\D)\in\OPDBMs$, from \crefDef{def:med-PDBM}~(\refc{PDBMii}) we know that~$\Dik \leq \Dij+\Djk$ is valid for~$\paramR$;
							therefore, $\Dik'\leq\Dij'+\Djk'$ is valid for~$\paramR$.
						  \item if~$i,j$ are different from $0$, $k=0$, we have~$\Dio'=\Dix+(1,\leq)$, $\Dij'=\Dij$ and~$\Djo'=\Djx+(1,\leq)$;
							since~$(\E,\D)\in\OPDBMs$, from \crefDef{def:med-PDBM}~(\refc{PDBMii}) we know that~$\Dix \leq \Dij+\Djx$ is valid for~$\paramR$;
							then from \crefLemma{lemma:sumvalid} $\Dix+(1,\leq)\leq\Dij+\Djx+(1,\leq)$ is valid for~$\paramR$ and therefore, $\Dio'\leq\Dij'+\Djo'$ is valid for~$\paramR$.
						  \item if~$i,k$ are different from~$0$,~$j=0$, we have~$\Dik'=\Dik$, $\Dio'=\Dix+(1,\leq)$ and~$\Dok'=\Dxk+(-1,\leq)$;
							we claim that
							\begin{equation}\label{eq:onze}
							 \Dik\leq\Dix+(1,\leq)+\Dxk+(-1,\leq)
							 \quad
							\end{equation}
							is valid for~$\paramR$, which is equivalent to $\Dik' \leq \Dio'+\Dok'$ is valid for~$\paramR$.
							We have
							\begin{equation}\label{eq:douze}
							(1,\leq)+(-1,\leq)=(1+-1,\leq \oplus \leq)=(0,\leq)
							\quad
							\end{equation}
							Since~$(\E,\D)\in\OPDBMs$ from \crefDef{def:med-PDBM}~(\refc{PDBMii}), we know that~$\Dik \leq \Dix+\Dxk$ is valid for~$\paramR$;
							combining with (\refc{eq:douze}) and since $\Dxk+(0,\leq)=\Dxk$,
							we obtain (\refc{eq:onze}) and therefore our result.

							\item if~$i$ is different from~$0$,~$j=k=0$, we have~$\Dio'=\Dix+(1,\leq)$, $\Djk'=\Doo'=(0,\leq)$;
							we have from \crefDef{def:validity}~(\refc{valideiiid}) that
							\[
							\Dix+(1,\leq)\leq\Dix+(1,\leq)
							\]
							is valid for~$\paramR$.
							Hence \crefLemma{lemmaDiiDijGeq} gives that
							\[
							\Dio'\leq\Dio'+\Doo'
							\]
							is valid for~$\paramR$.
						\item if~$j,k$ are different from~$0$,~$i=0$, we have~$\Dok'=\Dxk+(-1,\leq)$, $\Doj'=\Dxj+(-1,\leq)$ and $\Djk'=\Djk$;
							since~$(\E,\D)\in\OPDBMs$, from \crefDef{def:med-PDBM}~(\refc{PDBMii}) we know that~$\Dxk \leq \Dxj+\Djk$ is valid for~$\paramR$.
							Moreover we have that
							\[
							(-1,\leq)\leq (-1,\leq)
							\]
							is valid for~$\paramR$ so we have from\longVersion{ \crefDef{def:validity}~(\refc{valideiiid}) and} \crefLemma{lemma:sumvalid}
							\[
							\Dxk+(-1,\leq)\leq\Dxj+(-1,\leq)+\Djk
							\]
							is valid for~$\paramR$.
							Hence $\Dok' \leq \Doj'+\Djk'$ is valid for $\paramR$.

						\item if~$j$ is different from~$0$,~$i=k=0$, we have~$\Doj'=\Dxj+(-1,\leq)$ and $\Djo'=\Djx+(1,\leq)$;
							since $(\E,\D)\in\OPDBMs$, from \crefLemma{lemmaDijDjiGeq} we know that~$(0,\leq) \leq \Dxj+\Djx$ is valid for~$\paramR$;
							moreover, we have that
							\[
							(1,\leq)+(-1,\leq)=(1+-1,\leq \oplus \leq)=(0,\leq)
							\]
							and $\Djx+(0,\leq)=\Djx$.
							Then we have from \crefLemma{lemma:sumvalid}
							\[
							(0,\leq)\leq\Dxj+(-1,\leq)+\Djx+(1,\leq)
							\]
							is valid for~$\paramR$ and therefore  $\Doo' \leq \Doj'+\Djo'$ is valid for $\paramR$.
						\item if~$k$ is different from~$0$,~$i=j=0$, we have~$\Dok'=\Dxk+(-1,\leq)$, $\Dij'=\Doo'=(0,\leq)$;
							we have from \crefDef{def:validity}~(\refc{valideiiid}) that
							\[
							\Dxk+(-1,\leq)\leq\Dxk+(-1,\leq)
							\]
							is valid for~$\paramR$.
							Hence, as~$\Dxk+(-1,\leq)+(0,\leq)=\Dxk+(-1,\leq)$ we have
							\[
							\Dok'\leq\Doo'+\Dok'
							\]
							is valid for~$\paramR$.
						\item if~$i=j=k=0$, we trivially have from\longVersion{ \crefDef{def:med-PDBM}~(\refc{PDBMii}) and} \crefLemma{lemmaDiiDijGeq}
							\[
							\Doo'\leq\Doo'+\Doo'
							\]
							is valid for~$\paramR$.

					\end{itemize}

		      \item if~$\clock_k\in\LFPd{\D}$ and~$\clock_i, \clock_j\in\Clock\setminus\LFPd{\D}$: $k\neq 0$ and
		      		\begin{itemize}
						  \item if~$i,j$ are different from $0$, we have~$\Dik'=\Dio'=\Dik+(1,\leq)$, $\Dij'=\Dij$ and~$\Djk'=\Djo'=\Djk+(1,\leq)$;
							since~$(\E,\D)\in\OPDBMs$, from \crefDef{def:med-PDBM}~(\refc{PDBMii}) we know that~$\Dik \leq \Dij+\Djk$ is valid for~$\paramR$;
							moreover, since we have $(1,\leq)\leq(1,\leq)$ is valid for~$\paramR$
							then from \crefLemma{lemma:sumvalid}
							\[
							\Dik+(1,\leq)\leq\Dij+\Djk+(1,\leq)
							\]
							is valid for~$\paramR$, therefore we have $\Dik'\leq\Dij'+\Djk'$ is valid for~$\paramR$.
						  \item if~$i\neq 0$,~$j=0$, we have~$\Dik'=\Dio'=\Dik+(1,\leq)$, $\Dio'=\Dik+(1,\leq)$ and~$\Dok'=(0,\leq)$;
						  	clearly
							\[
							(1,\leq)\leq(1,\leq)+(0,\leq)
							\]
							and
							\[
							\Dik\leq\Dik
							\]
							are valid for~$\paramR$, then from \crefLemma{lemma:sumvalid}
							we obtain~$\Dik'\leq\Dio'+\Dok'$ is valid for~$\paramR$.
						\item if~$i=0$,~$j\neq 0$, we have~$\Dok'=(0,\leq)$, $\Doj'=\Dkj+(-1,\leq)$ and $\Djk'=\Djo'=\Djk+(1,\leq)$;
							since~$(\E,\D)\in\OPDBMs$, from \crefLemma{lemmaDijDjiGeq}
							we know that~$(0,\leq) \leq \Dkj+\Djk$ is valid for~$\paramR$.
							Moreover we have that
							\[
							(1,\leq)+(-1,\leq)=(1+-1,\leq \oplus \leq)=(0,\leq)
							\]
							so we have from \crefLemma{lemma:sumvalid}
							\[
							(0,\leq)\leq\Dkj+\Djk+(0,\leq)
							\]
							is valid for~$\paramR$.
							Hence $\Dok' \leq \Doj'+\Djk'$ is valid for $\paramR$.
						\item if~$i=j=0$, we trivially have from\longVersion{ \crefDef{def:med-PDBM}~(\refc{PDBMii}) and} \crefLemma{lemmaDiiDijGeq}
							\[
							\Dok'\leq\Doo'+\Dok'
							\]
							is valid for~$\paramR$.
					\end{itemize}
		      \item if $\clock_j\in\LFPd{\D}$ and $\clock_i, \clock_k\in\Clock\setminus\LFPd{\D}$: $j\neq 0$ and
					\begin{itemize}
						  \item if~$i,k$ are different from $0$, we have~$\Dik'=\Dik$, $\Dij'=\Dio'=\Dij+(1,\leq)$ and~$\Djk'=\Dok'=\Djk+(-1,\leq)$;
							since~$(\E,\D)\in\OPDBMs$, from \crefDef{def:med-PDBM}~(\refc{PDBMii}) we know that~$\Dik \leq \Dij+\Djk$ is valid for~$\paramR$;
							moreover, since we have
							\[
							(1,\leq)+(-1,\leq)=(1+(-1),\leq \oplus \leq)=(0,\leq)
							\]
 							then as $\Dij+\Djk+(0,\leq)=\Dij+\Djk$,
							clearly $\Dik'\leq\Dij'+\Djk'$ is valid for~$\paramR$.
						  \item if~$i\neq 0$, $k=0$, we have~$\Dio'=\Dij+(1,\leq)$, $\Dij'=\Dio'=\Dij+(1,\leq)$ and~$\Djo'=(0,\leq)$;
						   	From \crefDef{def:validity}~(\refc{valideiiid}) we trivially have that
							$\Dij+(1,\leq)\leq\Dij+(1,\leq)$ is valid for~$\paramR$ and therefore, $\Dio'\leq\Dij'+\Djo'$ is valid for~$\paramR$.
						  \item if~$i=0$, $k\neq 0$, we have~$\Dok'=\Djk+(-1,\leq)$, $\Doj'=(0,\leq)$ and $\Djk'=\Dok'=\Djk+(-1,\leq)$;
							since~$(\E,\D)\in\OPDBMs$, from \crefDef{def:med-PDBM}~(\refc{PDBMii}) we know that~$\Dok \leq \Doj+\Djk$ is valid for~$\paramR$.
						   	From \crefDef{def:validity}~(\refc{valideiiid}) we trivially have that
						   	$\Djk+(-1,\leq)\leq\Djk+(-1,\leq)$ is valid for~$\paramR$.
							As~$(-1,\leq)+(0,\leq)=(-1,\leq)$, we have $\Dok' \leq \Doj'+\Djk'$ is valid for~$\paramR$.
						\item if~$i=k=0$, we have~$\Doj'=(0,\leq)$ and $\Djo'=(0,\leq)$;
							As we have
							\[
							(0,\leq)+(0,\leq)=(0,\leq)
							\]
							we clearly have that~$\Doo' \leq \Doj'+\Djo'$ is valid for $\paramR$.
					\end{itemize}

		      \item if $\clock_j, \clock_k\in\LFPd{\D}$ and $\clock_i\in\Clock\setminus\LFPd{\D}$: $j\neq0, k\neq 0$ and
					\begin{itemize}
						  \item if~$i$ is different from $0$, we have~$\Dik'=\Dio'=\Dik+(-1,\leq)$, $\Dij'=\Dio'=\Dik+(-1,\leq)$ and~$\Djk'=(0,\leq)$;
							we have that $(-1,\leq)+(0,\leq)=(-1,\leq)$ and
							\[
							\Dik+(-1,\leq)\leq\Dik+(-1,\leq)
							\]
							holds from \crefDef{def:validity}~(\refc{valideiiid}).
							Therefore,~$\Dik'\leq\Dij'+\Djk'$.
						  \item if~$i=0$, we have~$\Dok'=(0,\leq)$, $\Doj'=(0,\leq)$ and $\Djk'=(0,\leq)$;
							since~$(\E,\D)\in\OPDBMs$ from \crefDef{def:med-PDBM}~(\refc{PDBMii}), we know that~$\Dok \leq \Doj+\Djk$.
							As we have
							\[
							(0,\leq)+(0,\leq)=(0,\leq)
							\]
							we clearly have that $\Dok' \leq \Doj'+\Djk'$ is valid for~$\paramR$.
					\end{itemize}
		    \item if $\clock_i\in\LFPd{\D}$ and $\clock_j,\clock_k\in\Clock\setminus\LFPd{\D}$: $i\neq 0$ and
					\begin{itemize}
						  \item if~$j,k$ are different from $0$, we have~$\Dik'=\Dok'=\Dik+(-1,\leq)$, $\Dij'=\Doj'=\Dij+(-1,\leq)$ and~$\Djk'=\Djk$;
							since~$(\E,\D)\in\OPDBMs$, from \crefDef{def:med-PDBM}~(\refc{PDBMii}) we know that~$\Dik \leq \Dij+\Djk$ is valid for~$\paramR$;
							moreover, since we have
							\[
							(-1,\leq)\leq(-1,\leq)
							\]
							is valid for~$\paramR$ then from \crefLemma{lemma:sumvalid} we have~$\Dik'\leq\Dij'+\Djk'$ is valid for~$\paramR$.
						  \item if~$j\neq 0$, $k=0$, we have~$\Dio'=(0,\leq)$, $\Dij'=\Doj'=\Dij+(-1,\leq)$ and~$\Djo'=\Dji+(1,\leq)$;
						  	from \crefLemma{lemmaDijDjiGeq} we know that $(0,\leq)\leq \Dij+\Dji$ is valid for~$\paramR$.
						  	Moreover, we have
							\[
							(1,\leq)+(-1,\leq)=(1+(-1),\leq \oplus \leq)=(0,\leq)
							\]
							then
							\[
							(0,\leq)\leq\Dij+\Dji+(0,\leq)
							\]
							is valid for~$\paramR$ and therefore, $\Dio'\leq\Dij'+\Djo'$ is valid for~$\paramR$.
						  \item if~$j=0$, $k\neq 0$, we have~$\Dik'=\Dik$, $\Dio'=(1,\leq)$ and~$\Dok'=\Dik+(-1,\leq)$;
							we have that
							\[
							(1,\leq)+(-1,\leq)=(1+(-1),\leq \oplus \leq)=(0,\leq)
							\]
							and from \crefDef{def:validity}~(\refc{valideiiid}) that
							\[
							 \Dik\leq \Dik+(0,\leq)
							\]
							is valid for~$\paramR$, which gives us our result.
 						\item if~$i$ is different from~$0$,~$j=k=0$, we have~$\Dio'=(0,\leq)$, $\Djk'=\Doo'=(0,\leq)$;
							we have from \crefDef{def:validity}~(\refc{valideiiid}) that
							\[
							(0,\leq)\leq (0,\leq)
							\]
							is valid for~$\paramR$.
							Hence
							\[
							\Dio'\leq\Dio'+\Doo'
							\]
							is valid for~$\paramR$.
					\end{itemize}

		      \item if $\clock_i, \clock_k\in\LFPd{\D}$ and $\clock_j\in\Clock\setminus\LFPd{\D}$: $i\neq 0, k\neq 0$ and
		      			\begin{itemize}
						  \item if~$j\neq 0$, we have~$\Dik'=(0,\leq)$, $\Dij'=\Doj'=\Dij+(-1,\leq)$ and~$\Djk'=\Djo'=\Dji+(1,\leq)$;
							since~$(\E,\D)\in\OPDBMs$, from \crefLemma{lemmaDijDjiGeq}
							we know that~$(0,\leq) \leq \Dij+\Dji$ is valid for~$\paramR$;
							we have
							\[
							(1,\leq)+(-1,\leq)=(1+(-1),\leq \oplus \leq)=(0,\leq)
							\]
							and therefore from \crefLemma{lemma:sumvalid},~$\Dik'\leq\Dij'+\Djk'$ is valid for $\paramR$.
						  \item if~$j=0$, we have~$\Dik'=(0,\leq)$, $\Dio'=(0,\leq)$ and~$\Dok'=(0,\leq)$;
							we have that $(0,\leq)+ (0,\leq)=(0,\leq)$ and from \crefDef{def:validity}~(\refc{valideiiid})
							\[
							(0,\leq)\leq (0,\leq)
							\]
							is valid for~$\paramR$.
 							Therefore we obtain our result.
							\end{itemize}

		      \item if $\clock_i, \clock_j\in\LFPd{\D}$ and $\clock_k\in\Clock\setminus\LFPd{\D}$: $i\neq 0, j\neq 0$ and
					\begin{itemize}
						 \item if~$k\neq 0$, we have~$\Dik'=\Dok'=\Dik+(-1,\leq)$, $\Dij'=(0,\leq)$ and~$\Djk'=\Dok'=\Dik+(-1,\leq)$;
						      we have that
						      \[
						      \Dik\leq\Dik
						      \]
						      is valid for~$\paramR$ and from \crefLemma{lemma:sumvalid}
						      \[
						      (-1,\leq)\leq(-1,\leq)
						      \]
						      is valid for~$\paramR$. Therefore, $\Dik'\leq\Dij'+\Djk'$ is valid for~$\paramR$.
						  \item if~$k=0$, we have~$\Dio'=(0,\leq)$, $\Dij'=(0,\leq)$ and~$\Djo'=(0,\leq)$;
						      we have that $(0,\leq)+ (0,\leq)=(0,\leq)$ and from \crefDef{def:validity}~(\refc{valideiiid})
							\[
							(0,\leq)\leq (0,\leq)
							\]
							is valid for~$\paramR$: therefore~$\Dio'\leq\Dij'+\Djo'$ is valid for~$\paramR$.
					\end{itemize}

		      \item if $\clock_i,\clock_j, \clock_k\in\LFPd{\D}$:
		     				$i,j,k$ are different from $0$, we have~$\Dik'=(0,\leq)$, $\Dij'=(0,\leq)$ and~$\Djk'=(0,\leq)$;
							we have that $(0,\leq)+ (0,\leq)=(0,\leq)$ and from \crefDef{def:validity}~(\refc{valideiiid})
							\[
							(0,\leq)\leq (0,\leq)
							\]
							is valid for~$\paramR$: therefore, $\Dik'\leq\Dij'+\Djk'$ is valid for~$\paramR$.
	\end{enumerate}

\medskip

 \noindent \textbf{proof that \crefDef{def:med-PDBM}~(\refc{PDBMiii}) holds}

\medskip

Finally, there is at least one clock~$\clock_i\in\LFPd{\D}$ \st{} $\Doi=\Dio=(0,\leq)$.
Hence condition \crefDef{def:med-PDBM}~(\refc{PDBMiii}) holds.

 To conclude the proof, we set $\E_i'=\E_i+1$ if~$\clock_i\in\LFPd{\D}$ and $\E_j'=\E_j$ if~$\clock_j\in\Clock\setminus\LFPd{\D}$
 We denote by $(\E,\D')$ the obtained \mPDBM{} and $(\E',\D')\in\OPDBMs$.
 \end{preuve}

 \subsubsection{$\rightarrow$ \crefDef{definition:pointPDBM} to \crefDef{def:med-PDBM} type (\refc{PDBMiii})}

 \newcommand{\lemmaStablePointPDBM}{%
Let $\paramR$ be a parameter region and $(\E,\D)\in\CPr$, then $\TEF_{<}\big((\E,\D)\big)\in\OPDBMs$ respecting condition~\ref{PDBMiv}.
}
\begin{clm}[$\CPr$ becomes \OPDBMs{} after $\TEF_<$]%
\label{TECPr}
  \lemmaStablePointPDBM{}
\end{clm}

\begin{preuve}

 Suppose~$(\E,\D)\in\CPr$.
 Since, in~$\paramR$, we know which parameters have the largest fractional part, we can determine $\LFPd{\D}$ from \crefLemma{lemmaLargestFracClock}.
 If more than one clock belong to $\LFPd{\D}$ then their valuations have the same fractional part.

 Indeed, from \crefDef{def:largestclock} if $\clock_i, \clock_j\in\LFPd{\D}$ then both $(0,\leq)\leq\Dij$ and $(0,\leq)\leq\Dji$ are valid for~$\paramR$, and from \crefDef{def:med-PDBM}~(\refc{PDBMv})
 we must have~$\Dij=\Dji=(0,\leq)$.

 Let~$\pval\in\paramR$. Let $\clock_i\in\LFPd{\D}$ and $\clockval\in(\E,\pval(\D))$.
 By letting time elapse, $\partieFrac(\clockval(\clock_i))$ is the first that might reach $1$.
 Moreover, for all~$\clock_j\in\Clock\setminus\LFPd{\D}$, $\partieFrac(\clockval(\clock_j))$ cannot reach~$1$ before~$\partieFrac(\clockval(\clock_i))$.
 We are going to construct a new~$(\E',\D')=\TEF_<((\E,\D))$ which is an~\oPDBM{} respecting condition~\ref{PDBMiv}.
 While detailing the procedure of~$\TEF_<$, we are going to prove that \crefDef{def:med-PDBM}~(\refc{PDBMi}) and~(\refc{PDBMv}) hold for $(\E', \D')$. Further we will prove that~(\refc{PDBMii}) and~(\refc{PDBMiv}) also hold.

\medskip

 \noindent \textbf{proof that \crefDef{def:med-PDBM}~(\refc{PDBMi}) holds}

\medskip

According to the definition of~$\TEF_<$ (\cref{algorithm:TEup}\ea{réf dynamique manquante})
the first step is to set a new upper bound
 \[
  \Dio'=(1,<)
  \quad
  \mbox{for all } \clock_i\in\LFPd{\D}
 \]
 and obviously $(0,\leq)\leq\Dio'\leq (1,<)$ is valid for~$\paramR$.
 Then we set new upper bounds for all other clock $\clock_j\in\Clock\setminus\LFPd{\D}$ by setting
 \[
  \Djo'=\Dji+(1,<)\text{.}
 \]

 Indeed, $\Dji$ is the constraint on the lower bound of $\clockval(\clock_j)-\clockval(\clock_i)$
 and since the upper bound of~$\clock_i$ has increased, this gives the new upper bound of $\clock_j$.
 Note that since~$\clock_i\in\LFPd{\D}$, from \crefDef{def:largestclock} we have for all clock~$\clock_j$ that~$(-1,<)\leq\Dji\leq (0,\leq)$ is valid for~$\paramR$.
 Precisely, $\dji \in \{0,-\param_1,\param_2-\param_1, \param_1-1-\param_2,\param_1-1\}$ for some $\param_1,\param_2\in\Param$
 where $\param_2\leq\param_1$ \valid{}.
 Hence as $\dji+1 \in \{1,1-\param_1,\param_2+1-\param_1, \param_1-\param_2,\param_1\}$,
 we have that $\djo'\in\PLT$, $\compOpLeq_{j0'}=\compOpLeq_{j0'}\oplus{<}={<}$ and $(0,\leq)\leq\Djo'\leq (1,<)$ is valid for~$\paramR$.
 \longVersion{Note that we cannot have $(\dji,\compOpLeq_{ji})=(-1,<)$ because even if~$(\dij,\compOpLeq_{ij})=(1,<)$, since~$(\E,\D)\in\OPDBMs$ we do not have have $0\leq\Dji+\Dij$ is valid for~$\paramR$ from \crefDef{def:med-PDBM}~(\refc{PDBMii}) and \crefLemma{lemmaDijDjiGeq}.}

 Secondary we set for all clock $\clock$ regardless of whether they are in $\LFPd{\D}$ %
 \[
  \Dox'=\Dox+(0,<)\text{.}
 \]
  Since some time elapsed, lower bounds of all clocks are increased.
  Moreover, from \crefDef{definition:pointPDBM}~(\refc{PDBM2i}) as~$(-1,<)\leq\Dox\leq (0,\leq)$ is valid for~$\paramR$,
  $(-1,<)\leq\Dox'\leq (0,\leq)$ is also valid for~$\paramR$.

\medskip

 \noindent \textbf{proof that \crefDef{def:med-PDBM}~(\refc{PDBMv}) holds}

\medskip

  Third we set for all clocks $\clock, \clockk$ regardless of whether they are in $\LFPd{\D}$
  \[
  \Dxy'=\Dxy
  \]
  since no fractional part has reached~$1$, constraints on differences of clocks and integer parts remain unchanged.
  As it is the case in $(\E,\D)$, \crefDef{def:med-PDBM}~(\refc{PDBMv}) holds.

\medskip

 \noindent \textbf{proof that \crefDef{def:med-PDBM}~(\refc{PDBMvi}) holds}

\medskip

   For all~$\clock_i$:
\begin{itemize}
 \item if~$\clock_i\in\LFPd{\D}$, $\Dio'=(1,<)$, $\Doi'=\Doi+(0,<)$ hence~$\dio'\neq\dzeroi'$ and~$\compOpLeq_{i0'}\compOpLeq_{0i'}={<}$, condition \crefDef{def:med-PDBM}~(\refc{PDBMvi}) holds;
 \item if~$\clock_i\in\Clock\setminus\LFPd{\D}$, $\clock\in\LFPd{\D}$, $\Dio'=\Dix+(1,<)$, $\Doi'=\Doi+(0,<)$ hence as~$(0,\leq)\leq\Dio'$ \valid{} and~$\Doi'\leq(0,\leq)$ \valid{},
 we have~$\dio'\neq\dzeroi'$ and~$\compOpLeq_{i0'}\compOpLeq_{0i'}={<}$ and condition \crefDef{def:med-PDBM}~(\refc{PDBMvi}) holds.
\end{itemize}
  For all~$\clock_i, \clock_j$:
  \begin{itemize}
   \item if~$\clock_i, \clock_j\in\Clock\setminus\LFPd{\D}$, $\Dij'=\Dij$ and~$\Dji'=\Dji$, condition \crefDef{def:med-PDBM}~(\refc{PDBMvi}) holds as it holds for~$\Dij$ and~$\Dji$.
   \item if~$\clock_i\in\Clock\setminus\LFPd{\D}$, $\clock_j\in\LFPd{\D}$, $\Dio'=\Dij+(1,<)$, $\Doi'=\Doi+(0,<)$ hence as~$(0,\leq)\leq\Dio'$ \valid{} and~$\Doi'\leq(0,\leq)$ \valid{},
    we have~$\dio'\neq\dzeroi'$ and~$\compOpLeq_{i0'}\compOpLeq_{0i'}={<}$, condition \crefDef{def:med-PDBM}~(\refc{PDBMvi}) holds. The case~$\clock_j\in\Clock\setminus\LFPd{\D}$, $\clock_i\in\LFPd{\D}$ is treated similarly.
    \item if~$\clock_i, \clock_j\in\LFPd{\D}$, $\Dij'=\Dji'=(0,\leq)$, hence~$\dij'=-\dji'=0$ and~$\compOpLeq_{ij'}\compOpLeq_{ji'}=\leq$ and condition \crefDef{def:med-PDBM}~(\refc{PDBMvi}) holds.
  \end{itemize}

\medskip

 \noindent \textbf{proof that \crefDef{def:med-PDBM}~(\refc{PDBMii}) holds}

\medskip

 Now we prove that \crefDef{def:med-PDBM}~(\refc{PDBMii}) holds, \ie{} for all clocks $\clock_i,\clock_j,\clock_k$ valid conditions such as $\Dij'\leq\Dik'+\Dkj'$ remain valid in $\paramR$.
 Indeed, when time elapses, all clocks have the same behavior, hence the difference between two clocks does not change without an update.
 Precisely, for all clocks $\clock_i,\clock_j,\clock_k$, are valid for $\paramR$:
	\begin{enumerate}
		      \item if~$\clock_i,\clock_j,\clock_k\in\Clock\setminus\LFPd{\D}$: let $\clock\in\LFPd{\D}$ and
					\begin{itemize}
						  \item if~$i,j,k$ are different from $0$, we have~$\Dik'=\Dik$, $\Dij'=\Dij$ and~$\Djk'=\Djk$;
							since~$(\E,\D)\in\CPr$ from \crefDef{definition:pointPDBM} (\refc{PDBM2ii}), we know that~$\Dik \leq \Dij+\Djk$ is valid for~$\paramR$;
							therefore, $\Dik'\leq\Dij'+\Djk'$ is valid for~$\paramR$.
						  \item if~$i,j$ are different from $0$, $k=0$, we have~$\Dio'=\Dix+(1,<)$, $\Dij'=\Dij$ and~$\Djo'=\Djx+(1,<)$;
							since~$(\E,\D)\in\CPr$, from \crefDef{definition:pointPDBM} (\refc{PDBM2ii}) we know that~$\Dix \leq \Dij+\Djx$ is valid for~$\paramR$;
							then from \crefLemma{lemma:sumvalid} $\Dix+(1,<)\leq\Dij+\Djx+(1,<)$ is valid for~$\paramR$ and therefore, $\Dio'\leq\Dij'+\Djo'$ is valid for~$\paramR$.
						  \item if~$i,k$ are different from~$0$,~$j=0$, we have~$\Dik'=\Dik$, $\Dio'=\Dix+(1,<)$ and~$\Dok'=\Dok+(0,<)$;
							we claim that
							\begin{equation}\label{eq:treize}
							 \Dik\leq\Dix+(1,<)+\Dok+(0,<)
							 \quad
							\end{equation}
							is valid for~$\paramR$, which is equivalent to $\Dik' \leq \Dio'+\Dok'$ is valid for~$\paramR$.
							Since~$(\E,\D)\in\CPr$, from \crefDef{definition:pointPDBM} (\refc{PDBM2i}) we know that
							\begin{equation}\label{eq:quatorze}\Dxo \leq (1,<)\end{equation}
							is valid for~$\paramR$; moreover we have
							\begin{equation}\label{eq:quinze}
							(1,<)+(0,<)=(1+0,< \oplus <)=(1,<)\text{.}
							\quad
							\end{equation}
							Since~$(\E,\D)\in\CPr$, from \crefDef{definition:pointPDBM} (\refc{PDBM2ii}) we know that~$\Dxk \leq \Dxo+\Dok$ is valid for~$\paramR$;
							combining with (\refc{eq:quatorze}) and (\refc{eq:quinze}) %
							we obtain~$\Dxk\leq (1,<)+\Dok+(0,<)$ is valid for~$\paramR$.
 							As~$\Dix\leq\Dix$ is valid for~$\paramR$, using \crefLemma{lemma:sumvalid}
 							we obtain
 							\begin{equation}\label{eq:seize}
 							\Dix+\Dxk\leq \Dix+(1,<)+\Dok+(0,<)
							\quad
							\end{equation}
 							is valid for~$\paramR$.
 							Now, since~$(\E,\D)\in\CPr$, from \crefDef{definition:pointPDBM}~(\refc{PDBM2ii})
 							we know that~$\Dik \leq \Dix+\Dxk$ is valid for~$\paramR$
 							and combining with (\refc{eq:seize}) we obtain (\refc{eq:treize}) and therefore our result.
 						\item if~$i$ is different from~$0$,~$j=k=0$, we have~$\Dio'=\Dix+(1,<)$, $\Djk'=\Doo'=(0,\leq)$;
							we have from \crefDef{def:validity}~(\refc{valideiiid}) that
							\[
							\Dix+(1,<)\leq\Dix+(1,<)
							\]
							is valid for~$\paramR$.
							Hence
							\[
							\Dio'\leq\Dio'+\Doo'
							\]
							is valid for~$\paramR$.
						\item if~$j,k$ are different from~$0$,~$i=0$, we have~$\Dok'=\Dok+(0,<)$, $\Doj'=\Doj+(0,<)$ and $\Djk'=\Djk$;
							since~$(\E,\D)\in\CPr$, from \crefDef{definition:pointPDBM} (\refc{PDBM2ii}) we know that~$\Dok \leq \Doj+\Djk$ is valid for~$\paramR$.
							Moreover we have that
							\[
							\Dok+(0,<) = (\dok,<)
							\quad
							\mbox{and}
							\quad
							\Doj+(0,<)+\Djk	= (\doj+\djk,<)
							\]
							so we have from \crefLemma{lemma:sumvalid}
							\[
							\Dok+(0,<)\leq\Doj+(0,<)+\Djk
							\]
							is valid for~$\paramR$.
							Hence $\Dok' \leq \Doj'+\Djk'$ is valid for $\paramR$.
						\item if~$j$ is different from~$0$,~$i=k=0$, we have~$\Dik'=\Doo'=(0,\leq)$, $\Doj'=\Doj+(0,<)$ and $\Djo'=\Djx+(1,<)$;
							since $(\E,\D)\in\CPr$, from \crefDef{definition:pointPDBM} (\refc{PDBM2ii})
							we know that~$\Dox \leq \Doj+\Djx$ is valid for~$\paramR$;
							moreover from \crefLemma{lemma:sumvalid},
							\[
							\Dox+(0,<) \leq \Doj+(0,<)+\Djx
							\]
							is valid for~$\paramR$.
							As we have
							\[
							(1,<)+(0,<)=(1+0,< \oplus <)=(1,<)
							\]
							we obtain from \crefLemma{lemma:sumvalid} that
							\[
							\Dox+(1,<)\leq\Doj+\Djx+(1,<)
							\]
							is valid for~$\paramR$.
							Recall that from \crefLemma{lemmaDijDjiGeq} $(0,\leq)\leq\Dox+\Dxo$ is valid for~$\paramR$.
							Since from \crefDef{definition:pointPDBM}~(\refc{PDBM2i}) $\Dxo\leq (1,<)$ is valid for~$\paramR$,
							we have $(0,\leq)\leq\Dox+(1,<)$ is valid for~$\paramR$.
							Therefore~$\Doo' \leq \Doj'+\Djo'$ is valid for~$\paramR$.
						\item if~$k$ is different from~$0$,~$i=j=0$, we have~$\Dik'=\Djk'=\Dok'=\Dok+(0,<)$, $\Dij'=\Doo'=(0,\leq)$;
							we have from \crefDef{def:validity}~(\refc{valideiiid}) that
							\[
							\Dok+(0,<)\leq\Dok+(0,<)
							\]
							is valid for~$\paramR$.
							Hence from \crefLemma{lemmaDiiDijGeq}
							\[
							\Dok'\leq\Doo'+\Dok'
							\]
							is valid for~$\paramR$.
						\item if~$i=j=k=0$, we trivially have
							\[
							\Doo'\leq\Doo'+\Doo'
							\]
							is valid for~$\paramR$.

					\end{itemize}
		      \item if~$\clock_k\in\LFPd{\D}$ and~$\clock_i, \clock_j\in\Clock\setminus\LFPd{\D}$: $k\neq 0$ and
		      		\begin{itemize}
						  \item if~$i,j$ are different from $0$, we have~$\Dik'=\Dik$, $\Dij'=\Dij$ and~$\Djk'=\Djk$;
							since~$(\E,\D)\in\CPr$, from \crefDef{definition:pointPDBM} (\refc{PDBM2ii})
							we know that~$\Dik \leq \Dij+\Djk$ is valid for~$\paramR$;
							therefore, $\Dik'\leq\Dij'+\Djk'$ is valid for~$\paramR$.
						  \item if~$i\neq 0$,~$j=0$, we have~$\Dik'=\Dik$, $\Dio'=\Dik+(1,<)$ and~$\Dok'=\Dok+(0,<)$;
						  	we claim that $\Dik\leq\Dik+(1,<)+\Dok+(0,<)$, \ie{}
							\begin{equation}\label{eq:dixsept}
							 0\leq (1,<)+\Dok+(0,<)
							 \quad
							\end{equation}
							is valid for~$\paramR$, which is from \crefLemma{lemma:sumvalid} equivalent to $\Dik' \leq \Dio'+\Dok'$ is valid for~$\paramR$.
							We have
							\begin{equation}\label{eq:dixhuit}
							(1,<)+(0,<)=(1+0,< \oplus <)=(1,<)\text{.}
							\quad
							\end{equation}
							Since~$(\E,\D)\in\CPr$, from \crefDef{definition:pointPDBM}~(\refc{PDBM2ii})
							we know that~$0\leq\Dok+\Dko$ is valid for~$\paramR$
							and from \crefDef{definition:pointPDBM}~(\refc{PDBM2i})
							that~$\Dko \leq (1,<)$ is valid for~$\paramR$;
							combining with (\refc{eq:dixhuit}) we obtain (\refc{eq:dixsept}) and therefore our result.

							\item if~$i=0$,~$j\neq 0$, we have~$\Dok'=\Dok+(0,<)$, $\Doj'=\Doj+(0,<)$ and $\Djk'=\Djk$;
							since~$(\E,\D)\in\CPr$, from \crefDef{definition:pointPDBM} (\refc{PDBM2ii})
							we know that~$\Dok \leq \Doj+\Djk$ is valid for~$\paramR$.
							Moreover we have that~$(0,<)\leq(0,<)$ is valid for~$\paramR$
							so we have from \crefLemma{lemma:sumvalid}
							\[
							\Dok+(0,<)\leq\Doj+(0,<)+\Djk
							\]
							is valid for~$\paramR$.
							Hence $\Dok' \leq \Doj'+\Djk'$ is valid for $\paramR$.

							\item if~$i=j=0$, from \crefDef{definition:pointPDBM}~(\refc{PDBM2ii}) we trivially have
							\[
							\Dok'\leq\Doo'+\Dok'
							\]
							is valid for~$\paramR$.
					\end{itemize}

		      \item if $\clock_j\in\LFPd{\D}$ and $\clock_i, \clock_k\in\Clock\setminus\LFPd{\D}$: $j\neq 0$ and
					\begin{itemize}
						  \item if~$i,k$ are different from $0$, we have~$\Dik'=\Dik$, $\Dij'=\Dij$ and~$\Djk'=\Djk$;
							since~$(\E,\D)\in\CPr$, from \crefDef{definition:pointPDBM} (\refc{PDBM2ii})
							we know that~$\Dik \leq \Dij+\Djk$ is valid for~$\paramR$;
							therefore, $\Dik'\leq\Dij'+\Djk'$ is valid for~$\paramR$.
						  \item if~$i\neq 0$, $k=0$, we have~$\Dio'=\Dij+(1,<)$, $\Dij'=\Dij$ and~$\Djo'=(1,<)$;
						   	from \crefDef{def:validity}~(\refc{valideiiid}) we trivially have that
							$\Dij+(1,<)\leq\Dij+(1,<)$ \valid{} and therefore, $\Dio'\leq\Dij'+\Djo'$ is valid for~$\paramR$.
						\item if~$i=0$, $k\neq 0$, we have~$\Dok'=\Dok+(0,<)$, $\Doj'=\Doj+(0,<)$ and $\Djk'=\Djk$;
							since~$(\E,\D)\in\CPr$, from \crefDef{definition:pointPDBM} (\refc{PDBM2ii})
							we know that~$\Dok \leq \Doj+\Djk$ \valid{}.
							Moreover we have that~$(0,<)\leq(0,<)$ \valid{}
							so we have
							\[
							\Dok+(0,<)\leq\Doj+(0,<)+\Djk
							\]
							holds from \crefDef{def:validity}~(\refc{valideiiid}).
							Hence $\Dok' \leq \Doj'+\Djk'$ is valid for $\paramR$.
						\item if~$i=k=0$, we have~$\Doj'=\Doj+(0,<)$ and $\Djo'=(1,<)$;
							since $(\E,\D)\in\CPr$, from \crefLemma{lemmaDijDjiGeq} we know that~$0 \leq \Doj+\Djo$ \valid{},
							from \crefDef{definition:pointPDBM}~(\refc{PDBM2i}) we know that $\Djo\leq 1$ \valid{}
							which means $0\leq\Doj+(1,<)$ \valid{}.
							As we have
							\[
							(1,<)+(0,<)=(1+0,< \oplus <)=(1,<)
							\]
							we obtain that
							\[
							(0,\leq)\leq\Doj+(0,<)+(1,<)
							\]
							\valid{} and therefore  $\Doo' \leq \Doj'+\Djo'$ is valid for~$\paramR$.
					\end{itemize}

		      \item if $\clock_j, \clock_k\in\LFPd{\D}$ and $\clock_i\in\Clock\setminus\LFPd{\D}$: $j\neq0, k\neq 0$ and
					\begin{itemize}
						  \item if~$i$ is different from $0$, we have~$\Dik'=\Dik$, $\Dij'=\Dij$ and~$\Djk'=\Djk$;
							since~$(\E,\D)\in\CPr$, from \crefDef{definition:pointPDBM} (\refc{PDBM2ii})
							we know that~$\Dik \leq \Dij+\Djk$ \valid{};
							therefore, $\Dik'\leq\Dij'+\Djk'$ \valid{}.
					\item if~$i=0$, we have~$\Dok'=\Dok+(0,<)$, $\Doj'=\Doj+(0,<)$ and $\Djk'=\Djk$;
							since~$(\E,\D)\in\CPr$, from \crefDef{definition:pointPDBM} (\refc{PDBM2ii})
							we know that~$\Dok \leq \Doj+\Djk$ \valid{}.
							Moreover we have that $(0,<)\leq(0,<)$ \valid{}
							so we have from \crefLemma{lemma:sumvalid}
							\[
							\Dok+(0,<)\leq\Doj+(0,<)+\Djk
							\]
							\valid{}.
							Hence $\Dok' \leq \Doj'+\Djk'$ is valid for~$\paramR$.
					\end{itemize}
		      \item if $\clock_i\in\LFPd{\D}$ and $\clock_j,\clock_k\in\Clock\setminus\LFPd{\D}$: $i\neq 0$ and
					\begin{itemize}
						  \item if~$j,k$ are different from~$0$, we have~$\Dik'=\Dik$, $\Dij'=\Dij$ and~$\Djk'=\Djk$;
							since~$(\E,\D)\in\CPr$, from \crefDef{definition:pointPDBM} (\refc{PDBM2ii})
							we know that~$\Dik \leq \Dij+\Djk$ \valid{};
							therefore, $\Dik'\leq\Dij'+\Djk'$ \valid{}.
						  \item if~$j\neq 0$, $k=0$, we have~$\Dio'=(1,<)$, $\Dij'=\Dij$ and~$\Djo'=\Dji+(1,<)$;
						  	since~$(\E,\D)\in\CPr$, from \crefLemma{lemmaDijDjiGeq} we know that $0\leq \Dij+\Dji$.
							Then from \crefLemma{lemma:sumvalid}
							\[
							(1,<)\leq\Dij+\Dji+(1,<)
							\]
							\valid{} and therefore, $\Dio'\leq\Dij'+\Djo'$ is valid for~$\paramR$.
						  \item if~$j=0$, $k\neq 0$, we have~$\Dik'=\Dik$, $\Dio'=(1,<)$ and~$\Dok'=\Dok+(0,<)$;
							we claim that
							\[
							 \Dik\leq (1,<)+\Dok+(0,<)
							\]
							\valid{}, which is equivalent to $\Dik' \leq \Dio'+\Dok'$ \valid{}.
							Since~$(\E,\D)\in\CPr$ from \crefDef{definition:pointPDBM} (\refc{PDBM2ii}),
							we know that~$\Dik \leq \Dio+\Dok$ \valid{};
							moreover, from \crefDef{definition:pointPDBM} (\refc{PDBM2i}),
							we know that~$\Dio \leq (1,<)$ \valid{}.
							We have
							\[
							(1,<)+(0,<)=(1+0,< \oplus <)=(1,<)
							\]
							We obtain that
							\[
							\Dik\leq \Dio+\Dok\leq (1,<)+\Dok=(1,<)+\Dok+(0,<)
							\]
 							\valid{} and therefore our result.
 						\item if~$i$ is different from~$0$,~$j=k=0$, we have~$\Dio'=(1,<)$, $\Djk'=\Doo'=(0,\leq)$;
							from \crefDef{def:validity}~(\refc{valideiiid}) we have that
							\[
							(1,<)\leq (1,<)
							\]
							\valid{}.
							Hence
							\[
							\Dio'\leq\Dio'+\Doo'
							\]
							\valid{}.
					\end{itemize}

		      \item if $\clock_i, \clock_k\in\LFPd{\D}$ and $\clock_j\in\Clock\setminus\LFPd{\D}$: $i\neq 0, k\neq 0$ and
		      			\begin{itemize}
						  \item if~$j\neq 0$, we have~$\Dik'=\Dik$, $\Dij'=\Dij$ and~$\Djk'=\Djk$;
							since~$(\E,\D)\in\CPr$, from \crefDef{definition:pointPDBM} (\refc{PDBM2ii})
							we know that~$\Dik \leq \Dij+\Djk$ \valid{};
							therefore, $\Dik'\leq\Dij'+\Djk'$ \valid{}.
						  \item if~$j=0$, we have~$\Dik'=\Dik$, $\Dio'=(1,<)$ and~$\Dok'=\Dok+(0,<)$;
							we claim that
							\[
							 \Dik\leq (1,<)+\Dok+(0,<)
							\]
							\valid{}, which is equivalent to $\Dik' \leq \Dio'+\Dok'$ \valid{}.
							Since~$(\E,\D)\in\CPr$ from \crefDef{definition:pointPDBM} (\refc{PDBM2ii}),
							we know that~$\Dik \leq \Dio+\Dok$ \valid{};
							moreover, from \crefDef{definition:pointPDBM} (\refc{PDBM2i}),
							we know that~$\Dio \leq (1,<)$ \valid{}.
							We have
							\[
							(1,<)+(0,<)=(1+0,< \oplus <)=(1,<)
							\]
							We obtain that
							\[
							\Dik\leq \Dio+\Dok\leq (1,<)+\Dok=(1,<)+\Dok+(0,<)
							\]
 							\valid{} and therefore our result.
		      			\end{itemize}

		      \item if $\clock_i, \clock_j\in\LFPd{\D}$ and $\clock_k\in\Clock\setminus\LFPd{\D}$: $i\neq 0, j\neq 0$ and
					\begin{itemize}
						  \item if~$k\neq 0$, we have~$\Dik'=\Dik$, $\Dij'=\Dij$ and~$\Djk'=\Djk$;
							since~$(\E,\D)\in\CPr$, from \crefDef{definition:pointPDBM} (\refc{PDBM2ii}),
							we know that~$\Dik \leq \Dij+\Djk$ \valid{};
							therefore, $\Dik'\leq\Dij'+\Djk'$ \valid{}.
						  \item if~$k=0$, since both $\clock_i, \clock_j\in\LFPd{\D}$ we have~$\Dij'=\Dij=(0,\leq)$, $\Dio'=(1,<)$ and~$\Djo'=(1,<)$;
							trivially $(1,<)\leq(0,\leq)+(1,<)$ \valid{} and therefore, $\Dio'\leq\Dij'+\Djo'$ \valid{}.
					\end{itemize}

		      \item if $\clock_i,\clock_j, \clock_k\in\LFPd{\D}$:
		     				$i,j,k$ are different from $0$, we have~$\Dik'=\Dik$, $\Dij'=\Dij$ and~$\Djk'=\Djk$;
						since~$(\E,\D)\in\CPr$, from \crefDef{definition:pointPDBM} (\refc{PDBM2ii}) we know that~$\Dik \leq \Dij+\Djk$ \valid{};
						therefore, $\Dik'\leq\Dij'+\Djk'$ \valid{}.
	\end{enumerate}

\medskip

 \noindent \textbf{proof that \crefDef{def:med-PDBM}~(\refc{PDBMiv}) holds}

\medskip

 Finally, for~$\clock_i\in\LFPd{\D}$, $\Dio'=(1,<)$ and for all clock~$j$ \st{} $\Doj'=(0,\compOpLeq)$, then we have~${\compOpLeq}={<}$.
 Condition \crefDef{def:med-PDBM}~(\refc{PDBMiv}) is satisfied.

 We set~$\E'=\E$ and denote by $(\E,\D')$ the obtained \mPDBM{}, which is $(\E,\D')\in\OPDBMs$.
\end{preuve}

}
\begin{preuve}
\versionProofIn{
\prooflemmaTEstable{}
}
\versionProofOut{
Although we perform some additions such as~$\Dji+(1,<)$, we do not create new expressions that are not in~$\PLT$.
In fact, this addition is performed on a negative term (\eg{} $\partieFrac(\param)-1$), as~$\clock_i$ is a clock with the largest fractional part and adding~$1$ transforms it into another term of~$\PLT$. The intuition is similar when performing additions such as~$\Dij+(-1,\leq)$: as~$\clock_i$ is a clock with the largest fractional part, $\dij$ is a positive term. The canonical form is also preserved by the last setting operations of the algorithm, as in the update operator. Therefore~$\TEF((\E,\D))$ is a \mPDBM{}.
	\longVersion{See \cref{appendix:proofTEstable} for details.}%
}
\end{preuve}

\longVersion{
	Note that, by \crefLemma{lemmaTEstable} $(\E',\D')$ is a \mPDBM{}.
	\oPDBMs{} are stable under $\TEF_<$ and $\TEF_=$, switching the condition they respect (\ref{PDBMiii},~\ref{PDBMiv}).
	Applying $\TEF_<$ on a \pointPDBM{} transforms it into an \oPDBM{}.
}

The following proposition proves that time elapsing behaves as we expect.

\newcommand{\propTimeElapsingLong}{%
 Let $\paramR$ be a parameter region and $(\E,\D)\in\PDBMRp$. Let $\pval\in\paramR$.
 There exists $\clockval'\in\TEF((\E, \valuate{\D}{\pval}))$ iff
 there exist $\clockval\in(\E, \valuate{\D}{\pval})$ and a delay~$\delta$ \st{} $\clockval'=\clockval+\delta$.
 }
\begin{prop}[semantics of~\mPDBM{} under~$\TEF$]%
\label{ssiTimeElapsing}
	\propTimeElapsingLong{}
\end{prop}

\newcommand{\proofpropTimeElapsingLong}{
 Note that this proof is inspired by~\cite[Proof of Lemma 3.13]{HRSV02}. We treat first~\oPDBMs{} and then~\pointPDBMs{}.

 \begin{clm}\label{presentfuturPDBM}
  Let $(\E,\D)\in\OPDBMs$. If~$(\E,\D)$ satisfies condition \crefDef{def:med-PDBM}~(\refc{PDBMiv})
  it has been obtained after applying \cref{algorithm:TEup}\ea{réf dynamique manquante}
  on another \oPDBM{} satisfying condition \crefDef{def:med-PDBM}~(\refc{PDBMiii}) or a \pointPDBM{}.

  Let $(\E,\D)\in\OPDBMs$. If~$(\E,\D)$ satisfies condition \crefDef{def:med-PDBM}~(\refc{PDBMiii})
  it has been obtained after applying \cref{algorithm:TEeq}\ea{réf dynamique manquante}
  on another \oPDBM{} satisfying condition \crefDef{def:med-PDBM}~(\refc{PDBMiv}) or
  after a non-parametric update applied on another \oPDBM{} or a \pointPDBM{}.
\end{clm}
 \begin{preuve}
  Let $(\E,\D)\in\OPDBMs$ and suppose~$(\E,\D)$ satisfies condition \crefDef{def:med-PDBM}~(\refc{PDBMiv}).
  Since for all~$\clockk$, if~$\doy=0$ we have~$\compOpLeq_{0y}={<}$, from \refClaim{T1toT2reset} and \refClaim{T2toT2reset}
  it cannot be the result of a non-parametric update where there is at least a clock~$\clock$ update and~$\Dxo=\Dox=(0,\leq)$.
  From \refClaim{TE2mPDBM} it cannot be the result of \cref{algorithm:TEeq}\ea{réf dynamique manquante},
  as there must be at least a clock~$\clock$ \st{} $\Dxo=\Dox=(0,\leq)$.
  Then it is the result either from \refClaim{TEmPDBM} of \cref{algorithm:TEup}\ea{réf dynamique manquante}
  applied on an \oPDBM{} satisfying condition \crefDef{def:med-PDBM}~(\refc{PDBMiii}),
  or from \refClaim{TECPr} of \cref{algorithm:TEup}\ea{réf dynamique manquante}
  applied on a \pointPDBM{}.

  Let $(\E,\D)\in\OPDBMs$ and suppose~$(\E,\D)$ satisfies condition \crefDef{def:med-PDBM}~(\refc{PDBMiii}).
  Since there is at least a clock~$\clockk$ \st{} $\Dyo=\Doy=(0,\leq)$, from \cref{TEmPDBM,TECPr} it cannot be the result of \cref{algorithm:TEup}\ea{réf dynamique manquante},
  as for all~$\clock$, if~$\dox=0$ we must have~$\compOpLeq_{ox}={<}$.
  Then it is the result of either from \refClaim{TE2mPDBM} of \cref{algorithm:TEeq}\ea{réf dynamique manquante}
  applied on an \oPDBM{} satisfying condition \crefDef{def:med-PDBM}~(\refc{PDBMiv})
  or from \cref{T1toT2reset,T2toT2reset} of \cref{algorithm:reset}\ea{réf dynamique manquante}
  applied on an \oPDBM{} or a \pointPDBM{}.
 \end{preuve}

\bigskip

 \ea{transforme en lemme et mets une petite preuve}

	  Let $\paramR$ be a parameter region and $(\E,\D)\in\PDBMRp$.
	  We have to consider two different cases: $(\E,\D)\in\OPDBMs$ and $(\E,\D)\in\CPr$.

	  \newcommand{\lemmaTimeElapsing}{%
		    Let $\paramR$ be a parameter region and $(\E,\D)\in\OPDBMs$. Let $\pval\in\paramR$.
		    There is $\clockval'\in\TEF((\E, \valuate{\D}{\pval}))$ iff
		    there is $\clockval\in(\E, \valuate{\D}{\pval})$ and a delay~$\de$ \st{} $\clockval'=\clockval+\de$.
		    }
	  \begin{clm}\label{ssiTEF}

		  \lemmaTimeElapsing{}
	  \end{clm}

	  \ea{attention, le $d$ prete \`a confusion car il y a d\'ej\`a la matrice ! $t$ ?} 

	  \begin{preuve}%
	  \label{proofssiTEF}
		    Let $\paramR$ a parameter region and $(\E,\D)\in\OPDBMs$. Let $\pval\in\paramR$.
		      \begin{itemize}[$\Longrightarrow$]

		      \item \textbf{\oPDBM{} respecting \crefDef{def:med-PDBM}~(\refc{PDBMiii})}

          \medskip

		      Let~$\pval\in\paramR$. Consider $(\E',\D')=\TEF((\E, \D))$ respecting condition \crefDef{def:med-PDBM}~(\refc{PDBMiii}), \ie{}
		      suppose there is $\clock_{i}$ \st{} $\Dio'=-\Doi'=(0,\leq)$.
		      Let~$\clockval'\in(\E', \pval(\D'))$, for this~$\clock_i$ we have~$\clockval'(\clock_{i})=0$.
			We need to find a value $\de$ \st{} $\clockval'-\de\in(\E, \valuate{\D}{\pval})$
		      which is equivalent to prove for all $\clock_i,\clock_j$
		      \[
		       \partieFrac(\clockval'(\clock_j))-\de-\partieFrac(\clockval'(\clock_i))+\de\compOpLeq_{ji}\pval(\dji)
		       \]
		       and
		       \[
		       \partieFrac(\clockval'(\clock_i))-\de-\partieFrac(\clockval'(\clock_j))+\de\compOpLeq_{ij}\pval(\dij)
		       \]
		       and
		       \[
		        -\partieFrac(\clockval'(\clock_j))+\de\compOpLeq_{0j}\pval(\doj)
		       \quad
		       \text{and}
		       \quad
		       \partieFrac(\clockval'(\clock_j))-\de\compOpLeq_{j0}\pval(\djo)\text{.}
		      \]

			 In this proof we are going to define a~$\de$ which is different from~$0$,
			 and give it an upper bound in order to show that constraints in~$(\E,\D)$
			 are satisfied while going backward of$~\de$ units of time from~$\clockval'$.

			 First we will prove that for all clock~$j$, its constraints of lower bound~$\Doj$ and upper bound~$\Djo$ are satisfied.
			 Second we will prove that for all~$i$, bounds on their difference~$\Dij$ and~$\Dji$ are also satisfied.

			We want to show that we have to go a little backward in time from~$\clockval'$
			to ensure the upper bounds~$\Djo$ of~$(\E,\D)$ hold.
			For this purpose, we are going to prove that for all~$\clock_j$
			\[
			 \Djo\leq\Djo'
			\]
			is valid for~$\paramR$. Intuitively this means upper bounds of clocks in~$(\E',\D')$ are greater than in~$(\E,\D)$,
			which is consistent as time is elapsing.

			As~$(\E',\D')$ respects \crefDef{def:med-PDBM}~(\refc{PDBMiii}) and precisely $(\E',\D')=\TEF_=((\E, \D))$,
			we know~$(\E,\D)$ is respecting condition \crefDef{def:med-PDBM}~(\refc{PDBMiv}) from \refClaim{TEmPDBM}.
			As $\partieFrac(\clockval'(\clock_{i}))=0$ it was in~$(\E,\D)$ a clock with the largest fractional part,
			\ie{} $\clock_i\in\LFPd{\D}$ and $\Dio=(1,<)$.

			By definition of~$\TEF_<$ (cf.\ \cref{algorithm:TEup}\ea{réf dynamique manquante}), %
			in~$(\E,\D)$ which is the \oPDBM{} obtained after the application of $\TEF_<$ on another \mPDBM{} (see \refClaim{presentfuturPDBM}),
			for each $\clock_j\in\Clock\setminus\LFPd{\D}$, $\Djo=\Dji+(1,<)$
			and for all~$\clock_{j} \in \Clock $, we have~$\Djo$ is of the form $(\djo,<)$ for some~$\djo$.

			By definition of~$\TEF_=$ applied to~$(\E,\D)$ (cf.\ \cref{algorithm:TEeq}\ea{réf dynamique manquante}), %
			in~$(\E',\D')$, for each~$\clock_j\in\Clock\setminus\LFPd{\D}$,
			$\Djo'=\Dji+(1,\leq)$, \ie{} $\djo=\djo'$.
			Hence by \crefDef{def:validity}~(\refc{valideiiid}) and as~$\compOpLeq_{j0'}$ is either~$\leq$ or~$<$, we have
			\[
			 (\djo,<)=\Djo\leq\Djo'=(\djo,\compOpLeq_{j0'})
			\]
			is valid for~$\paramR$.
			Next we define the largest amount of time so that all upper bounds of $(\E,\D)$ are satisfied.

  			We claim that for all~$\clock_j$, $\partieFrac(\clockval'(\clock_j))-\pval(\djo)\leq 0$.
 			\ea{petit commentaire car c'est contre intuitif}
 			Indeed, remark that by applying \cref{algorithm:TEup}\ea{réf dynamique manquante} then 3,
			constraints on upper bounds of clocks
 			in~$(\E,\D)$ and~$(\E',\D')$ differ only by their~$\compOpLeq$. As for~$i\in\LFPd{\D}$
 			and~$j\in\Clock\setminus\LFPd{\D}$ it we have~$\Djo=\Dji+(1,<)$ in~$(\E,\D)$
 			and~$\Djo'=\Dji+(1,\leq)$ in~$(\E',\D')$, so~$\djo=\djo'$.
 			Since for any~$\clock$, its fractional part is less or equal to its upper bound in~$\D$ and therefore in~$\D'$,
 			any difference between a fractional part and its upper bound is either negative or null.
 			For all~$\clock$, since $\partieFrac(\clockval'(\clock))\compOpLeq_{x0'}\pval(\dxo')$
 			we have $\partieFrac(\clockval'(\clock))-\pval(\dxo')\compOpLeq_{x0'}0$.
 			Since $\pval(\dxo')=\pval(\dxo)$, $\partieFrac(\clockval'(\clock))-\pval(\dxo)\compOpLeq_{x0'}0$,
 			therefore we have our result.

			Now we claim that we have to go at least an~$\epsilon>0$ backward in time to ensure all bounds of~$(\E,\D)$ are met.
			Let~$\clock_j\in\Clock\setminus\LFPd{\D}$. As
			\[
			\partieFrac(\clockval'(\clock_j))\compOpLeq_{j0'}\pval(\djo)
			\]
			we have
			\begin{itemize}
				\item either $\compOpLeq_{j0'}= {<}$ and we already have
				$\partieFrac(\clockval'(\clock_j))<\pval(\djo)$,
				\item or $\compOpLeq_{j0'}= {\leq}$ and for any~$\epsilon>0$
				we have~$\partieFrac(\clockval'(\clock_j))-\epsilon<\pval(\djo)$.
			\end{itemize}

			It is also true for each~$\clock_i\in\LFPd{\D}$: after applying~$\TEF_<$ recall that we have~$\Dio=(1,<)$.
			We can take $\epsilon>0$ and define~$\partieFrac(\clockval(\clock_i))=1-\epsilon$,
			so we have~$\partieFrac(\clockval(\clock_i))<\pval(\dio)$.

			Now that we know we have to go a little backward in time (at least an~$\epsilon>0$)
			so upper bounds of~$(\E,\D)$ are satisfied, we are going to give an upper bound to~$\epsilon$
			so that all lower bounds~$\Doj$ of~$(\E,\D)$ are also satisfied.

			Let
			\[
			 t_1=\min\limits_{\clock\in\Clock}\{\partieFrac(\clockval'(\clock))+\pval(\dox)\}
			\]
			We want to prove that $t_1>0$.

			Let us prove that for all~$\clock_j$, $\Doj'\leq\Doj$ is valid for~$\paramR$.
			Recall that for $\clock_i\in\LFPd{\D}$, we have that $\Dio=(1,<)$.
			Moreover, from \crefDef{def:med-PDBM}~(\refc{PDBMii}) $\Dij\leq\Dio+\Doj$ \valid{}, then we have
			\[
			 \Dij\leq(1,<)+\Doj
			\]
			is valid for~$\paramR$.
			Recall that after applying \cref{algorithm:TEeq}\ea{réf dynamique manquante},
			$\Doj'=\Dij+(-1,\leq)$.
			By \crefDef{def:validity}~(\refc{valideiiid}) we have $(-1,\leq)\leq(-1,\leq)$.
			We invoke \crefLemma{lemma:sumvalid} which gives
			\begin{equation}\label{eq:dixneuf}
			\Dij+(-1,\leq)\leq(1,<)+\Doj+(-1,\leq)=\Doj+(0,<)\text{ \valid{}.}
			\quad
			\end{equation}
			As, from \crefDef{def:validity}~(\refc{valideiiid}) we have~$\Doj+(0,<)\leq\Doj$ \valid{},
			we infer (\refc{eq:dixneuf}) and it gives
			\[
			 \Doj'\leq\Doj\text{ \valid{}.}
			\]

			Since~$\clockval'\in(\E',\pval(\D'))$ we have $-\partieFrac(\clockval'(\clock_j))\compOpLeq_{0j'}\pval(\doj')$,
			\[
			 0\compOpLeq_{0j'}\partieFrac(\clockval'(\clock_j))+\pval(\doj')\text{.}
			\]
			Then we have that
			\[
			 0\compOpLeq_{0j'}\partieFrac(\clockval'(\clock_j))+\pval(\doj')\leq\partieFrac(\clockval'(\clock_j))+\pval(\doj)
			\]
			where,
			\begin{itemize}
			      \item either from \crefDef{def:validity}~(\refc{valideiiie}) $\doj'<\doj$;
			      \item or from \crefDef{def:validity}~(\refc{valideiiid}), $\doj'\leq\doj$
				    and then $\compOpLeq_{0j'}=\compOpLeq_{0j}= {<}$.
				    Indeed as~$\Doj'\leq\Doj$ \valid{}, and since~$(\E,\D)$ is the
				    \oPDBM{} obtained after the application of $\TEF_<$ (cf.\ \cref{algorithm:TEup}\ea{réf dynamique manquante}) %
				    on another \mPDBM{} (see \refClaim{presentfuturPDBM}),
				    we have~$\compOpLeq_{0j}= {<}$.
			\end{itemize}
			To conclude we have that for all~$\clock_j$ either
			\[
			 0\compOpLeq_{0j'}\partieFrac(\clockval'(\clock_j))+\pval(\doj')<\partieFrac(\clockval'(\clock_j))+\pval(\doj)
			\]
			or
			\[
			  0<\partieFrac(\clockval'(\clock_j))+\pval(\doj')\leq\partieFrac(\clockval'(\clock_j))+\pval(\doj)\text{.}
			\]

			As~$t_1$ is by definition the minimum value of an expression~$\partieFrac(\clockval'(\clock_j))+\pval(\doj)$ for a given~$\clock_j$,
			which as we just proved are all strictly positive, we have that for all~$\clock_j$
			\[
			 0< t_1\leq\partieFrac(\clockval'(\clock_j))+\pval(\doj)\text{.}
			\]
			We proved that~$t_1>0$,
			so we can set~$\de=\frac{t_1}{2}$ (therefore $\de>0$).

			More intuitively $\de$ is the value right in the middle of the least and the largest amount of time
			\st{} we can go backward in time from~$\clockval'$ and respect all constraints defined in~$(\E,\pval(\D))$.

			Now we are going to prove that for any clock $\clock_j$, its constraints on lower and upper bounds are satisfied, \ie{}
			\[
			-\pval(\doj)\compOpLeq_{0j}\partieFrac(\clockval'(\clock_j))-\de\compOpLeq_{j0}\pval(\djo)\text{.}
			\]
			First as $\de<t_1$, we have
			\[
			 -\partieFrac(\clockval'(\clock_j))+\de<-\partieFrac(\clockval'(\clock_j))+t_1\leq-\partieFrac(\clockval'(\clock_j))+\partieFrac(\clockval'(\clock_j))+\pval(\doj)=\pval(\doj)
			\]
			which is $-\pval(\doj)<\partieFrac(\clockval'(\clock_j))-\de$.
			Since~$(\E,\D)$ is the \oPDBM{} obtained after the application of $\TEF_<$ (cf.\ \cref{algorithm:TEeq}\ea{réf dynamique manquante}) %
			on another \mPDBM{} (see \refClaim{presentfuturPDBM}),
			we have~$\compOpLeq_{0j}= {<}$ so~$-\pval(\doj)\compOpLeq_{0j}\partieFrac(\clockval'(\clock_j))-\de$.
			Secondary as $0 < \de$, we have
			\[
			 \partieFrac(\clockval'(\clock_j))-\de<\partieFrac(\clockval)'(\clock_j)-0\leq\partieFrac(\clockval'(\clock_j))-\partieFrac(\clockval'(\clock_j))+\pval(\djo)=\pval(\djo)
			\]
			which is $\partieFrac(\clockval'(\clock_j))-\de<\pval(\djo)$.
			Since~$(\E,\D)$ is the \oPDBM{} obtained after the application of $\TEF_<$ (cf.\ \cref{algorithm:TEeq}\ea{réf dynamique manquante}) %
			on another \mPDBM{} (see \refClaim{presentfuturPDBM}),
			we have~$\compOpLeq_{j0}= {<}$ so~$\partieFrac(\clockval'(\clock_j))-\de\compOpLeq_{j0}\pval(\djo)$

		      Now we prove that constraints defined in~$(\E,\D)$ on differences of clocks are also satisfied
		      by going back of~$\de$ units of time from~$\clockval'$.

		      Recall that in $(\E',\D')$ we have for all clock~$\clock_j$,
		      \[
		      \Dji'=\Djo'=\Dji+1
		       \quad
		       \text{and}
		       \quad
		       \Dij'=\Doj'=-1+\Dij\text{.}
		      \]

		      In addition by definition of $\TEF_=$, for $\clock_i\in\LFPd{\D}$, $\E_{\clock_i}=\E_{\clock_i}'-1$
		      and for $\clock_j\in\Clock\setminus\LFPd{\D}$, $\E_{\clock_j}=\E_{\clock_j}'$.

		      We already treated the case whether~$i$ or~$j$ are $0$, now suppose~$i,j$ are both different from~$0$.
		      \begin{itemize}
			      \item if~$\clock_i,\clock_j\in\Clock\setminus\LFPd{\D}$: let $\clock\in\LFPd{\D}$ and
				   recall that after applying \cref{algorithm:TEeq}\ea{réf dynamique manquante},
				   $\Dij'=\Dij$, $\Dji'=\Dji$;
				  we have that $\partieFrac(\clockval'(\clock_j))-\partieFrac(\clockval'(\clock_i))\compOpLeq_{ij'}\dji'=\dji$,
				  and therefore $\partieFrac(\clockval'(\clock_j))-\de-\partieFrac(\clockval'(\clock_i))+\de\compOpLeq_{ji}\dji$.

				We also have that $\partieFrac(\clockval'(\clock_i))-\partieFrac(\clockval'(\clock_j))\compOpLeq_{ij'}\dij'=\dij$,
				therefore $\partieFrac(\clockval'(\clock_i))-\de-\partieFrac(\clockval'(\clock_j))+\de\compOpLeq_{ij}\dij$;

				 \item if~$\clock_i\in\LFPd{\D}$ and $\clock_j\in\Clock\setminus\LFPd{\D}$:
				 recall that after applying \cref{algorithm:TEeq}\ea{réf dynamique manquante},
				 $\Djo'=\Dji+(1,\leq)$, and~$\Doj'=\Dij+(-1,\leq)$.
				 Observe that as we added~$\leq$ which is the neutral element of the addition~$\oplus$ between two operators~$\compOpLeq$,
				 we have~$\compOpLeq_{j0'}=\compOpLeq_{ji}$ and~$\compOpLeq_{0j'}=\compOpLeq_{ij}$.
				Note that as~$\clock_i\in\LFPd{\D}$, in~$(\E',\D')$ we have~$\Doi'=(0,\leq)=\Dio'$
				which means~$\partieFrac(\clockval'(\clock_i))=0$.
				Going backward in time of~$\de$ units of time from~$\clockval'(\clock_i)$
				means that~$\partieFrac(\clockval(\clock_i))=1-\de$.

				 We have that
				\[
				\partieFrac(\clockval'(\clock_j))\compOpLeq_{j0'}\pval(\djo')=\pval(\dji)+1
				\]
				hence $\partieFrac(\clockval'(\clock_j))-1\compOpLeq_{ji}\pval(\dji)$ which is equivalent to
				\[
				\partieFrac(\clockval'(\clock_j))-\de-1+\de\compOpLeq_{ji}\pval(\dji)\text{.}
				\]
				The same way we have
				\[
				-\partieFrac(\clockval'(\clock_j))\compOpLeq_{0j'}\pval(\doj')=\pval(\dij)-1
				\]
				hence $1-\partieFrac(\clockval'(\clock_j))\compOpLeq_{ij}\pval(\dij)$ which is equivalent to
				\[
				1-\de-\partieFrac(\clockval'(\clock_j))+\de\compOpLeq_{ij}\pval(\dij)\text{.}
				\]

			\end{itemize}

				To conclude, we define for all $\clock_j$ \st{} $\Doj'\neq(0,\leq)$ and $\Djo'\neq(0,\leq)$
				\[
				\clockval(\clock_j)=\clockval'(\clock_j)-\de
				\]
				and for all $\clock_i$ \st{} $\Doi'=(0,\leq)=\Dio'$
				\[
				\clockval(\clock_i)=(\clockval'(\clock_i)-1)+1-\de
				\]\mr{c'est la meme chose, mais ca facilite la compr\'ehension ?} 
				and clearly, $\clockval\in(\E,\pval(\D))$.

        \medskip

		      \item \textbf{\oPDBM{} respecting \crefDef{def:med-PDBM}~(\refc{PDBMiv})}

          \medskip

		      Let~$\pval\in\paramR$.
		      Consider $(\E',\D')=\TEF((\E, \D))$ respecting condition \crefDef{def:med-PDBM}~(\refc{PDBMiv}), \ie{}
		      suppose there is at least an~$\clock_i$ \st{} $\Dio'=(1,<)$
		      and for all~$j$ \st{} $\Doj=(0,\compOpLeq_{0j})$, then we have ${\compOpLeq_{0j}} = {<}$.
		      Let $\clockval'\in(\E',\pval(\D'))$.

		      We need to find a value~$\de$ \st{} $\clockval'-\de\in(\E, \valuate{\D}{\pval})$
		       which is equivalent to prove for all~$\clock_i,\clock_j$
		      \[
		       \partieFrac(\clockval'(\clock_j))-\de-\partieFrac(\clockval'(\clock_i))+\de\compOpLeq_{ji}\pval(\dji)
		       \]
		       and
		       \[
		       \partieFrac(\clockval'(\clock_i))-\de-\partieFrac(\clockval'(\clock_j))+\de\compOpLeq_{ij}\pval(\dij)
		       \]
		       and
		       \[
		        -\partieFrac(\clockval'(\clock_j))+\de\compOpLeq_{0j}\pval(\doj)
		       \quad
		       \text{and}
		       \quad
		       \partieFrac(\clockval'(\clock_j))-\de\compOpLeq_{j0}\pval(\djo)\text{.}
		      \]

		      As done previously we are going to define a~$\de$ which is different from~$0$ so we satisfy condition \crefDef{def:med-PDBM}~(\refc{PDBMiii}),
		      and show that constraints in~$(\E,\D)$
		      are satisfied while going backward of~$\de$ units of time from~$\clockval'$.

		      We define the largest and the least amount of time so that all upper bounds of~$(\E,\D)$ are satisfied.
 			Let
 			\[
 			 t_0=\max\limits_{\clock\in\Clock}\{0,\partieFrac(\clockval'(\clock))-\pval(\dxo)\}
 			\]

		      and
		      \[
		      t_1=\min\limits_{\clock\in\Clock}\{\partieFrac(\clockval'(\clock))+\pval(\dox)\}\text{.}
		      \]
		      We want to prove that $t_0= t_1>0$.
		      For this purpose, let us first show that for all~$i,j$
		      we have~$\partieFrac(\clockval'(\clock_j))-\pval(\djo')\leq\partieFrac(\clockval'(\clock_i))+\pval(\dzeroi')$,
		      which is~$t_0\leq t_1$.

		      First note that for all~$i,j$
		      \[
		      \partieFrac(\clockval'(\clock_j))-\partieFrac(\clockval'(\clock_i))\compOpLeq_{ji'}\pval(\dji')\text{.}
		      \]
		      By applying~$\TEF_<$ (\cref{algorithm:TEup}\ea{réf dynamique manquante})
		      to~$(\E,\D)$, we have that~$\Dji'=\Dji$, \ie{} $(\dij,\compOpLeq_{ij})=(\dij',\compOpLeq_{ij'})$,
		      and from \crefDef{def:med-PDBM}~(\refc{PDBMii}) we have that~$\Dji\leq\Djo+\Doi$ \valid{}.

		      Hence, we have from \crefDef{def:validity}~(\refc{valideiiid}) that either $\pval(\dji)<\pval(\djo)+\pval(\dzeroi)$ or
		      $\pval(\dji)\leq\pval(\djo)+\pval(\dzeroi)$ and $\compOpLeq_{ji}=\compOpLeq_{j0} \oplus \compOpLeq_{0i}$ or $\compOpLeq_{ji}= {<}$ and $\compOpLeq_{j0} \oplus \compOpLeq_{0i}= {\leq}$.

		      We can then write that
		      \[
		       \partieFrac(\clockval'(\clock_j))-\partieFrac(\clockval'(\clock_i))(\compOpLeq_{j0} \oplus \compOpLeq_{0i})\pval(\djo)+\pval(\dzeroi)
		      \]
		      which is equivalent to
		      \[
		       \partieFrac(\clockval'(\clock_j))-\pval(\djo)(\compOpLeq_{j0} \oplus \compOpLeq_{0i})\partieFrac(\clockval'(\clock_i))+\pval(\dzeroi)
		      \]
		      so we obtain our result, as~$(\compOpLeq_{j0} \oplus \compOpLeq_{0i})$ is either~$\leq$ or~$<$.

		      Now, recall that~$(\E,\D)$ respects condition \crefDef{def:med-PDBM}~(\refc{PDBMiii})
		      so we have at least an $\clock$ \st{} $\Dxo=\Dox=(0,\leq)$.

		      For this clock~$\clock$ we have that~$\partieFrac(\clockval'(\clock))=\partieFrac(\clockval'(\clock))-\pval(\dxo)\leq t_0$
		      and that~$t_1\leq\partieFrac(\clockval'(\clock))+\pval(\dox)=\partieFrac(\clockval'(\clock))$.

		      Hence $t_0=t_1=\partieFrac(\clockval'(\clock))$.

		      As $\compOpLeq_{x0}= {\leq}$,
		      we have~$(\compOpLeq_{x0} \oplus \compOpLeq_{0i})=\compOpLeq_{0i}$
		      and $(\compOpLeq_{j0} \oplus \compOpLeq_{0x})=\compOpLeq_{j0}$, which gives
		      \[
		       \partieFrac(\clockval'(\clock))=\partieFrac(\clockval'(\clock))-\pval(\dxo)\compOpLeq_{0i}\partieFrac(\clockval'(\clock_i))+\pval(\dzeroi)
		      \]
		      and
		      \[
		       \partieFrac(\clockval'(\clock_j))-\pval(\djo)\compOpLeq_{j0}\partieFrac(\clockval'(\clock))+\pval(\dox)=\partieFrac(\clockval'(\clock))\text{.}
		      \]

		      Moreover in~$(\E',\D')$ we have that~$\partieFrac(\clockval'(\clock))\compOpLeq_{0x'}\pval(\dox')$.
		      Since~$(\E',\D')$ respects condition \crefDef{def:med-PDBM}~(\refc{PDBMiv}), if $\Dox'=(0,\compOpLeq_{0x'})$ then
		      $\compOpLeq_{0x'}={<}$.
		      Hence~$0<\partieFrac(\clockval'(\clock))$ and
		      \[
		       0<t_0=t_1\text{.}
		      \]

		      Let $\de=t_0=t_1$.
			More intuitively $\de$ is the value right in the middle of the least and the largest amount of time
			\st{} we can go backward in time from $\clockval'$ and respect all constraints defined in $(\E,\pval(\D))$.

			First we have
			\[
			 -\partieFrac(\clockval'(\clock_j))+\de\leq-\partieFrac(\clockval'(\clock_j))+t_1\compOpLeq_{j0}-\partieFrac(\clockval'(\clock_j))+\partieFrac(\clockval'(\clock_j))+\pval(\doj)=\pval(\doj)
			\]
			which is $-\pval(\doj)\compOpLeq_{j0}\partieFrac(\clockval'(\clock_j))-\de$.

			Secondary we have
			\[
			 \partieFrac(\clockval'(\clock_j))-\de\leq\partieFrac(\clockval)'(\clock_j)-t_0\compOpLeq_{0j}\partieFrac(\clockval'(\clock_j))-\partieFrac(\clockval'(\clock_j))+\pval(\djo)=\pval(\djo)
			\]
			which is $\partieFrac(\clockval'(\clock_j))-\de\compOpLeq_{0j}\pval(\djo)$.

		      Now we prove that constraints defined in~$(\E,\D)$ on differences of clocks are also satisfied
		      by going back of~$\de$ units of time from~$\clockval'$

		      Recall that in $(\E',\D')$ from the definition of \cref{algorithm:TEup}\ea{réf dynamique manquante}
		      we have for all clocks~$\clock_i,\clock_j$,
		      \[
		      \Dji'=\Dji
		       \quad
		       \text{and}
		       \quad
		       \Dij'=\Dij\text{.}
		      \]
		      Since we already treated the case whether $i$ or $j$ are $0$, now suppose~$i,j$ are both different from~$0$.
		      We have that $\partieFrac(\clockval'(\clock_j))-\partieFrac(\clockval'(\clock_i))\compOpLeq_{ji'}\pval(\dji')=\pval(\dji)$,
		      and therefore as $\compOpLeq_{ji'}=\compOpLeq_{ji}$,
		      \[
		      \partieFrac(\clockval'(\clock_j))-\de-\partieFrac(\clockval'(\clock_i))+\de\compOpLeq_{ji}\pval(\dji)\text{.}
		      \]
		      We also have that $\partieFrac(\clockval'(\clock_i))-\partieFrac(\clockval'(\clock_j))\compOpLeq_{ij'}\pval(\dij')=\pval(\dij)$,
		      therefore as $\compOpLeq_{ij'}=\compOpLeq_{ij}$,
		      \[
		      \partieFrac(\clockval'(\clock_i))-\de-\partieFrac(\clockval'(\clock_j))+\de\compOpLeq_{ij}\pval(\dij)\text{.}
		      \]

				To conclude, we define for all $\clock_j$ %
				\[
				\clockval(\clock_j)=\clockval'(\clock_j)-\de
				\]
				and clearly, $\clockval\in(\E,\pval(\D))$.

		    \end{itemize}

        \bigskip

		     Conversely, let $\clockval\in(\E, \valuate{\D}{\pval})$,
		      \begin{itemize}[$\Longleftarrow$]
		      \item \textbf{\oPDBM{} respecting \crefDef{def:med-PDBM}~(\refc{PDBMiv})}

          \medskip

		      Suppose in~$(\E,\D)$ there is at least an $\clock_i$ \st{} $\Dio=(1,<)$
		      and for all $j$ \st{} $\Doj=(0,\compOpLeq)$, we have ${\compOpLeq} = {<}$.
		      Let~$\clock_i$ be such a clock and~$\pval\in\paramR$.%

		      Now consider $(\E',\D')=\TEF((\E,\D))$.
		      We need to find a value $\de$ \st{} $\clockval+\de\in(\E', \valuate{\D'}{\pval})$.
		       which is equivalent to prove for all $\clock_i,\clock_j$
		      \[
		       \partieFrac(\clockval(\clock_j))+\de-\partieFrac(\clockval(\clock_i))-\de\compOpLeq_{ji'}\pval(\dji')
		       \]
		       and
		       \[
		       \partieFrac(\clockval(\clock_i))+\de-\partieFrac(\clockval(\clock_j))-\de\compOpLeq_{ij'}\pval(\dij')
		       \]
		       and
		       \[
		        -\partieFrac(\clockval(\clock_j))-\de\compOpLeq_{0j'}\pval(\doj')
		       \quad
		       \text{and}
		       \quad
		       \partieFrac(\clockval(\clock_j))+\de\compOpLeq_{j0'}\pval(\djo')\text{.}
		      \]

		      As done previously we are going to define a~$\de$ which is different from~$0$,
		      and show that constraints in~$(\E,\D)$
		      are satisfied while going forward of~$\de$ units of time from~$\clockval$.

		      Recall that~$\clock_i\in\LFPd{\D}$ and let~$\de=1-\partieFrac(\clockval(\clock_i))$
		      which we will prove is the exact amount of time so that all upper bounds of $(\E',\D')$ are satisfied.
		      Let
		      \[
		      t_0=\max\limits_{\clock\in\Clock}\{-\partieFrac(\clockval(\clock))-\partieFrac(\pval(\dox'))\}
		      \]

		      and
		      \[
		      t_1=\min\limits_{\clock\in\Clock}\{\partieFrac(\pval(\dxo'))-\partieFrac(\clockval(\clock))\}\text{.}
		      \]

		      Recall that since~$(\E,\D)$ respects condition \crefDef{def:med-PDBM}~(\refc{PDBMiv}),
		      for all $j$ \st{} $\Doj=(0,\compOpLeq_{0j})$, we have ${\compOpLeq_{0j}} = {<}$.
		      Hence as~$-\partieFrac(\clockval(\clock_i))<\pval(\doj)$, $\partieFrac(\clockval(\clock_i))\neq 0$.
		      Using the same reasoning as before, we are going to prove that~$t_0\leq \de\leq t_1$.

		      First we will prove that~$t_0\leq \de$.
		      Consider~$\clock_i\in\LFPd{\D}$. For all clock~$\clock_j$, since~$\clockval\in(\E,\pval(\D))$
		      we have~$\partieFrac(\clockval(\clock_i))-\partieFrac(\clockval(\clock_j))\compOpLeq_{ij}\partieFrac(\pval(\dij))$.

		      From \cref{algorithm:TEeq}\ea{réf dynamique manquante}
		      applied to~$(\E,\D)$ and since~$\clock_i\in\LFPd{\D}$ we obtain in~$(\E',\D')$ that~$\Doj'=\Dij+(-1,\leq)$.
		      Clearly we have~$\compOpLeq_{0j'}={\compOpLeq_{ij}}\oplus{\leq}={\compOpLeq_{ij}}$.
		      It gives that
		      \[
		      \partieFrac(\clockval(\clock_i))-\partieFrac(\clockval(\clock_j))-1(\compOpLeq_{ij}\oplus\leq)\partieFrac(\pval(\dij))-1
		      \]
		      which is equivalent to
		      $\partieFrac(\clockval(\clock_i))-\partieFrac(\clockval(\clock_j))-1\compOpLeq_{0j'}\partieFrac(\pval(\doj'))$
		      which is equivalent to
		      \[
		       \partieFrac(\clockval(\clock_i))-1\compOpLeq_{0j'}\partieFrac(\pval(\doj'))+\partieFrac(\clockval(\clock_j))\text{.}
		      \]
		      This gives us our first result.

		      Second we will prove that~$\de\leq t_1$.
		      Consider~$\clock_i\in\LFPd{\D}$. For all clock~$\clock_j$, from \crefDef{def:med-PDBM}~(\refc{PDBMii})
		      we have~$\partieFrac(\clockval(\clock_j))-\partieFrac(\clockval(\clock_i))\compOpLeq_{ji}\partieFrac(\pval(\dji))$.
		      We have
		      \[
		      \partieFrac(\clockval(\clock_j))-\partieFrac(\clockval(\clock_i))+1\compOpLeq_{ji}\partieFrac(\pval(\dji))+1\text{.}
		      \]
		      From \cref{algorithm:TEeq}\ea{réf dynamique manquante}
		      applied to~$(\E,\D)$ and since~$\clock_i\in\LFPd{\D}$ we obtain in~$(\E',\D')$ that~$\Djo'=\Dji+(1,\leq)$.
		      Clearly we have~$\compOpLeq_{j0'}=\compOpLeq_{ji}\oplus\leq=\compOpLeq_{ji}$.
		      Then we can write that
		      $\partieFrac(\clockval(\clock_j))-\partieFrac(\clockval(\clock_i))+1\compOpLeq_{j0'}\partieFrac(\pval(\djo'))$
		      which is equivalent to
		      \[
		       1-\partieFrac(\clockval(\clock_i))\compOpLeq_{j0'}\partieFrac(\pval(\djo'))-\partieFrac(\clockval(\clock_j))\text{.}
		      \]
		      This gives us our second result.

			Now for all clock~$\clock_j$, we obtain two results.
			First we have
			\[
			 -\partieFrac(\clockval(\clock_j))-\de \compOpLeq_{0j'} -\partieFrac(\clockval(\clock_j))-t_1 \leq -\partieFrac(\clockval(\clock_j))+\partieFrac(\clockval(\clock_j))+\pval(\doj')=\pval(\doj')
			\]
			which is $-\pval(\doj')\compOpLeq_{0j'}\partieFrac(\clockval(\clock_j))+\de$.

			Secondary we have
			\[
			 \partieFrac(\clockval(\clock_j))+\de \compOpLeq_{j0'} \partieFrac(\clockval(\clock_j))+t_0 \leq \partieFrac(\clockval(\clock_j))-\partieFrac(\clockval(\clock_j))+\pval(\djo')=\pval(\djo')
			\]
			which is $\partieFrac(\clockval(\clock_j))+\de\compOpLeq_{j0'}\pval(\djo')$.

			Since we already treated the case whether $i$ or $j$ are $0$, now suppose~$i,j$ are both different from~$0$.

			Note that if both~$\clock_i, \clock_j\in\LFPd{\D}$, as~$\partieFrac(\clockval(\clock_i))=\partieFrac(\clockval(\clock_j))$,
			$\Dij=\Dij'=(0,\leq)$ and~$\Dji=\Dji'=(0,\leq)$ from \crefDef{def:largestclock}.
			Hence~$\partieFrac(\clockval(\clock_i))+\de-\partieFrac(\clockval(\clock_j))-\de\compOpLeq_{ij'}\partieFrac(\pval(\dij'))$
			and~$\partieFrac(\clockval(\clock_j))+\de-\partieFrac(\clockval(\clock_j))-\de\compOpLeq_{ji'}\partieFrac(\pval(\dji'))$.

			The same way, if both~$\clock_i, \clock_j\not\in\LFPd{\D}$ we have~$\Dij=\Dij'$ and~$\Dji=\Dji'$ and again our result.
			If either~$\clock_i$ or~$\clock_j$ is in~$\LFPd{\D}$, the case is similar to~$\Doj'$ or~$\Dio'$.

		      Finally, we define~$\clockval'=\clockval+\de$ and~$\clockval'\in(\E',\pval(\D'))$.

          \medskip

		      \item \textbf{\oPDBM{} respecting \crefDef{def:med-PDBM}~(\refc{PDBMiii})}

          \medskip

		      Suppose in~$(\E,\D)$ there is at least an $\clock_j$ \st{} $\Djo=\Doj=(0,\leq)$
		      Let~$\pval\in\paramR$, and~$\clock_i\in\LFPd{\D}$.

		      Now consider $(\E',\D')=\TEF((\E,\D))$.
		      We need to find a value $\de$ \st{} $\clockval+\de\in(\E', \valuate{\D'}{\pval})$.
		       which is equivalent to prove for all $\clock_i,\clock_j$
		      \[
		       \partieFrac(\clockval(\clock_j))+\de-\partieFrac(\clockval(\clock_i))-\de\compOpLeq_{ji'}\pval(\dji')
		       \]
		       and
		       \[
		       \partieFrac(\clockval(\clock_i))+\de-\partieFrac(\clockval(\clock_j))-\de\compOpLeq_{ij'}\pval(\dij')
		       \]
		       and
		       \[
		        -\partieFrac(\clockval(\clock_j))-\de\compOpLeq_{0j'}\pval(\doj')
		       \quad
		       \text{and}
		       \quad
		       \partieFrac(\clockval(\clock_j))+\de\compOpLeq_{j0'}\pval(\djo')\text{.}
		      \]

		      As done previously we are going to define a~$\de$ which is different from~$0$,
		      and show that constraints in~$(\E,\D)$
		      are satisfied while going forward of~$\de$ units of time from~$\clockval$.

		      Let
		      \[
		      t_0=\max\limits_{\clock\in\Clock}\{0,-\partieFrac(\clockval(\clock))-\partieFrac(\pval(\dox'))\}
		      \]

		      and
		      \[
		      t_1=\min\limits_{\clock\in\Clock}\{\partieFrac(\pval(\dxo'))-\partieFrac(\clockval(\clock))\}\text{.}
		      \]

		      We want to prove that $t_0\leq t_1$.
		      For this purpose, we are going to prove for all clocks~$i,j$ that
		      $-\partieFrac(\clockval(\clock_j))-\pval(\djo')\leq\pval(\dzeroi')-\partieFrac(\clockval(\clock_i))$.

		      First note that
		      \[
		      \partieFrac(\clockval(\clock_j))-\partieFrac(\clockval(\clock_i))\compOpLeq_{ji}\pval(\dji)
		      \]
		      By definition of $\TEF_<$ applied to $(\E,\D)$, we have that $\Dji'=\Dji$,
		      and from \crefDef{def:med-PDBM}~(\refc{PDBMii}) we have that $\Dji'\leq\Djo'+\Doi'$.

		      Hence, we have from \crefDef{def:validity}~(\refc{valideiiid}) that either $\dji'<\djo'+\dzeroi'$ or
		      $\dji'=\djo'+\dzeroi'$ and $\compOpLeq_{ji'}=\compOpLeq_{j0'} \oplus \compOpLeq_{0i'}$ or $\compOpLeq_{ji'}= {<}$ and $\compOpLeq_{j0'} \oplus \compOpLeq_{0i'}= {\leq}$.

		      We can then write that
		      \[
		       \partieFrac(\clockval(\clock_j))-\partieFrac(\clockval(\clock_i))(\compOpLeq_{j0'} \oplus \compOpLeq_{0i'})\pval(\djo')+\pval(\dzeroi')
		      \]
		      which is equivalent to
		      \[
		       -\partieFrac(\clockval(\clock_i))-\pval(\dzeroi')(\compOpLeq_{j0'} \oplus \compOpLeq_{0i'})\pval(\djo')-\partieFrac(\clockval(\clock_j))
		      \]

		      Now we prove that~$t_0=0$.
		      Clearly from \crefDef{def:med-PDBM} for any clock~$i$ we have that
		      $-\partieFrac(\clockval(\clock_i))\compOpLeq_{0i}\pval(\dzeroi)$ which is equivalent to
		      $-\partieFrac(\clockval(\clock_i))-\pval(\dzeroi)\compOpLeq_{0i} 0$.

		      Hence if as~$(\E,\D)$ there is at least an $\clock_j$ \st{} $\Djo=\Doj=(0,\leq)$, for this clock~$j$
		      we have~$-\partieFrac(\clockval(\clock_j))-\pval(\doj)=0$.

		      By definition of $\TEF_<$ applied to $(\E,\D)$, we have that $\Doi'=\Doi+(0,<)$.
		      In order to respect the constraint~$-\partieFrac(\clockval(\clock_i))-\de\compOpLeq_{0i'} \pval(\dzeroi')$
		      which is, as~$\compOpLeq_{0i'}=<$, $-\partieFrac(\clockval(\clock_i))-\de< \pval(\dzeroi')$
		      and especially for~$j$ where $\pval(\doj')=0$ we have to find a~$\de>0$.

		      In order to find an upper bound for~$\de$, we are going to prove that~$t_1>0$.
		      From \crefDef{def:med-PDBM}~(\refc{PDBMii}) we have in~$(\E,\D)$ that for any clocks~$i,j$
		      $\Djo\leq\Dji+\Dio$. Let~$\clock_i\in\LFPd{\D}$.
		      From \crefDef{def:med-PDBM}~(\refc{PDBMi}), we have that~$\Dio\leq(1,<)$.
		      This gives that $\Dji+\Dio\leq\Dji+(1,<)$.

		      By definition of $\TEF_<$ applied to $(\E,\D)$, we have that $\Djo'=\Dji+(1,<)$.
		      Hence we have~$\Djo\leq\Djo'$.

		      Now as~$\partieFrac(\clockval(\clock_i))\compOpLeq_{i0}\pval(\dio)$
		      we can write~$\partieFrac(\clockval(\clock_i))\compOpLeq_{i0'}\pval(\dio')$
		      and then~$0 \compOpLeq_{i0'}\pval(\dio')-\partieFrac(\clockval(\clock_i))$
		      where~$\compOpLeq_{i0'}=<$, which prove our result.

		      We define~$\de=\frac{t_1}{2}$, therefore~$t_0<\de<t_1$.
		      Now for all clock~$\clock_j$, we obtain two results.
			First we have
			\[
			 -\partieFrac(\clockval(\clock_j))-\de<-\partieFrac(\clockval(\clock_j))-t_1\compOpLeq_{0j'}-\partieFrac(\clockval(\clock_j))+\partieFrac(\clockval(\clock_j))+\pval(\doj')=\pval(\doj')
			\]
			which is $-\pval(\doj')\compOpLeq_{0j}\partieFrac(\clockval(\clock_j))+\de$ as~$\compOpLeq_{0j'}={<}$.

			Secondary we have
			\[
			 \partieFrac(\clockval(\clock_j))+\de<\partieFrac(\clockval(\clock_j))+t_0\compOpLeq_{j0'}\partieFrac(\clockval(\clock_j))-\partieFrac(\clockval(\clock_j))+\pval(\djo')=\pval(\djo')
			\]
			which is $\partieFrac(\clockval(\clock_j))+\de\compOpLeq_{j0}\pval(\djo')$ as~$\compOpLeq_{0j'}={<}$.

		      Now we prove that constraints defined in~$(\E',\D')$ on differences of clocks are also satisfied
		      by going forward of~$\de$ units of time from~$\clockval$

		      Recall that in $(\E',\D')$ from the definition of \cref{algorithm:TEup}\ea{réf dynamique manquante}
		      we have for all clock~$\clock_j$,
		      \[
		      \Dji'=\Dji
		       \quad
		       \text{and}
		       \quad
		       \Dij'=\Dij\text{.}
		      \]
		      Since we already treated the case whether $i$ or $j$ are $0$, now suppose~$i,j$ are both different from~$0$.
		      We have that $\partieFrac(\clockval(\clock_j))-\partieFrac(\clockval(\clock_i))\compOpLeq_{ji}\pval(\dji)=\pval(\dji')$,
		      and therefore as $\compOpLeq_{ji'}=\compOpLeq_{ji}$,
		      \[
		      \partieFrac(\clockval(\clock_j))+\de-\partieFrac(\clockval(\clock_i))-\de\compOpLeq_{ji'}\pval(\dji')\text{.}
		      \]
		      We also have that $\partieFrac(\clockval(\clock_i))-\partieFrac(\clockval(\clock_j))\compOpLeq_{ij}\pval(\dij)=\pval(\dij')$,
		      therefore as $\compOpLeq_{ij'}=\compOpLeq_{ij}$,
		      \[
		      \partieFrac(\clockval(\clock_i))+\de-\partieFrac(\clockval(\clock_j))-\de\compOpLeq_{ij'}\pval(\dij')\text{.}
		      \]

		      Finally, we define~$\clockval'=\clockval+\de$ and~$\clockval'\in(\E',\pval(\D'))$. \qedhere
		      \end{itemize}
	  \end{preuve}

	  \newcommand{\lemmaTimeElapsingPointPDBM}{%
		    Let $\paramR$ be a parameter region and $(\E,\D)\in\CPr$. Let $\pval\in\paramR$.
		    There is $\clockval'\in\TEF((\E, \valuate{\D}{\pval}))$
		    iff there is $\clockval\in(\E, \valuate{\D}{\pval})$ and a delay~$\de$ \st{} $\clockval'=\clockval+\de$.
		    }
	  \begin{clm}\label{ssiTEFPointPDBM}

		  \lemmaTimeElapsingPointPDBM{}
	  \end{clm}

	  \begin{preuve}
	   ($\Longleftarrow$) \textbf{for \pointPDBMs{}}

\medskip

		      Let~$\pval\in\paramR$.
		      Consider $(\E',\D')=\TEF((\E, \D))$ respecting condition \crefDef{def:med-PDBM}~(\refc{PDBMiv}), \ie{}
		      suppose there is at least an~$\clock_i$ \st{} $\Dio'=(1,<)$
		      and for all~$j$ \st{} $\Doj=(0,\compOpLeq_{0j})$, then we have ${\compOpLeq_{0j}} = {<}$.
		      Let $\clockval'\in(\E',\pval(\D'))$.

		      We need to find a value~$\de$ \st{} $\clockval'-\de\in(\E, \valuate{\D}{\pval})$
		       which is equivalent to prove for all~$\clock_i,\clock_j$
		      \[
		       \partieFrac(\clockval'(\clock_j))-\de-\partieFrac(\clockval'(\clock_i))+\de\compOpLeq_{ji}\pval(\dji)
		       \]
		       and
		       \[
		       \partieFrac(\clockval'(\clock_i))-\de-\partieFrac(\clockval'(\clock_j))+\de\compOpLeq_{ij}\pval(\dij)
		       \]
		       and
		       \[
		        -\partieFrac(\clockval'(\clock_j))+\de\compOpLeq_{0j}\pval(\doj)
		       \quad
		       \text{and}
		       \quad
		       \partieFrac(\clockval'(\clock_j))-\de\compOpLeq_{j0}\pval(\djo)\text{.}
		      \]

		      As done previously we are going to define a~$\de$ which is different from~$0$,
		      and show that constraints in~$(\E,\D)$
		      are satisfied while going backward of~$\de$ units of time from~$\clockval'$.

		      We define the largest and the least amount of time so that all upper bounds of~$(\E,\D)$ are satisfied.
 			Let
 			\[
 			 t_0=\max\limits_{\clock\in\Clock}\{0,\partieFrac(\clockval'(\clock))-\pval(\dxo)\}
 			\]

		      and
		      \[
		      t_1=\min\limits_{\clock\in\Clock}\{\partieFrac(\clockval'(\clock))+\pval(\dox)\}\text{.}
		      \]
		      We want to prove that $t_0= t_1>0$.
		      For this purpose, let us first show that for all~$i,j$
		      we have~$\partieFrac(\clockval'(\clock_j))-\pval(\djo')\leq\partieFrac(\clockval'(\clock_i))+\pval(\dzeroi')$,
		      which is~$t_0\leq t_1$.

		      First note that for all~$i,j$
		      \[
		      \partieFrac(\clockval'(\clock_j))-\partieFrac(\clockval'(\clock_i))\compOpLeq_{ji'}\pval(\dji')\text{.}
		      \]
		      By applying~$\TEF_<$ (\cref{algorithm:TEup}\ea{réf dynamique manquante})
		      to~$(\E,\D)$, we have that~$\Dji'=\Dji$, \ie{} $(\dij,\compOpLeq_{ij})=(\dij',\compOpLeq_{ij'})$,
		      and from \crefDef{definition:pointPDBM}~(\refc{PDBM2ii}) we have that~$\Dji\leq\Djo+\Doi$ \valid{}.

		      Hence, we have from \crefDef{def:validity}~(\refc{valideiiid}) that either $\pval(\dji)<\pval(\djo)+\pval(\dzeroi)$ or
		      $\pval(\dji)\leq\pval(\djo)+\pval(\dzeroi)$ and $\compOpLeq_{ji}=\compOpLeq_{j0} \oplus \compOpLeq_{0i}$ or $\compOpLeq_{ji}= {<}$ and $\compOpLeq_{j0} \oplus \compOpLeq_{0i}= {\leq}$.

		      We can then write that
		      \[
		       \partieFrac(\clockval'(\clock_j))-\partieFrac(\clockval'(\clock_i))(\compOpLeq_{j0} \oplus \compOpLeq_{0i})\pval(\djo)+\pval(\dzeroi)
		      \]
		      which is equivalent to
		      \[
		       \partieFrac(\clockval'(\clock_j))-\pval(\djo)(\compOpLeq_{j0} \oplus \compOpLeq_{0i})\partieFrac(\clockval'(\clock_i))+\pval(\dzeroi)
		      \]
		      so we obtain our result, as~$(\compOpLeq_{j0} \oplus \compOpLeq_{0i})$ is either~$\leq$ or~$<$.

		      Now, recall that in~$(\E,\D)$ for all~$\clock$ we have~$\dox=-\dxo$ and~$\compOpLeq_{0\clock}=\compOpLeq_{\clock 0}$.

		      For any clock~$\clock$ we have that~$\partieFrac(\clockval'(\clock))-\pval(\dxo)\leq t_0$
		      and that~$t_1\leq\partieFrac(\clockval'(\clock))+\pval(\dox)=\partieFrac(\clockval'(\clock))-\pval(\dxo)$.

		      Hence $t_0=t_1$.

		      As for all~$\clock$,~$\compOpLeq_{x0}= {\leq}$,
		      we have for all~$i,j$ that~$(\compOpLeq_{x0} \oplus \compOpLeq_{0i})=\compOpLeq_{0i}$
		      and $(\compOpLeq_{j0} \oplus \compOpLeq_{0x})=\compOpLeq_{j0}$, which gives
		      \[
		       t_1\compOpLeq_{0i}\partieFrac(\clockval'(\clock_i))+\pval(\dzeroi)
		      \]
		      and
		      \[
		       \partieFrac(\clockval'(\clock_j))-\pval(\djo)\compOpLeq_{j0}t_0\text{.}
		      \]

		      Moreover in~$(\E',\D')$ we have that~$\partieFrac(\clockval'(\clock))\compOpLeq_{0x'}\pval(\dox')$.
		      From \refClaim{presentfuturPDBM},~$(\E',\D')$ is obtained after applying \cref{algorithm:TEup}\ea{réf dynamique manquante}
		      and therefore~$\compOpLeq_{0x'}={<}$.
		      Hence~$0<\partieFrac(\clockval'(\clock))$ and
		      \[
		       0<t_0=t_1\text{.}
		      \]

		      Let $\de=t_0=t_1$.
			More intuitively $\de$ is the value right in the middle of the least and the largest amount of time
			\st{} we can go backward in time from $\clockval'$ and respect all constraints defined in $(\E,\pval(\D))$.

			First we have
			\[
			 -\partieFrac(\clockval'(\clock_j))+\de\leq-\partieFrac(\clockval'(\clock_j))+t_1\compOpLeq_{j0}-\partieFrac(\clockval'(\clock_j))+\partieFrac(\clockval'(\clock_j))+\pval(\doj)=\pval(\doj)
			\]
			which is $-\pval(\doj)\compOpLeq_{j0}\partieFrac(\clockval'(\clock_j))-\de$.

			Secondary we have
			\[
			 \partieFrac(\clockval'(\clock_j))-\de\leq\partieFrac(\clockval)'(\clock_j)-t_0\compOpLeq_{0j}\partieFrac(\clockval'(\clock_j))-\partieFrac(\clockval'(\clock_j))+\pval(\djo)=\pval(\djo)
			\]
			which is $\partieFrac(\clockval'(\clock_j))-\de\compOpLeq_{0j}\pval(\djo)$.

		      Now we prove that constraints defined in~$(\E,\D)$ on differences of clocks are also satisfied
		      by going back of~$\de$ units of time from~$\clockval'$

		      Recall that in $(\E',\D')$ from the definition of \cref{algorithm:TEup}
		      we have for all clocks~$\clock_i,\clock_j$,
		      \[
		      \Dji'=\Dji
		       \quad
		       \text{and}
		       \quad
		       \Dij'=\Dij\text{.}
		      \]

		      Since we already treated the case whether $i$ or $j$ are $0$, now suppose~$i,j$ are both different from~$0$.
		      We have that $\partieFrac(\clockval'(\clock_j))-\partieFrac(\clockval'(\clock_i))\compOpLeq_{ji'}\pval(\dji')=\pval(\dji)$,
		      and therefore as $\compOpLeq_{ji'}=\compOpLeq_{ji}$,
		      \[
		      \partieFrac(\clockval'(\clock_j))-\de-\partieFrac(\clockval'(\clock_i))+\de\compOpLeq_{ji}\pval(\dji)\text{.}
		      \]
		      We also have that $\partieFrac(\clockval'(\clock_i))-\partieFrac(\clockval'(\clock_j))\compOpLeq_{ij'}\pval(\dij')=\pval(\dij)$,
		      therefore as $\compOpLeq_{ij'}=\compOpLeq_{ij}$,
		      \[
		      \partieFrac(\clockval'(\clock_i))-\de-\partieFrac(\clockval'(\clock_j))+\de\compOpLeq_{ij}\pval(\dij)\text{.}
		      \]

				To conclude, we define for all $\clock_j$ %
				\[
				\clockval(\clock_j)=\clockval'(\clock_j)-\de
				\]
				and clearly, $\clockval\in(\E,\pval(\D))$.

        \medskip

	   ($\Longrightarrow$) \textbf{for \pointPDBMs{}}

     \medskip

		      Assume in~$(\E,\D)\in\CPr$.
		      Let~$\pval\in\paramR$, and~$\clock_i\in\LFPd{\D}$.

		      Now consider $(\E',\D')=\TEF((\E,\D))$.
		      We need to find a value $\de$ \st{} $\clockval+\de\in(\E', \valuate{\D'}{\pval})$.
		       which is equivalent to prove for all $\clock_i,\clock_j$
		      \[
		       \partieFrac(\clockval(\clock_j))+\de-\partieFrac(\clockval(\clock_i))-\de\compOpLeq_{ji'}\pval(\dji')
		       \]
		       and
		       \[
		       \partieFrac(\clockval(\clock_i))+\de-\partieFrac(\clockval(\clock_j))-\de\compOpLeq_{ij'}\pval(\dij')
		       \]
		       and
		       \[
		        -\partieFrac(\clockval(\clock_j))-\de\compOpLeq_{0j'}\pval(\doj')
		       \quad
		       \text{and}
		       \quad
		       \partieFrac(\clockval(\clock_j))+\de\compOpLeq_{j0'}\pval(\djo')\text{.}
		      \]

		      As done previously we are going to define a~$\de$ which is different from~$0$,
		      and show that constraints in~$(\E,\D)$
		      are satisfied while going forward of~$\de$ units of time from~$\clockval$.

		      Let
		      \[
		      t_0=\max\limits_{\clock\in\Clock}\{0,-\partieFrac(\clockval(\clock))-\partieFrac(\pval(\dox'))\}
		      \]

		      and
		      \[
		      t_1=\min\limits_{\clock\in\Clock}\{\partieFrac(\pval(\dxo'))-\partieFrac(\clockval(\clock))\}\text{.}
		      \]

		      We prove that~$t_1\leq t_0$.

		      for any clock~$i$ we have that~$\Dio=(\partieFrac(\param),\leq)$ and $\Dio=(-\partieFrac(\param),\leq)$ \ie{}~$\dzeroi=-\dio$
		      for some~$\param$, hence~$-\partieFrac(\clockval(\clock_i))-\pval(\dzeroi)=-\partieFrac(\clockval(\clock_i))+\pval(\dio)$.

		      By definition of $\TEF_<$ applied to $(\E,\D)$, we have that $\Doi'=\Doi+(0,<)$.
		      In order to respect the constraint~$-\partieFrac(\clockval(\clock_i))-\de\compOpLeq_{0i'} \pval(\dzeroi')$
		      which is, as~$\compOpLeq_{0i'}=<$, $-\partieFrac(\clockval(\clock_i))-\de< \pval(\dzeroi')$,
		      we have to find a~$\de>0$.

		      In order to find an upper bound for~$\de$, we are going to prove that~$t_1>0$.
		      From \crefDef{definition:pointPDBM}~(\refc{PDBM2ii}) we have in~$(\E,\D)$ that for any clocks~$i,j$
		      $\Djo\leq\Dji+\Dio$. Let~$\clock_i\in\LFPd{\D}$.
		      From \crefDef{definition:pointPDBM}~(\refc{PDBM2i}), we have that~$\Dio\leq(1,<)$.
		      This gives that $\Dji+\Dio\leq\Dji+(1,<)$.

		      By definition of $\TEF_<$ applied to $(\E,\D)$, we have that $\Djo'=\Dji+(1,<)$.
		      Hence we have~$\Djo\leq\Djo'$.

		      Now as~$\partieFrac(\clockval(\clock_i))\compOpLeq_{i0}\pval(\dio)$
		      we can write~$\partieFrac(\clockval(\clock_i))\compOpLeq_{i0'}\pval(\dio')$
		      and then~$0 \compOpLeq_{i0'}\pval(\dio')-\partieFrac(\clockval(\clock_i))$
		      where~$\compOpLeq_{i0'}=<$, which prove our result.

		      We define~$\de=\frac{t_1}{2}$, therefore~$t_0<\de<t_1$.
		      Now for all clock~$\clock_j$, we obtain two results.
			First we have
			\[
			 -\partieFrac(\clockval(\clock_j))-\de<-\partieFrac(\clockval(\clock_j))-t_1\compOpLeq_{0j'}-\partieFrac(\clockval(\clock_j))+\partieFrac(\clockval(\clock_j))+\pval(\doj')=\pval(\doj')
			\]
			which is $-\pval(\doj')\compOpLeq_{0j}\partieFrac(\clockval(\clock_j))+\de$ as~$\compOpLeq_{0j'}={<}$.

			Secondary we have
			\[
			 \partieFrac(\clockval(\clock_j))+\de<\partieFrac(\clockval(\clock_j))+t_0\compOpLeq_{j0'}\partieFrac(\clockval(\clock_j))-\partieFrac(\clockval(\clock_j))+\pval(\djo')=\pval(\djo')
			\]
			which is $\partieFrac(\clockval(\clock_j))+\de\compOpLeq_{j0}\pval(\djo')$ as~$\compOpLeq_{0j'}={<}$.

		      Now we prove that constraints defined in~$(\E',\D')$ on differences of clocks are also satisfied
		      by going forward of~$\de$ units of time from~$\clockval$

		      Recall that in $(\E',\D')$ from the definition of \cref{algorithm:TEup}\ea{réf dynamique manquante} we have for all clock~$\clock_j$,
		      \[
		      \Dji'=\Dji
		       \quad
		       \text{and}
		       \quad
		       \Dij'=\Dij\text{.}
		      \]

		      Since we already treated the case whether $i$ or $j$ are $0$, now suppose~$i,j$ are both different from~$0$.
		      We have that $\partieFrac(\clockval(\clock_j))-\partieFrac(\clockval(\clock_i))\compOpLeq_{ji}\pval(\dji)=\pval(\dji')$,
		      and therefore as $\compOpLeq_{ji'}=\compOpLeq_{ji}$,
		      \[
		      \partieFrac(\clockval(\clock_j))+\de-\partieFrac(\clockval(\clock_i))-\de\compOpLeq_{ji'}\pval(\dji')\text{.}
		      \]
		      We also have that $\partieFrac(\clockval(\clock_i))-\partieFrac(\clockval(\clock_j))\compOpLeq_{ij}\pval(\dij)=\pval(\dij')$,
		      therefore as $\compOpLeq_{ij'}=\compOpLeq_{ij}$,
		      \[
		      \partieFrac(\clockval(\clock_i))+\de-\partieFrac(\clockval(\clock_j))-\de\compOpLeq_{ij'}\pval(\dij')\text{.}
		      \]

		      Finally, we define~$\clockval'=\clockval+\de$ and~$\clockval'\in(\E',\pval(\D'))$. \qedhere
	  \end{preuve}
}

\begin{preuve}
\versionProofIn{
\propTimeElapsingLong{}
}
\versionProofOut{
	This proof is quite technical. Intuitively, we bound the difference of each upper bound~$\pval(\dio)$ and~$\clockval(\clock_i)$ and each lower bound~$\pval(\dzeroi)$ and~$\clockval(\clock_i)$. This allows us to take a delay~$\delta$ inside these bounds that allows us to reach the next~\mPDBM{}.
	\longVersion{See \cref{proofssiTimeElapsing} for details.}%
}
\end{preuve}

\begin{figure*}[h!]
\vspace{-9em}
\hspace{-3em}
\includegraphics[width=1\linewidth]{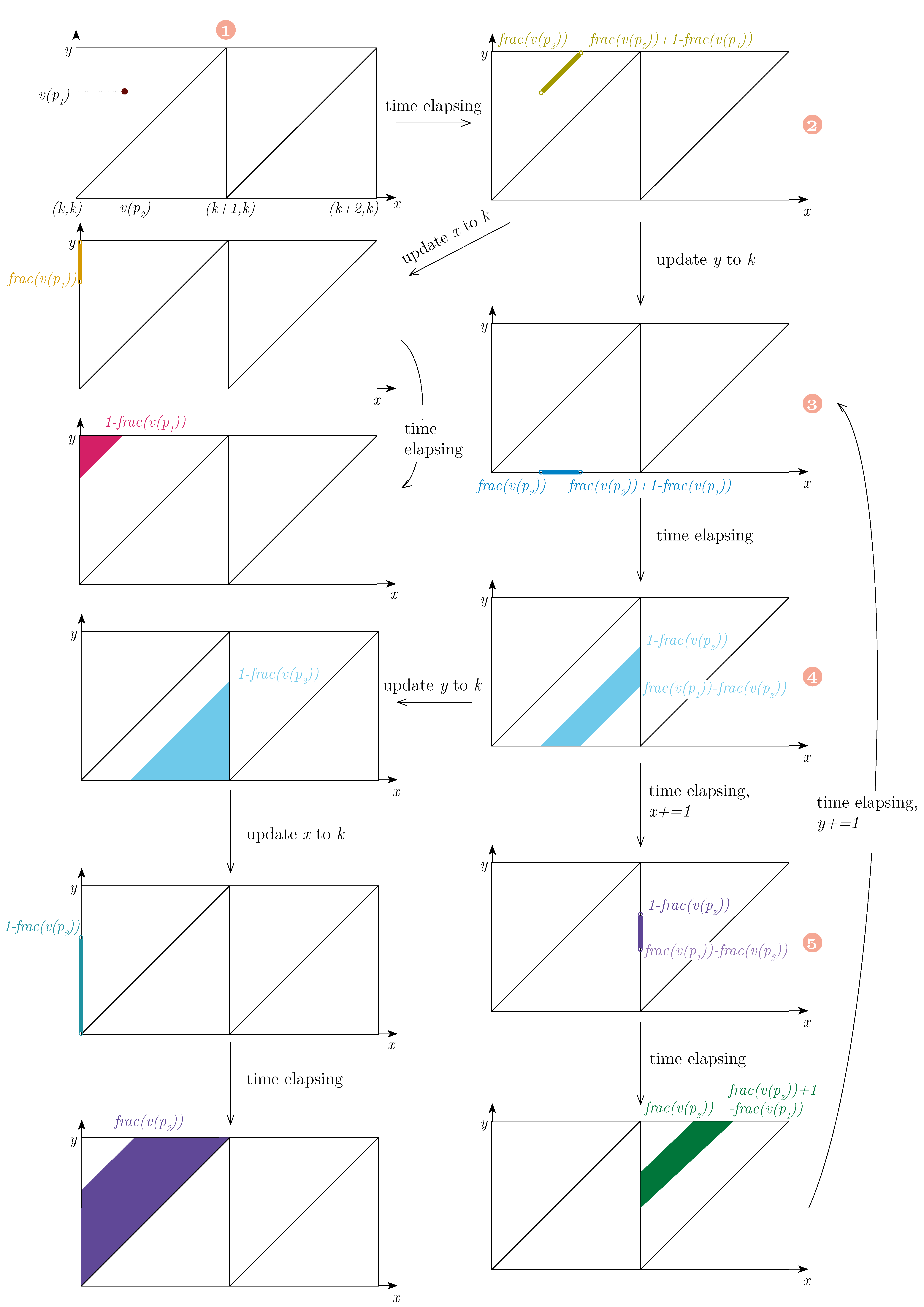}
\caption{Representation of \mPDBMs{} in two dimensions with two clocks~$\clock, \clockk$, two parameters~$\param_1, \param_2$ and~$\pval$ \st{} $\lfloor\pval(\param_1)\rfloor=\lfloor\pval(\param_2)\rfloor$ and $\partieFrac(\pval(\param_1))>\partieFrac(\pval(\param_2))$.}%
\label{fig:pPDBMs}
\end{figure*}

\medskip
\noindent\emph{Running example:} \cref{fig:pPDBMs} represents graphically different \mPDBMs{} obtained after an update $\resetfun(\clock)=\pval(\param_2)$
 and $\resetfun(\clockk)=\pval(\param_1)$ (figure \textcolor{red}{1}). %
 Time elapsing before~$\clockk\in\LFP$ reaches the next integer gives the \oPDBM{} satisfying condition~\ref{PDBMiv} (figure \textcolor{red}{2})
 	{\scriptsize
\[\hspace{-1.5em}(\E,\D)=\Big(	\begin{pmatrix}
	k \\
 	k
	\end{pmatrix},
	\begin{pmatrix}
							& \mathbf{0} 									& \mathbf{\clock} 								& \mathbf{\clockk}\\
	\mathbf{0} 				& (0,\leq) 										& (-\partieFrac(\param_2),<) 						& (-\partieFrac(\param_1),<)\\
	\mathbf{\clock}				& (\partieFrac(\param_2)+1-\partieFrac(\param_1),<) 	& (0,\leq) 										& (-\partieFrac(\param_1)+\partieFrac(\param_2),\leq) \\
	\mathbf{\clockk} 			& (1,<) 										& (\partieFrac(\param_1)-\partieFrac(\param_2),\leq) 	& (0,\leq)
	\end{pmatrix}
\Big)\] }%

\noindent  After an update of~$\clockk$ to~$k$ prior to reaching~$k+1$, the \oPDBM{} satisfying condition~\ref{PDBMiii} obtained is (figure \textcolor{red}{3})
 	{\scriptsize
\[\hspace{-1.5em}(\E,\D)=\Big(	\begin{pmatrix}
	k \\
 	k
	\end{pmatrix},
	\begin{pmatrix}
							& \mathbf{0} 									& \mathbf{\clock} 								& \mathbf{\clockk}\\
	\mathbf{0} 				& (0,\leq) 										& (-\partieFrac(\param_2),<) 						& (0,\leq)\\
	\mathbf{\clock}				& (\partieFrac(\param_2)+1-\partieFrac(\param_1),<) 	& (0,\leq) 										& (\partieFrac(\param_2)+1-\partieFrac(\param_1),<) \\
	\mathbf{\clockk} 			& (0,\leq) 										& (-\partieFrac(\param_2),<)						& (0,\leq)
	\end{pmatrix}
\Big)\] }%

 \noindent Time elapsing before~$\clock\in\LFP$ reaches the next integer gives the \oPDBM{} satisfying condition~\ref{PDBMiv} (figure \textcolor{red}{4})
	{\scriptsize
\[\hspace{-1.5em}(\E,\D)=\Big(	\begin{pmatrix}
	k \\
 	k
	\end{pmatrix},
	\begin{pmatrix}
							& \mathbf{0} 						& \mathbf{\clock} 						& \mathbf{\clockk}\\
	\mathbf{0} 				& (0,\leq) 							& (-\partieFrac(\param_2),<) 				& (0,<)\\
	\mathbf{\clock}				& (1,<) 							& (0,\leq) 								& (\partieFrac(\param_2)+1-\partieFrac(\param_1),<) \\
	\mathbf{\clockk} 			& (1-\partieFrac(\param_2),<) 			& (-\partieFrac(\param_2),<)				& (0,\leq)
	\end{pmatrix}
\Big)\] }%

 \noindent When~$\clock\in\LFP$ reaches~$k+1$, the \oPDBM{} satisfying condition~\ref{PDBMiii} obtained is (figure \textcolor{red}{5})
	{\scriptsize
\[\hspace{-1.5em}(\E,\D)=\Big(	\begin{pmatrix}
	k+1\\
 	k
	\end{pmatrix},
	\begin{pmatrix}
							& \mathbf{0} 						& \mathbf{\clock} 					& \mathbf{\clockk}\\
	\mathbf{0} 				& (0,\leq) 							& (0,\leq) 							& (-\partieFrac(\param_1)+\partieFrac(\param_2),<)\\
	\mathbf{\clock}				& (0,\leq) 							& (0,\leq) 							& (-\partieFrac(\param_1)+\partieFrac(\param_2),<) \\
	\mathbf{\clockk} 			& (1-\partieFrac(\param_2),<) 			& (1-\partieFrac(\param_2),<)			& (0,\leq)
	\end{pmatrix}
\Big)\] }%

\longVersion{
\subsection{Non-parametric guard}\label{section:npg}
}
\shortVersion{\textbf{Non-parametric guard.}}

From~\cite[Section 4.2]{AD94} we have that either every clock valuation of a clock region satisfies a guard, or none of them does.
Note that a \mPDBM{} for $\paramR$ is contained into a clock region of \cref{regionAD} (see \cref{proof:ssiNPguard} for more details), therefore
we have that if~$\clockval\in(\E,\pval(\D))$ satisfies a non-parametric guard~$\guard$,
then for all~$\clockval'\in(\E,\pval(\D))$ we also have~$\clockval'$ satisfies~$\guard$.

Let~$\pval\in\paramR$. We define $\pval\in\guardF(\guard, \E,\D)$ iff for all~$\clockval\in(\E,\pval(\D))$, $\clockval\models\guard$.
As any two~$\pval,\pval'\in\paramR$ satisfy the same constraints,
the following lemma is straightforward
\newcommand{\ssiNPguard}{
  Let~$(\E, \D)$ be a \mPDBM{} for~$\paramR$ and $\pval\in\paramR$.
  Let~$\guard$ be a non-parametric guard.
  If $\pval\in\guardF(\guard, \E,\D)$, then for all~$\pval'\in\paramR$, $\pval'\in\guardF(\guard, \E,\D)$.
}

  \begin{lem}\label{ssiNPguard}
\ssiNPguard{}
 \end{lem}

\newcommand{\proofssiNPguard}{%
\label{proof:ssiNPguard}

Our idea is to \shortVersion{consider}\longVersion{define} a clock region ``larger'' than our \mPDBM{} (following \cref{regionAD}) and \shortVersion{such}\longVersion{show} that, even for this (larger) clock region, either all clock valuations satisfy the guard---or none does.

\begin{defi}\label{definition:PDBMClockRegion}
 Let $\paramR$ be a parameter region, $\pval\in\paramR$. Let $(\E,\D)$ be a \mPDBM{} for $\paramR$.
 We define the \emph{clock region containing~$(\E, \valuate{\D}{\pval})$}, denoted by $\itsRegion{(\E, \valuate{\D}{\pval})}{\clockR}$, as follows:
 for all $\clockval\in \itsRegion{(\E, \valuate{\D}{\pval})}{\clockR}$, for all clocks $\clock_{i}, \clock_{j}$,
 \begin{itemize}
  \item if $\E_{\clock_{i}}<\CONSTMAX$, $\floor{\valuate{\clock_{i}}{\clockval}}=\E_{\clock_{i}}$,
  else if $\E_{\clock_{i}}=\infty$, $\valuate{\clock_{i}}{\clockval}\geq \CONSTMAX$;
  \item if $(0,\leq)<\Dij$ \valid{} and $\E_{\clock_{i}}<\CONSTMAX$,
  $\partieFrac(\valuate{\clock_{j}}{\clockval})< \partieFrac(\valuate{\clock_{i}}{\clockval})$;
    \item if $(0,\leq)=\Dij$ \valid{} and $\E_{\clock_{i}}<\CONSTMAX$,
  $\partieFrac(\valuate{\clock_{j}}{\clockval})= \partieFrac(\valuate{\clock_{i}}{\clockval})$;
  \item if $\Dio=\Doi=(0,\leq)$ and $\E_{\clock_{i}}<\CONSTMAX$,
  $\partieFrac(\valuate{\clock_{i}}{\clockval})=0$;
  \item if $\Dio\neq (0,\leq), \Doi\neq (0,\leq)$ and $\E_{\clock_{i}}<\CONSTMAX$,
  $\partieFrac(\valuate{\clock_{i}}{\clockval})\neq 0$.
 \end{itemize}
\end{defi}

\begin{clm}\label{lemma:containing-clock-regions}
	Let~$(\E, \D)$ be a \mPDBM{} for~$\paramR$ and $\pval\in\paramR$.
	We have $(\E,\pval(\D))\subseteq\itsRegion{(\E, \valuate{\D}{\pval})}{\clockR}$.
\end{clm}

\begin{preuve}
 Clock regions of \crefDef{regionAD} define constraints on clocks of the form~$0=\partieFrac(\clock)$,
 $0<\partieFrac(\clock)<1$, $0=\partieFrac(\clock)-\partieFrac(\clockk)$ and~$0<\partieFrac(\clock)-\partieFrac(\clockk)<1$ for some~$\clock, \clockk$,
 and~$\lfloor\clock\rfloor=k$ for some integer~$k$.
 Let~$(\E, \D)$ be a \mPDBM{} for~$\paramR$ and $\pval\in\paramR$. It defines constraints
\[\bigwedge_{i, j \in {[0, \ClockCard]}^2}
	\partieFrac(\clock_i)-\partieFrac(\clock_j)\compOpLeq_{i,j}\pval(\dij)
	\quad
	\land
	\quad
	\bigwedge_{i \in [1, \ClockCard]} \lfloor\clock_i\rfloor = \E_i\textit{.}
\]

Clearly, if $\clockval\in(\E,\pval(\D))$ satisfies~$\lfloor\clock_i\rfloor = \E_i$ then it satisfies the same constraint defined in~$\itsRegion{(\E, \valuate{\D}{\pval})}{\clockR}$.

Consider the constraints~$\partieFrac(\clock_i)-\partieFrac(\clock_j)\compOpLeq_{i,j}\pval(\dij)$ and~$\partieFrac(\clock_j)-\partieFrac(\clock_i)\compOpLeq_{j,i}\pval(\dji)$.
\begin{itemize}
 \item Suppose~$i, j$ are both different from~$0$. From \crefDef{def:med-PDBM}~(\refc{PDBMvi}) and~\crefDef{definition:pointPDBM},
 either~$\dij=\dji$ and then $\compOpLeq_{i,j}={\leq}=\compOpLeq_{j,i}$, then if~$\dij=\dji=0$ it satisfies the same constraint defined in~$\itsRegion{(\E, \valuate{\D}{\pval})}{\clockR}$,
 or~$\dij$ and~$\dji$ are different from~$0$, as they are elements of~$\PLT$ which are strictly smaller than~$1$, it satisfies either~$0<\partieFrac(\clock_i)-\partieFrac(\clock_j)<1$
 or~$0=\partieFrac(\clock_i)-\partieFrac(\clock_j)$ in~$\itsRegion{(\E, \valuate{\D}{\pval})}{\clockR}$.
 Finally if~$\dij\neq\dji$, then $\compOpLeq_{i,j}={<}=\compOpLeq_{j,i}$ and it satisfies~$0<\partieFrac(\clock_i)-\partieFrac(\clock_j)<1$ in~$\itsRegion{(\E, \valuate{\D}{\pval})}{\clockR}$.
 \item Suppose~$i$ is different from~$0$ and~$j=0$.
 From \crefDef{def:med-PDBM}~(\refc{PDBMvi}) and~\crefDef{definition:pointPDBM},
 either~$\dio=\dzeroi$ and then $\compOpLeq_{i,0}={\leq}=\compOpLeq_{0,i}$, then if~$\dio=\dzeroi=0$ it satisfies the same constraint defined in~$\itsRegion{(\E, \valuate{\D}{\pval})}{\clockR}$,
 or~$\dio$ and~$\dzeroi$ are different from~$0$, as they are elements of~$\PLT$ which are strictly smaller than~$1$, it satisfies either~$0<\partieFrac(\clock_i)<1$
 or~$0=\partieFrac(\clock_i)$ in~$\itsRegion{(\E, \valuate{\D}{\pval})}{\clockR}$.
 Finally if~$\dio\neq\dzeroi$, then $\compOpLeq_{i,0}={<}=\compOpLeq_{0,i}$ and it satisfies~$0<\partieFrac(\clock_i)<1$ in~$\itsRegion{(\E, \valuate{\D}{\pval})}{\clockR}$.
 \item The case~$j$ is different from~$0$ and~$i=0$ is similar.
 \item Suppose both~$i,j$ are~$0$, the constraint is not taken into account as we have no~$\clock_0$ in~$\itsRegion{(\E, \valuate{\D}{\pval})}{\clockR}$.
\end{itemize}
Finally, we have that if~$\clockval\in(\E,\pval(\D))$ then~$\clockval\in\itsRegion{(\E, \valuate{\D}{\pval})}{\clockR}$.
\end{preuve}

\medskip
Now we come back to the proof of the main lemma:

Let~$(\E, \D)$ be a \mPDBM{} for~$\paramR$ and $\pval\in\paramR$. It defines constraints
\[\bigwedge_{i, j \in {[0, \ClockCard]}^2}
	\partieFrac(\clock_i)-\partieFrac(\clock_j)\compOpLeq_{i,j}\pval(\dij)
	\quad
	\land
	\quad
	\bigwedge_{i \in [1, \ClockCard]} \lfloor\clock_i\rfloor = \E_i\textit{.}
\]
Moreover, let~$\guard$ be a non-parametric guard. It defines constraints for a finite number of integer constants~$k_i$ with~$i\in I\subseteq [1, \ClockCard]$
\[\bigwedge_{i \in I}
	\partieFrac(\clock_i)\leq 0
	\quad
	\land
	\quad
\bigwedge_{i \in I}
	-\partieFrac(\clock_i)\leq 0
	\quad
	\land
	\quad
	\bigwedge_{i \in I} \lfloor\clock_i\rfloor \compOp k_i\textit{.}
\]

The intersection between the two is given by the conjunction of those constraints.
We project this intersection on parameter variables (by elimination of clock variables) and we prove that the intersection does not create new constraints on parameters different from those we already have in~$(\E,\pval(\D))$ (and therefore in~$\paramR$).
For some set of clocks~$I\subseteq [1, \ClockCard]$ and~$i\in I$,
suppose we have the constraints~$\partieFrac(\clock_i)\leq 0$ and~$-\partieFrac(\clock_i)\leq 0$ in~$\guard$.
When eliminating~$\clock_i$ in any constraint of the form~$\partieFrac(\clock_i)-\partieFrac(\clock_j)\compOpLeq_{i,j}\pval(\dij)$,
it is clear that we proceed on~$\PLT$ to the operation $(0,\leq)+(\dij,\compOpLeq_{i,j})=(0+\dij,\leq\oplus\compOpLeq_{i,j})=(\dij,\compOpLeq_{i,j})$.
The same way on any constraint of the form~$\partieFrac(\clock_i)\compOpLeq_{i,0}\pval(\dio)$, eliminating~$\clock_i$
gives the constraint~$(0,\leq)+(\dio,\compOpLeq_{i,0})=(\dio,\compOpLeq_{i,0})$.
Hence it does not create new inequalities not belonging to~$\paramR$.

Now suppose~$\pval\in\guardF(\guard, \E,\D)$. We have that all~$\clockval\in(\E,\pval(\D))$ satisfy~$\guard$.

\noindent As no new constraints not in~$\PLT$ have been created, all~$\pval'\in\paramR$ respect the same constraints on their fractional part and integer part as~$\pval$
and therefore, $(\E,\pval'(\D))$ is contained in the same clock region as~$(\E,\pval(\D))$ is,
\ie{} $\itsRegion{(\E, \valuate{\D}{\pval})}{\clockR}=\itsRegion{(\E, \valuate{\D}{\pval'})}{\clockR}$.
Finally, $\pval'\in\guardF(\guard, \E,\D)$.
}
\begin{proof}
\versionProofIn{
\proofssiNPguard{}
}
\versionProofOut{
See \cref{appendix:proof:ssiNPguard}.
}
\end{proof}

\longVersion{
\subsection{Parametric guard}\label{section:pg}
}
\shortVersion{\textbf{Parametric guard.}}

\longVersion{As for the previous result, using}\shortVersion{Using} a projection on parameters\longVersion{ \ie{} eliminating clocks,} does not create new constraints on parameters
	that are not already in a parameter region~$\paramR$.
	Indeed, a parametric guard~$\guard$ only adds new constraints of the form~$\clock\compOp\param$ which gives\longVersion{,
	when eliminating clocks in both a \mPDBM{} $(\E,\D)$ and a parametric guard,}
	again a comparison between elements of~$\PLT$. Therefore, these new constraints already belong to~$\PLT$ and we can decide whether the set of clock valuations satisfying
	these constraints is non-empty \ie{} given~$\pval\in\paramR$,~$\pval(\guard)$ is satisfied by some clock valuation~$\clockval\in(\E,\pval(\D))$.
	This is a key point in the overall process of proving the decidability of our \RPGRtoPPTAs{}.

Note that there will also be additional constraints involving clocks (with other clocks, constants or parameters), but they will not be relevant as we immediately update all clocks, therefore replacing these constraints with new constraints encoding the clock updates.

Let~$\pval\in\paramR$. We define $\pval\in\guardpF(\guard,\E, \D)$ iff there is a $\clockval~{\in(\E,\valuate{\D}{\pval})}$ \st{} $\clockval\models\valuate{\guard}{\pval}$.\footnote{%
	Remark that here is why our construction works for EF-emptiness, but cannot be used for, \eg{}, AF-emptiness (``is there a parameter valuation such that all runs reach a goal location~$\loc$''): unlike~$\guardF(\guard,\E, \D)$, not all clock valuations in a \mPDBM{}~$(\E,\pval(\D))$ can satisfy a parametric guard if~$\pval\in\guardpF(\guard,\E, \D)$.
}
Again, as any two~$\pval,\pval'\in\paramR$ satisfy the same constraints,
the following lemma is straightforward
\newcommand{\ssiPguard}{
  Let~$(\E, \D)$ be a \mPDBM{} for~$\paramR$ and $\pval\in\paramR$.
  Let~$\guard$ be a parametric guard.
  If $\pval\in\guardpF(\guard, \E,\D)$, then for all~$\pval'\in\paramR$, $\pval'\in\guardpF(\guard, \E,\D)$.
}
 \begin{lem}\label{ssiPguard}
\ssiPguard{}
 \end{lem}

 \newcommand{\proofssiPguard}{%
 \label{proof:ssiPguard}
 Let~$(\E, \D)$ be a \mPDBM{} for~$\paramR$ and $\pval\in\paramR$.
 Let~$\guard$ be a parametric guard and suppose $\pval\in\guardpF(\guard, \E,\D)$.
 After applying a projection on parameters, we obtain constraints on elements of~$\PLT$.
 By hypothesis, all these constraints are satisfied by~$\pval$.
 Suppose~$\pval'\in\paramR$.
 By definition of our parameter regions, and since $\pval$ and $\pval'$ both belong to~$\paramR$, $\pval'$ satisfies the same constraints on elements of~$\PLT$.
 Therefore, the same constraints is satisfied by~$\pval'$ and~$\pval'\in\guardpF(\guard, \E,\D)$.
}

\begin{proof}
\versionProofIn{
\proofssiPguard{}
}
\versionProofOut{
See \cref{appendix:proof:ssiPguard}.
}
\end{proof}

\medskip

Now that we have defined useful operations on \mPDBMs{}, we are going, given a parameter region~$\paramR$, to construct a finite region automaton in which for any run, there is an equivalent concrete run in the \RPGRtoPPTA{}.

\section{Parametric region automaton}\label{section:region}

Let $(\E,\D)\in\PDBMRp$, we say $(\E',\D')\in \Succ((\E,\D)) \Leftrightarrow \exists~{i\geq 0}$
\st{} $(\E', \valuate{\D'}{\pval})=\TEF^i((\E, \D))$.
In other words, $(\E',\D')$ is obtained after applying~$\TEF((\E,\D))$ a finite number of times.
$\Succ((\E,\D))$ is also called the \emph{time successors} of~$(\E,\D)$.

In order to finitely simulate an \RPGRtoPPTA{}, we create a parametric region automaton.

\begin{defi}[Parametric region automaton]\label{definition:parametricRA}
    Let $\paramR$ be a parameter region.
For an \RPGRtoPPTA{} $\A=(\Actions, \Loc, \locinit, \Clock, \Param, \Edges)$,
    given %
    $(\E_{0},\D_{0})$ the initial~\mPDBM{} where all clocks are~$0$, the \emph{parametric region automaton} $\RA$ over $\paramR$
		is the tuple $(\Loc', \Actions, \Loc_{0}', \Edges')$ where:
	\begin{ienumeration}
		\item $\Loc'=\Loc\times \PDBMRp$
		\item $\Loc_{0}'= (\loc_{0}, (\E_0, \D_0) ) $
		\item $\Edges'=\{\big((\loc,(\E,\D)),\action,(\loc',(\E',\D'))\big)\in \Loc'\times\Actions\times \Loc' \mid \text {either } \exists \edge=\langle \loc, \guard, \action, \resetfunnp, \loc' \rangle\in \Edges$,
		$\guard$ is a non-parametric guard,
		$\exists (\E'',\D'')\in\Succ((\E,\D))$, $\paramR\subseteq\guardF(g,(\E'',\D''))$
		and $(\E',\D')=\resetF(\E'', \D'', \resetfunnp)$ is an \oPDBM{},
		or $\exists \edge=\langle \loc, \guard, \action, \resetfun, \loc' \rangle\in \Edges$, $\guard$ is a parametric guard,
		$\exists (\E'',\D'')\in\Succ((\E,\D))$, $\paramR\subseteq\guardpF(g,(\E'',\D''))$
		and $(\E',\D')=\resetpF(\E'', \D'', \resetfun)$ is a \pointPDBM{}.\}
	\end{ienumeration}
\end{defi}

\noindent
    Let $\paramR$ be a parameter region, \A{} be an \RPGRtoPPTA{} and $\RA = (\Loc', \Sigma, \Loc_{0}', \Edges')$\longVersion{ its parametric region automaton over $\paramR$}.
    A run in $\RA$ is an untimed sequence\\
    $\sigma:(\loc_{0},(\E_{0}, \D_{0}))\edge_{0}(\loc_{1}, (\E_{1}, \D_{1}))\edge_{1}\cdots (\loc_{i},(\E_{i}, \D_{i}))\edge_{i}(\loc_{i+1},(\E_{i+1}, \D_{i+1}))\edge_{i+1}\cdots$
    such that for all $i$ we have $\big((\loc_{i},(\E_{i}, \D_{i})),\action_{i},(\loc_{i+1},(\E_{i+1}, \D_{i+1}))\big)\in \Edges'$,
    which we also write $(\loc_{i},(\E_{i}, \D_{i}))\overset{\edge_{i}}{\longrightarrow}(\loc_{i+1},(\E_{i+1}, \D_{i+1}))$.
    Note that we label our transitions with the edges of the \RPGRtoPPTA{}.

\mr{parler qu'on a pas AF}
\section{Decidability of EF-emptiness and synthesis}\label{section:decidability}

Using our construction of the parametric region automaton $\RA$ for a given \RPGRtoPPTA{} $\A$, we state the next proposition.

\newcommand{\theoremCompletGuardRationalLong}{
Let~$\paramR$ be a parameter region. %
 Let~$\A$ be an \RPGRtoPPTA{} and~$\RA$ its parametric region automaton over $\paramR$.
There is a run~$\sigma:(\loc_{0},(\E_{0},\D{}_{0}))\overset{\edge_{0}}{\longrightarrow}(\loc_{1}, (\E{}_{1},\D{}_{1}))\overset{\edge_{1}}{\longrightarrow}\cdots (\loc_{f-1}, (\E{}_{f-1},\D{}_{f-1}))\overset{\edge_{f-1}}{\longrightarrow}(\loc_{f}, (\E{}_{f},\D{}_{f}))$
in~$\RA$ iff for all~$\pval\in\paramR$ there is a run~$\rho:(\loc_{0},\clockval_{0})\longuefleche{\edge_{0}}(\loc_{1}, \clockval_{1})\longuefleche{\edge_{1}}\cdots (\loc_{f-1},\clockval_{f-1})\longuefleche{\edge_{f-1}}(\loc_{f},\clockval_{f})$
in~$\valuate{\A}{\pval}$
\st{} for all~$0\leq i\leq f$, $\clockval_i\in(\E_i,\pval(\D_i))$.
}
\begin{prop}%
\label{completguardRational}
\theoremCompletGuardRationalLong{}
\end{prop}
\newcommand{\proofcompletguardRational}{%
\label{proofcompletguardRational}
\begin{itemize}
  \item[$\Leftarrow$]
By induction on the length of the run.

 Let $\pval\in \paramR$.
 As the basis for the induction, in the initial location $(\loc{}_{0},{\{0\}}^{\ClockCard})$ the only valuation is reachable by an empty run of $\valuate{\A}{\pval}$.
 Moreover ${\{0\}}^{\ClockCard}{\in}(\E_{0}, \valuate{\D_{0}}{\pval})$ the initial \mPDBM{} containing only $0$.
 Therefore the initial location
 $(\loc{}_{0},(\E_{0}, \valuate{\D_{0}}{\pval}))$ is reachable by an empty run of $\RA$.

For the induction step, suppose for all~$\pval$, there is run in~$\pval(\A)$ of length~$f-1$ we have our result.

Let $\pval\in \paramR$ and $\rho=(\loc_{0}, \clockval_{0})\overset{\edge_{0}}{\longrightarrow}\cdots\overset{\edge_{f-2}}{\longrightarrow}(\loc_{f-1}, \clockval_{f-1})\overset{\edge_{f-1}}{\longrightarrow}(\loc_{f}, \clockval_{f})$ be a run of $\valuate{\A}{\pval}$ of length $f$.
 By induction hypothesis, there is a run~$\sigma=(\loc_{0}, (\E_0,\D_0))\overset{\edge_{0}}{\longrightarrow}\cdots\overset{\edge_{f-2}}{\longrightarrow}(\loc_{f-1},(\E_{f-1},\D_{f-1}))$ in~$\RA$ and for all~$0\leq i\leq f-1$, $\clockval_i\in(\E_i, \pval(\D_i))$.

 Consider~$\edge_{f-1}$. By \crefDef{definition:parametricRA} of the parametric region automaton, it is also in its set of edges~$\Edges'$.
 Three cases show up:
 \begin{itemize}
 \item If~$\edge_{f-1}=\langle \loc_{f-1}, \action, \guard, \resetfunnp,\loc_f \rangle$ contains no parametric guard nor parametric update.
 Using \crefDef{definition:concretesemantics} there is a delay~$\delta$ (possibly $0$) \st{}~$(\loc_{f-1}, \clockval_{f-1})\longueflecheRel{\delta}(\loc_{f-1} \clockval_{f-1}')\longueflecheRel{\edge_{f-1}}(\loc_{f}, \clockval_{f})$ where~$\clockval_{f-1}'\models\guard$ and~$\clockval_{f}=\reset{\clockval_{f-1}'}{\resetfunnp}$.
As~$\clockval_{f-1}\in(\E_{f-1},\valuate{\D_{f-1}}{\pval})$ there is~$(\E_{f-1}',\D_{f-1}')\in\Succ((\E_{f-1},\D_{f-1}))$ \st{} from \cref{ssiTimeElapsing} we have~$\clockval_{f-1}'\in(\E_{f-1}',\pval(\D_{f-1}'))$.
As~$\clockval_{f-1}'\models\guard$ by construction of our \mPDBMs{} \longVersion{(see \cref{section:npg})} any other clock valuation belonging to~$(\E_{f-1}',\pval(\D_{f-1}'))$ satisfies~$\guard$.
Therefore~$\pval\in\guardF(\guard, \E_{f-1}',\D_{f-1}')$ and from \crefLemma{ssiNPguard}, $\paramR\subseteq\guardF(\guard, \E_{f-1}',\D_{f-1}')$.
Now, as~$\clockval_{f}=\reset{\clockval_{f-1}'}{\resetfunnp}$ consider the \oPDBM{} $(\E_f, \D_f)=\resetF((\E_{f-1}',\D_{f-1}'), \resetfunnp)$;
from \crefLemma{ssiReset} we have~$\clockval_f\in(\E_f, \pval(\D_f))$.
Finally there is an edge \[(\loc_{f-1},(\E_{f-1},\D_{f-1}))\overset{\edge_{f-1}}{\longrightarrow}(\loc_f,(\E_f, \D_f)).\]
	\item If~$\edge_{f-1}=\langle \loc_{f-1}, \action, \guard, \resetfun,\loc_f \rangle$ contains a parametric guard and a parametric update.
	Using \crefDef{definition:concretesemantics} there is a delay~$\delta$ (possibly $0$) \st{}~$(\loc_{f-1}, \clockval_{f-1})\longueflecheRel{\delta}(\loc_{f-1}, \clockval_{f-1}')\longueflecheRel{\edge_{f-1}}(\loc_{f}, \clockval_{f})$ where~$\clockval_{f-1}'\models\pval(\guard)$ and~$\clockval_{f}=\reset{\clockval_{f-1}'}{\pval(\resetfun)}$.
 As~$\clockval_{f-1}\in(\E_{f-1},\valuate{\D_{f-1}}{\pval})$ there is~$(\E_{f-1}',\D_{f-1}')\in\Succ((\E_{f-1},\D_{f-1}))$ \st{} from \cref{ssiTimeElapsing} we have~$\clockval_{f-1}'\in(\E_{f-1}',\pval(\D_{f-1}'))$.
 As~$\clockval_{f-1}'\models\pval(\guard)$, $\pval\in\guardpF(\guard, \E_{f-1}',\D_{f-1}')$ and from \crefLemma{ssiPguard}, $\paramR\subseteq\guardpF(\guard, \E_{f-1}',\D_{f-1}')$.
 Now, as~$\clockval_{f}=\reset{\clockval_{f-1}'}{\pval(\resetfun)}$ consider the \pointPDBM{} $(\E_f, \D_f)=\resetpF((\E_{f-1}',\D_{f-1}'), \resetfun)$;
 $(\E_f, \pval(\D_f))$ contains only one clock valuation, precisely defined by the fully parametric update~$\pval(\resetfun)$ so we have~$\clockval_f\in(\E_f, \pval(\D_f))$.
 Finally there is an edge $(\loc_{f-1},(\E_{f-1},\D_{f-1}))\overset{\edge_{f-1}}{\longrightarrow}(\loc_f,(\E_f, \D_f))$.
\item The case where~$\edge_{f-1}$ contains a non parametric guard and a parametric update is similar to the previous one.

\end{itemize}
	Finally, there is a run~$\sigma'=\sigma\overset{\edge_{f-1}}{\longrightarrow}(\loc_f,(\E_f, \D_f))$ of length~$f$ in~$\RA$ \st{} for all~$0\leq i\leq f$, $\clockval_i\in(\E_i, \pval(\D_i))$.

  \item [$\Rightarrow$]
	By induction on the length of the run.

	 Let $\pval\in \paramR$.
	 As the basis for the induction, the initial location $(\loc{}_{0},(\E_{0}, \valuate{\D_{0}}{\pval}))$ is reachable by an empty run of $\RA$.
	 Moreover, as~${\{0\}}^{\ClockCard}{\in}(\E_{0}, \valuate{\D_{0}}{\pval})$, the initial location
	 $(\loc{}_{0},{\{0\}}^{\ClockCard})$ is reachable by an empty run of $\pval(\A)$.

	For the induction step, suppose it is true for all run in~$\RA$ of length~$f-1$.

	Let~$\pval\in\paramR$ and $\sigma=(\loc_{0}, (\E_0,\D_0))\overset{\edge_{0}}{\longrightarrow}\cdots\overset{\edge_{f-2}}{\longrightarrow}(\loc_{f-1},(\E_{f-1},\D_{f-1}))\overset{\edge_{f-1}}{\longrightarrow}(\loc_f,(\E_f, \D_f))$ be a run of~$\RA$ of length~$f$.
 Consider~$\edge_{f-1}$. By \crefDef{definition:parametricRA} of the parametric region automaton, it is also in the set of edges~$\Edges$ of~$\A$.
 Two cases show up:
 \begin{itemize}
 \item If~$\edge_{f-1}=\langle \loc_{f-1}, \action, \guard, \resetfunnp,\loc_f \rangle$ contains no parametric guard nor parametric update.
 By induction hypothesis, there is a run $\rho=(\loc_{0}, \clockval_{0})\overset{\edge_{0}}{\longrightarrow}\cdots\overset{\edge_{f-2}}{\longrightarrow}(\loc_{f-1}, \clockval_{f-1})$ of $\valuate{\A}{\pval}$ of length $f-1$ \st{} for all~$0\leq i\leq f-1$, $\clockval_i\in(\E_i, \pval(\D_i))$.
 Using \crefDef{definition:parametricRA} there is~$(\E_{f-1}',\D_{f-1}')\in\Succ((\E_{f-1},\D_{f-1}))$,
 $\paramR\subseteq\guardF(\guard, \E_{f-1}',\D_{f-1}')$ and
$(\E_f, \D_f)=\resetF((\E_{f-1}',\D_{f-1}'), \resetfunnp)$.
From \cref{ssiTimeElapsing} we have~$\clockval_{f-1}'\in(\E_{f-1}',\pval(\D_{f-1}'))$ and a delay~$\delta$ \st{} $\clockval_{f-1}'=\clockval_{f-1}+\delta$.
As~$\paramR\subseteq\guardF(\guard, \E_{f-1}',\D_{f-1}')$
from \crefLemma{ssiNPguard} we have~$\pval\in\guardF(\guard, \E_{f-1}',\D_{f-1}')$ and $\clockval_{f-1}'\models\guard$.
 Moreover, since~$(\E_f, \D_f)=\resetF((\E_{f-1}',\D_{f-1}'), \resetfunnp)$, we define~$\clockval_{f}=\reset{\clockval_{f-1}'}{\resetfunnp}$
 and therefore from \crefLemma{ssiReset}, $\clockval_{f}\in(\E_f, \pval(\D_f))$.
 Finally there is an edge~$(\loc_{f-1}, \clockval_{f-1})\overset{\edge_{f-1}}{\longrightarrow}(\loc_f,\clockval_f)$ and a run~$\rho'=\rho\overset{\edge_{f-1}}{\longrightarrow}(\loc_f,\clockval_f)$ in~$\pval(\A)$ of length~$f$ \st{} for all~$0\leq i\leq f$, $\clockval_i\in(\E_i, \pval(\D_i))$.
\item If~$\edge_{f-1}=\langle \loc_{f-1}, \action, \guard, \resetfun,\loc_f \rangle$ contains a parametric guard and a parametric update.
Using \crefDef{definition:parametricRA} there is~$(\E_{f-1}',\D_{f-1}')\in\Succ((\E_{f-1},\D_{f-1}))$,
$\paramR\subseteq\guardpF(\guard, \E_{f-1}',\D_{f-1}')$ and
$(\E_f, \D_f)=\resetpF((\E_{f-1}',\D_{f-1}'), \resetfun)$.
From \crefLemma{ssiPguard} we can take~$\clockval_{f-1}'\in(\E_{f-1}',\pval(\D_{f-1}'))$ \st{} $\clockval_{f-1}'\models\pval(\guard)$.
Let~$\clockval_{f}=\reset{\clockval_{f-1}'}{\pval(\resetfun)}$.
Clearly, $(\E_f, \D_f)=\resetpF((\E_{f-1}',\D_{f-1}'), \resetfun)$ is a \pointPDBM{};
as~$(\E_f, \pval(\D_f))$ contains only one clock valuation precisely defined by the fully parametric update~$\pval(\resetfun)$, we have~$\clockval_f\in(\E_f, \pval(\D_f))$.
From \cref{ssiTimeElapsing} as~$\clockval_{f-1}'\in(\E_{f-1}',\pval(\D_{f-1}'))$ there is a delay~$\delta$ and a~$\clockval_{f-1}\in(\E_{f-1},\pval(\D_{f-1}))$ \st{} $\clockval_{f-1}'=\clockval_{f-1}+\delta$.
Using the induction hypothesis, there is a run
$\rho=(\loc_{0}, \clockval_{0})\overset{\edge_{0}}{\longrightarrow}\cdots\overset{\edge_{f-2}}{\longrightarrow}(\loc_{f-1}, \clockval_{f-1})$ of $\valuate{\A}{\pval}$ of length $f-1$ \st{} for all~$0\leq i\leq f-1$, $\clockval_i\in(\E_i, \pval(\D_i))$.
Finally there is an edge~$(\loc_{f-1}, \clockval_{f-1})\overset{\edge_{f-1}}{\longrightarrow}(\loc_f,\clockval_f)$
and a run~$\rho'=\rho\overset{\edge_{f-1}}{\longrightarrow}(\loc_f,\clockval_f)$ in~$\pval(\A)$ of length~$f$ \st{} for all~$0\leq i\leq f$, $\clockval_i\in(\E_i, \pval(\D_i))$.

\item The case where~$\edge_{f-1}$ contains a non parametric guard and a parametric update is similar to the previous one. \qedhere
\end{itemize}
\end{itemize}
}
\begin{preuve}
\versionProofIn{
\proofcompletguardRational{}
}
\versionProofOut{
We prove this result by induction on the length of the run. It is quite direct as we construct runs without parametric guards.
See \cref{appendix:proofcompletguardRational} for details.
}
\end{preuve}

\begin{figure*}[tb]

\scriptsize
{\centering

\scalebox{0.80}{

	\begin{subfigure}[c]{1\linewidth}
	 \hspace*{-7em}\begin{tikzpicture}[shorten >=1pt, node distance=2.5cm, on grid, auto]
	\node[initial]       (Y)   {$(\loc_{0}, \ClocksZero)$};
	\node   (Z) [right=of Y]    {$(\loc_{1}, \clockval_1)$};
	\node      (A) [right=of Z]    {$\cdots$};
	\node          (B)   [right=of A]        {$(\loc_{i}, \clockval_i)$};
	\node         (C)   [right=of B]        {$(\loc_{i+1}, \clockval_{i+1})$};
	\node         (D)   [right=of C]        {$\cdots$};
	\node         (E)   [right=of D]        {$(\loc_{j}, \clockval_j)$};
	\node[accepting]        (F)   [right=of E]        {$(\loc_{f}, \clockval_f)$};

	\path[->]
				(Y)  edge   node  [swap, above]   {$\edge_0$}   (Z)
				(Z)  edge   node  [swap, above]   {$\edge_1$}   (A)
				(A)  edge   node  [swap, above]   {$\edge_{i-1}$}   (B)
				(B) edge  node[swap, above]   {$\paraminline{\edge_{i}}$}   (C)
				(C) edge  node[swap, above]   {$\edge_{i+1}$}   (D)
				(D) edge  node[swap, above]   {$\edge_{j-1}$}   (E)
				(E) edge  node[swap, above]   {$\edge_{j}$}   (F);

	\end{tikzpicture}
	\caption{run of \A{} with one parametric transition~$\paraminline{\edge_{i}}$}%
	\label{figure:runR2PPTA}
	\end{subfigure}

  }

  \scalebox{0.80}{

	\begin{subfigure}[c]{1\linewidth}
	\hspace*{-8.3em}\begin{tikzpicture}[shorten >=1pt, node distance=2.5cm, on grid, auto]
	\node[initial, rectangle, rounded corners, minimum size=12pt, draw=black, fill=gray!10, inner sep=2pt]       (Y)   {$(\loc_{0}, (\E_0,\D_0))$};
	\node[rectangle, rounded corners, minimum size=12pt, draw=black, fill=gray!10, inner sep=2pt]   (Z) [right=of Y]    {$(\loc_{1}, (\E_1,\D_1))$};
	\node      (A) [right=of Z]    {$\cdots$};
	\node[rectangle, rounded corners, minimum size=12pt, draw=black, fill=gray!10, inner sep=2pt]          (B)   [right=of A]        {$(\loc_{i},(\E_i,\D_i))$};
	\node[rectangle, rounded corners, minimum size=12pt, draw=black, fill=gray!10, inner sep=2pt]       (C) [right=of B]   {$(\loc_{i+1}, (\E_{i+1},\D_{i+1}))$};
	\node   (F) [right=of C]    {$\cdots$};
	\node[rectangle, rounded corners, minimum size=12pt, draw=black, fill=gray!10, inner sep=2pt]       (D) [right=of F]    {$(\loc_{j}, (\E_j,\D_j))$};
	\node[accepting, rectangle, rounded corners, minimum size=12pt, draw=black, fill=gray!10, inner sep=2pt]           (E)   [right=of D]        {$(\loc_{f}, (\E_f,\D_f))$};

	\path[->]
				(Y)  edge   node  [swap, above]   {$\edge_0$}   (Z)
				(Z)  edge   node  [swap, above]   {$\edge_1$}   (A)
				(A)  edge   node  [swap, above]   {$\edge_{i-1}$}   (B)
				(B)  edge   node  [swap, above]   {$\paraminline{\edge_{i}}$}   (C)
				(C)  edge   node  [swap, above]   {$\edge_{i+1}$}   (F)
				 (F)  edge   node  [swap, above]   {$\edge_{j-1}$}   (D)
				 (D)  edge   node  [swap, above]   {$\edge_{j}$}   (E);

	\end{tikzpicture}
	\caption{run of $\RA$ with one parametric transition~$\paraminline{\edge_{i}}$}%
	\label{figure:runRAL1}
	\end{subfigure}
}

}
	\caption{A run in an \RPGRtoPPTA{} \A{} (above) and its equivalent run in~$\RA$ (below)}%
	\label{figure:RR2PPTAtoRALtoRA}

\end{figure*}
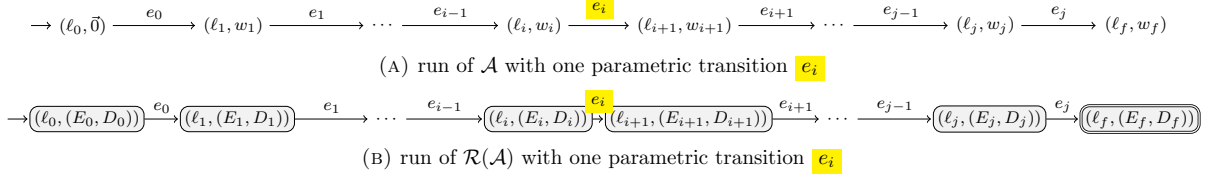

\begin{exa}
 Consider \cref{figure:RR2PPTAtoRALtoRA}.
 Let~$\A$ be an \RPGRtoPPTA{},~$\paramR$ a parameter region and~$\pval\in\paramR$.
 Suppose there is a run in~$\A$, starting from the initial location~$(\loc_0, \ClocksZero)$ reaching a goal location~$(\loc_f, \clockval_f)$.
 Along this run, all edges are non-parametric transitions but~$\paraminline{\edge_i}=\langle \loc_i, \guard, \action_i, \resetfun, \loc_{i+1} \rangle$.
 That is,~$\resetfun$ is a total parametric update,
 and~$\guard$ is a possibly parametric guard.

 The first part of this run, from~$(\loc_0, \ClocksZero)$ to~$(\loc_i, \clockval_i)$ %
    starts from~$(l_0, (\E_0,\D_0))$ where~$(\E_0,\D_0)$ is the~\mPDBM{} of the initial clock region $\{\vec{0}\}$,
 and ends in~$(\loc_i, (\E_i,\D_i))$. The second part of this run, from~$(\loc_{i+1}, \clockval_{i+1})$ to~$(\loc_f, \clockval_f)$
 starts from~$(l_{i+1}, (\E_{i+1},\D_{i+1}))$ where~$(\E_{i+1},\D_{i+1})$ is a~\pointPDBM{},
 and can reach~$(\loc_f, (\E_f,\D_f))$ and further ends in~$(\loc_s, (\E_s,\D_s))$.

 These runs %
 contain only non-parametric transitions, and as there is an edge in~\A{} from~$(\loc_i, \clockval_i)$ to~$(\loc_{i+1}, \clockval_{i+1})$,
 we have to bisimulate this run in~$\RA$
 containing the parametric transition~$\paraminline{\edge_i}$, where~$\resetpF((\E_{i},\D_{i}),\resetfun)$ gives~$(\E_{i+1},\D_{i+1})$.
\end{exa}

From \cref{completguardRational}, \longVersion{we deduce that }if there is a run reaching a goal location in an instantiated \RPGRtoPPTA{},
then for another parameter valuation in the same parameter region there is a run in the instantiated \RPGRtoPPTA{} with the same locations and transitions (but possibly different delays), reaching the same location.

\newcommand{\lemmaRunCompleteLong}{
	  Let $\A$ be an \RPGRtoPPTA{}.
	  Let $\paramR$ be a parameter region and~$\pval\in\paramR$.
	  If there is a run $\rho=(\loc_{0}, \clockval_{0})\longuefleche{\edge_{0}}\cdots\longuefleche{\edge_{i-1}}(\loc_{i}, \clockval_{i})$ in~$\valuate{\A}{\pval}$,
	    then for all $\pval'\in \paramR$
	    there is a run $\rho'=(\loc_{0}, \clockval_{0}')\longuefleche{\edge_{0}}\cdots\longuefleche{\edge_{i-1}}(\loc_{i}, \clockval_{i}')$ in $\pval'(\A)$
	    with for all~$0\leq j\leq i$, there is~$(\E_j, \D_j)\in\PDBMRp$ \st{} $\clockval_j\in(\E_j,\valuate{D_j}{\pval})$ and $\clockval_{j}'\in(\E_j,\valuate{D_j}{\pval'})$.

}
  \begin{thm}%
    \label{lemmaRunCompleteRational}
    \lemmaRunCompleteLong{}
    \end{thm}
    \newcommand{\prooflemmaRunCompleteRational}{%
    \label{proof:lemmaRunCompleteRational}

    	  Let $\pval\in\paramR$ and $\rho$ a run of $\pval(\A)$ reaching~$(\loc_i, \clockval_i)$.
    	  From \cref{completguardRational}, there is a run $\sigma$ in $\RA$
    		\st{} each clock valuation at a location~$j$ in~$\rho$ is in the \mPDBM{} $(\E_j, \D_j)$ at the same location in~$\sigma$.
    	  Still from \cref{completguardRational}, for all~$\pval'\in\paramR$ there is a run~$\rho'$ in~$\pval'(\A)$ reaching~$(\loc_i, \clockval_i')$
    	  \st{} each clock valuation at a location~$j$ in~$\rho'$ is in the \mPDBM{} $(\E_j, \D_j)$ at the same location in~$\sigma$ (note that possibly~$\pval=\pval'$).
    		Therefore, we have
		for all~$0\leq j\leq i$, there is~$(\E_j, \D_j)\in\PDBMRp$ \st{} $\clockval_j\in(\E_j,\valuate{D_j}{\pval})$ and $\clockval_{j}'\in(\E_j,\valuate{D_j}{\pval'})$
		and the expected result.
        }
      \begin{preuve}
        \versionProofIn{
        \prooflemmaRunCompleteRational{}
        }
        \versionProofOut{
        \prooflemmaRunCompleteRational{}
        }
    \end{preuve}
 Note that there is a finite number of \mPDBMs{} for each parameter region~$\paramR$.
 Let $(\E,\D)\in\PDBMRp$ and consider $\PLT$: $\D$ is an~${(\ClockCard+1)}^2$ matrix made of pairs~$(d, \compOpLeq)$
 where~$d\in\PLT$ and~${\compOpLeq\in}\{\leq, <\}$.
 Therefore the number of possible $\D$ is bounded by \[{\left(2\times\left(2+3\times\binom{M}{2}+4\times M\right)\right)}^{{(\ClockCard+1)}^2}.\]
 Moreover the number of $\E$ is unbounded, but only a finite subset of all values needs to be explored, \ie{} those smaller than $\CONSTMAX+1$: indeed, following classical works on timed automata~\cite{AD94,BDFP04}, (integer) values exceeding the largest constant used in the guards or the parameter bounds are equivalent.

\ea{petit commentaire qui explique quand meme comment on prouve le th\'eor\`eme}

To test EF-emptiness given an \RPGRtoPPTA{} $\A$ and a goal location~$\loc$, we first enumerate all parameter regions (which are a finite number),
and apply for each~$\paramR$ the following process:
we pick~$\pval\in\paramR$ (\eg{} using a linear programming algorithm~\cite{Kar84}).
Then, we consider~$\pval(\A)$ which is an updatable timed automaton and test the reachability of~$\loc$ in~$\pval(\A)$~\cite{BDFP04}.
Then EF-emptiness is false if and only if there is~$\pval$ and a run in~$\pval(\A)$ reaching~$\loc$.

\newcommand{\corollaireEFLong}{
 The EF-emptiness problem is PSPACE-complete for bounded~\RPGRtoPPTAs{}.
}

\begin{thm}\label{corollaryEFemptiness}
 \corollaireEFLong{}
\end{thm}
\newcommand{\proofcorollaryEFemptiness}{%
\label{proof:corollaryEFemptiness}

Since a TA is a special case of~\RPGRtoPPTA{} we have the PSPACE-hardness~\cite{AD94}.
Now, let $G$ be a set of goal locations of $\A$.
We build a non-deterministic Turing machine that:
	\begin{ienumeration}
	\item takes $\A$, $G$ and $\CONSTMAX$ as input;
	\item non-deterministically ``guesses'' a parameter region~$\paramR$;
	\item takes~$\pval\in\paramR$ and writes it to the tape;
	\item overwrites on the tape each parameter $\param$ by $\valuate{\param}{\pval}$
	giving the updatable TA $\valuate{\A}{\pval}$;
	\item solves reachability in $\valuate{\A}{\pval}$ for $G$;
	\item accepts iff the result of the previous step is ``yes''.
	\end{ienumeration}
	The machine accepts iff there is an integer valuation $\pval$ bounded by $\CONSTMAX$
	and a run in $\valuate{\A}{\pval}$ reaching a location $\loc\in G$.

	The size of the input is $|\A|+|G|+|\CONSTMAX|$, using $|.|$ to denote the size in bits
	of the different objects.
	Moreover, the number of parameter regions is bounded ($\ParamCard$ is the number of parameters in~$\A$)
	by~$\big(\ParamCard!\times 2^{\ParamCard}\times\prod_{\param\in\Param}(2M+2)\big)\times\big(2\times{(2+\ParamCard(3\frac{\ParamCard-1}{2}+4))}^3\big)$
	since they are constructed as the clock regions of~\cite{AD94}, the second part being the maximal number of constraints in a parameter region.
	Picking~$\pval$ at step (3) uses a PSPACE linear programming algorithm (\eg{}~\cite{Kar84}).
	Storing the valuation at step (4) uses at most $\ParamCard\times|\CONSTMAX|$
	additional bits, which is polynomial \wrt{} the size of the input.
	Step (5) also needs polynomial space from~\cite{BDFP04}.
	So globally this non-deterministic machine runs in polynomial space.
	Finally, by Savitch's theorem we have $\text{PSPACE}=\text{NPSPACE}$~\cite{Sav70}, and the expected result.
}
\begin{preuve}
\versionProofIn{
\proofcorollaryEFemptiness{}
}
\versionProofOut{
\proofcorollaryEFemptiness{}
}
\end{preuve}

Given a goal location~$\loc$ and a bounded \RPGRtoPPTA{} \A{}, we can exactly synthesize the parameter valuations~$\pval$ \st{} there is a run in~$\valuate{\A}{\pval}$ reaching~$\loc$ by enumerating each parameter region (of which there is a finite number) and test if~$\loc$ is reachable for one of its parameter valuations.
The result of the synthesis is the union of the parameter regions for which one valuation (and, from our results, all valuations in that region) indeed reaches the goal location in the instantiated TA\@.

\newcommand{\corollaireSyntheseLong}{
 Given a bounded~\RPGRtoPPTA{} \A{} and a goal location~$\loc$ we can effectively compute the set of parameter valuations~$\pval$ \st{} there is a run in~$\pval(\A)$ reaching~$\loc$.
}

\begin{cor}\label{corollaryEFsynthesis}
 \corollaireSyntheseLong{}
\end{cor}
\newcommand{\proofcorollaryEFsynthesis}{%
\label{proof:corollaryEFsynthesis}
The procedure to obtain synthesis is as follows.
We assume an \RPGRtoPPTA{} $\A$ and a goal location~$\loc$.
\begin{enumerate}
	\item enumerate all parameter regions (of which there is a finite number);
	\item for each~$\paramR$, pick a parameter valuation we pick~$\pval\in\paramR$ (\eg{} using a linear programming algorithm~\cite{Kar84});
	\item test the reachability of~$\loc$ in the updatable timed automaton~$\pval(\A)$, which is decidable~\cite{BDFP04};
	\item if $\loc$ is reachable in $\pval(\A)$, add $\paramR$ to the list of synthesized regions.
\end{enumerate}
We finally return the union of all regions $\paramR$ that reach~$\loc$.

The correctness immediately comes from \cref{lemmaRunCompleteRational,corollaryEFemptiness}.
\dl{c'est normal que le and soit en rouge dans la citation? À d'autres endroits aussi d'aillers...} 
\mr{ah oui j'ai trouvé pourquoi, il y avait une redéfinition de cref}%
}
\begin{preuve}
\versionProofIn{
\proofcorollaryEFsynthesis{}
}
\versionProofOut{
\proofcorollaryEFsynthesis{}
}
\end{preuve}

\begin{rem}
 By bounding parameter valuations in guards but not those used in updates, we still have a finite number of parameter regions.
 Indeed, an integer vector~$\E$ with components~$\E_\clock$ greater than~$\floor{\CONSTMAX}+1$ is equivalent to an
 integer vector~$\E'$ with~$\E_{\clock}'=\E_\clock$ if~$\E_\clock<\floor{\CONSTMAX}+1$ and~$\E_{\clock}'=\floor{\CONSTMAX}+1$
 if~$\E_\clock\geq\floor{\CONSTMAX}+1$.
 Moreover for all~$\param$, we have to replace each parameter valuation~$\pval$
 used in an update by~$\valuate{\param}{\pval}=\valuate{\param}{\pval'}$ if~$\valuate{\param}{\pval}\leq\CONSTMAX$
 and~$\valuate{\param}{\pval'}=\CONSTMAX+1$ if~$\valuate{\param}{\pval}>\CONSTMAX$.
 \end{rem}

\section{Parametric updates and stopwatches}\label{section:stopwatch}

In this section, we consider clocks in \RPGRtoPPTAs{} as \emph{stopwatches}~\cite{CL00}: stopwatches can be stopped and started again on transitions. In the general case, stopwatches bring back undecidability in timed automata~\cite{BBR06}. Similarly to the ``initialization'' constraint of~\cite{HKPV98} we allow stopwatches to be stopped and started again only on specific transitions.
We define~$\A$ an~\RPGRtoPPTA{} with stopwatches instead of clocks. We will prove by using our \mPDBMs{} structure that the \styleTCTL{EF}-emptiness problem is decidable under the condition that stopwatches can be stopped at each full update function, and started again at the next full update function.
Such a condition, as in \cref{def:RtoPPTA} is critical: allowing a partial (or empty) update of clocks ruins the efforts made to keep the set of \mPDBMs{} stable and allows accumulation of parameters, leading to the undecidability of the \styleTCTL{EF}-emptiness problem.

\mr{We separate stopwatches of~$\Clock$ into two sets: the set of stopwatches that continue and the set of stopwatches~$\Clock_s$ that are stopped. Note that stopped stopwatches of~$\Clock_s$ are removed from~$LFP$.}

\begin{defi}\label{def:PTAS}
	A stopwatch reset update-to-parameter PTA (\SPGRtoPPTAs{})
	$\A$ is a tuple $\A = (\Actions, \Loc, \locinit, \Clock, \Param, \Edges, \stpw)$, where:
  \begin{ienumeration}
		\item $\Actions$ is a finite set of actions,
		\item $\Loc$ is a finite set of locations,
		\item $\locinit \in \Loc$ is the initial location,
		\item $\Clock$ is a finite set of stopwatches,
		\item $\Param$ is a finite set of parameters,
		\item $\Edges$ is a finite set of edges $\edge = \langle\loc,\guard,\action,\resetfun,\loc'\rangle$
		where
		$\loc,\loc'\in \Loc$ are the source and target locations, $\guard$ is a parametric guard, $\action \in \Actions$ and
		$\resetfun : \Clock \rightharpoonup \grandn \cup \Param$ is a parametric update function.
    \item $\stpw : \Loc\to 2^\Clock$ assigns to each location a set of stopwatches that are stopped at this location.
  \end{ienumeration}
  Moreover,~$\resetfun$ is a total function whenever:
  \begin{ienumeration}
 		\item $\guard$ is a parametric guard,
 		\item $\resetfun(\clock) \in \Param$ for some~$\clock \in \Clock$, or
    \item $\stpw(\loc)\neq\stpw(\loc')$ for $\edge = \langle\loc,\guard,\action,\resetfun,\loc'\rangle$.
  \end{ienumeration}
\end{defi}

\noindent
The semantics is defined in a straightforward manner.
The update and time elapsing operators are defined in a similar manner as for \RPGRtoPPTA{}.

\newcommand{\corollaireEFStop}{
 The EF-emptiness problem is PSPACE-complete for bounded~\SPGRtoPPTAs{}.
}

\begin{thm}\label{corollaryEFemptinessS}
 \corollaireEFStop{}
\end{thm}
\begin{preuve}%
\label{proof:corollaryEFemptinessS}
To prove this result, we first remove stopped stopwatches from~\mPDBMs{}, and show that we can still reason as in~\cref{corollaryEFemptiness}.

Note that after a full update, stopped stopwatches satisfy constraints defined by a~$\CPr$.
These constraints, defined in~\crefDef{definition:pointPDBM}, are
\begin{enumerate}
  \item[(\ref{PDBM2i})]For all $i$, $(-1,<)\leq\Doi\leq (0,\leq)$ and $(0,\leq)\leq\Dio\leq (1,<)$ are valid for $\paramR$,
  \item[(\ref{PDBM2ii})] For all $i,j,k$, $\Dij\leq\Dik+\Dkj$ is valid for $\paramR$ (canonical form).
\end{enumerate}
Let~$\clock$ be a stopped stopwatch, we remove the column and the row corresponding to~$\clock$ in the~\pointPDBM{}, and we put these constraints aside.

(\ref{PDBM2i}) is still satisfied for stopwatches different from~$\clock$ as nothing changed.
As both the column and the row are removed for~$\clock$, constraints of the form~$\Dxj\leq\Dxk+\Dkj$ and~$\Dix\leq\Dik+\Dkx$ are removed as well. With the same argument, constraints of the form~$\Dij\leq\Dix+\Dxj$ are also removed. Only constraints of the form~$\Dij\leq\Dik+\Dkj$ where~$i,j,k$ are different from~$\clock$ remain, and therefore (\ref{PDBM2ii}) is still satisfied.
We apply the same reasoning for all stopped stopwatches.

After this modification, we still have a~\pointPDBM{}.
Lemmas \cref{lemmaStableReset,ssiReset,lemmaTEstable} and \cref{ssiTimeElapsing} are still applicable in this context.

Remains the comparison with non parametric guards~\cref{ssiNPguard} and parametric guards~\cref{ssiPguard}.

As stated, constraints on stopped stopwatches are put aside, call these constraints~$\C_\stpw$; as these constraints are the result of a full update it means there is one clock valuation that satisfies these constraints (it is not an interval, or a set of clock valuations). This argument is crucial in the following: we do not need to be able to bound difference of a stopped stopwatch and a running stopwatch (\ie{} $\Dxy$ in the \mPDBM{} for some stopped~$\clock$ and running~$\clockk$, or conversely) to compare separately the \mPDBM{} with a guard, parametric or not.

First, we compare the \mPDBM{} with a possibly parametric guard. Let~$\paramR$ a parameter region and~$\pval\in\paramR$.
Separately we compare the guard~$\guard$ with the constraints on stopped stopwatch~$\C_\stpw$:
\begin{itemize}
  \item If~$\guard$ is a non parametric guard, let~$\clock$ be a stopped stopwatch appearing in~$\guard$.
  \begin{itemize}
    \item If~$\clock$ has been updated to a constant, we can test whether~$\guard$ is satisfied by testing the conjunction of the constraints~$\pval(\C_\stpw)$ projected on~$\clock$ and~$\guard$ projected on~$\clock$.
    \item If~$\clock$ has been updated to a parameter, we can test the same conjunction as the order between parameters is defined in~$\paramR$.
  \end{itemize}
  \item If~$\guard$ is a parametric guard, let~$\clock$ be a stopped stopwatch appearing in~$\guard$.
  In both cases ($\clock$ updated to a parameter or a constant) we can test whether~$\pval(\guard)$ is satisfied by testing the conjunction of the constraints~$\pval(\C_\stpw)$ projected on~$\clock$ and~$\pval(\guard)$ projected on~$\clock$ as the order between parameters and constants is defined in~$\paramR$.
\end{itemize}
Therefore, \cref{ssiNPguard} and parametric guards~\cref{ssiPguard} are still applicable in this context.

With our modified structure of~\mPDBM{} with stopwatches, the core operators for clock updates, time elapsing and comparison with guards are still applicable. We obtain our result by a similar reasoning as \cref{corollaryEFemptiness}.
\end{preuve}

Similarly to \cref{corollaryEFsynthesis}, we obtain the following result:
\newcommand{\corollaireSyntheseLongS}{
 Given a bounded~\SRPGRtoPPTA{} \A{} and a goal location~$\loc$ we can effectively compute the set of parameter valuations~$\pval$ \st{} there is a run in~$\pval(\A)$ reaching~$\loc$.
}

\begin{cor}\label{corollaryEFsynthesisS}
 \corollaireSyntheseLongS{}
\end{cor}

 \section{Case study}\label{section:casestudy}

 We implemented \EFsynth{} for \RPGRtoPPTAs{} in \imitator{}, a parametric model checker for (extensions of) PTAs~\cite{AFKS12}.

 Our class is the first for which synthesis is possible over bounded rational parameters.
 We believe our formalism is useful to model several categories of case studies, notably distributed systems with a periodic (global) behavior for which the period is unknown: this can be encoded using a parametric guard while resetting all clocks---possibly to other parameters.

  Consider the \RPGRtoPPTA{} in \cref{figure:peer} with six locations, three clocks compared to parameters ($\styleclock{x}$, $\styleclock{y}$, $\styleclock{t}$), one constant ($max$) and six parameters ($\styleparam{\param}$, $\styleparam{\param_1}$, $\styleparam{\param_2}$, $\styleparam{v}$, $\styleparam{\param v_1}$, $\styleparam{\param v_2}$). 

   We consider the case of a network of peers exchanging transactions grouped by blocks, \eg{} a blockchain, using the Proof-of-Work %
	as a mean to validate new blocks to add.
 	In this simplified example, we consider a set of two peers (represented by~$\styleclock{x}$, $\styleclock{y}$) which have different computation power (represented by~$\styleparam{\param_1}$, $\styleparam{\param_2}$).
 	Peers write new transactions on the current block ($\styleact{newTx}$). If it is full ($\styleclock{t}=\styleparam{\param}$), both peers try to add a new block ($\styleact{newBlock}$) to write the transaction on it. We update~$\styleclock{x}$ to~$\styleparam{\param_1}$, $\styleclock{y}$ to~$\styleparam{\param_2}$, and~$\styleclock{t}$ to~$0$ as the peers have a different computation power, and they start ``mining'' the block (find a solution to a computation problem).
 	Either~$\styleclock{x}$ or~$\styleclock{y}$ will eventually offer a solution to the problem ($\styleact{blockSolution_x}$ if~$\styleclock{x} = max$ or $\styleact{blockSolution_y}$ if~$\styleclock{y} = max$). If~$\styleclock{y}$ offers a solution, $\styleclock{x}$ will check whether the solution is correct: $\styleclock{x}$ is updated to $\styleparam{\param v_1}$ to represent its speed to verify an offer. 
 	$\styleclock{x}$ can refuse the offer if the verification is too long ($\styleact{fakeBlock}$ if~$\styleclock{x}>\styleparam{v}$) therefore the mining step restarts. $\styleclock{x}$ can approve the offer ($\styleact{okBlock}$ if~$\styleclock{x}\leq\styleparam{v}$), $\styleclock{y}$ is rewarded and the block is added to the blockchain ($\styleact{addBlock}$).

 	 We are interested in a malicious peer~$\styleclock{x}$ that wants to avoid~$\styleclock{y}$ being rewarded for every new block. Therefore~$\styleclock{x}$ asks: \emph{``what are the possible computation power configurations and verification speed so that $\styleclock{y}$ can be rewarded''} ($EF(\styleloc{reward_y})$-synthesis), considered as a bug state in the automaton.

 	 We run this \RPGRtoPPTA{} using \imitator{}~\cite{AFKS12}\footnote{Experiments were conducted with \imitator{} 2.10.4 ``Butter Jellyfish'' on a 2.4\,GHz Intel Core i5 processor with 2 GiB memory. Computation time is less than 1~second.
   Sources, binaries, models and results are available at \url{imitator.fr/static/FORTE19/}}.
 We set~$max=30$ units of time and also the upper bound of~$\styleparam{\param}$ and~$1\geq\styleparam{v}> 0$ unit of time. \imitator{} computes a disjunction of constraints so that~$\styleloc{reward_y}$ is unreachable: we keep two relevant ones; 
 \begin{ienumeration}
 \item $\styleparam{\param_1}\geq\styleparam{\param_2}$:
 $\styleclock{x}$ has strictly more computation power than~$\styleclock{y}$ in which case~$\styleclock{x}$ always offers a block solution, or has the same computation power than~$\styleclock{y}$ in which case the systems blocks. $\styleclock{x}$ should invest heavily into hardware to keep its computation power high;
 \item $\styleparam{\param v_1} > \styleparam{v}$:
 the malicious peer~$\styleclock{x}$ is always faster to verify the solution offered by~$\styleclock{y}$ and refuses it. The blockchain is probably compromised.
 \end{ienumeration}

\noindent
 Using a parameter valuation respecting one of the previous constraints guarantees that~$\styleclock{y}$ is never rewarded.

\bigskip

\begin{rem}

 Even if we have to update all clocks whenever a parameter is met in a guard or in an update, the possibility to update clocks to unknown parameter offers an appreciable freedom in the range of system that can be modeled with \RPGRtoPPTA{}. Especially as our parameters can take unbounded rational values in updates and bounded rational values in guards.

 However, such a restriction restricts the behavior that can be modeled. Consider the PTA of \cref{figure:PTAAHV} for which the set of possible parameter valuations \st{}~$\styleloc{\loc_3}$ is reachable is
$\{\pval \mid \pval(\styleparam{\param_3}) = n\pval(\styleparam{\param_1}) + \pval(\styleparam{\param_2})$ for some $n \in \grandn \}$.
This set cannot be computed from a \RPGRtoPPTA{} as in the loop transition over~$\styleloc{\loc_2}$ both clocks~$\styleclock{\clock}$ and~$\styleclock{\clockk}$ must be updated.

 \begin{figure*}[tb!]
 \scalebox{0.85}{
 {\centering
 \small
 \hspace{7em}\begin{tikzpicture}[shorten >=1pt, node distance=4cm, on grid, auto]
 \node[location, initial]       (A)   {$\styleloc{\loc_1}$};
 \node[location, inner sep=0pt,inner sep=1pt]       (B)  [right=of A]     {$\styleloc{\loc_2}$};
 \node[location, inner sep=0pt,inner sep=1pt, final]       (C)  [right =of B]     {$\styleloc{\loc_3}$};

 \path[->]
 	(B)  edge[loop] node  [swap,align=center, above]   {$\styleclock{\clock}=\styleparam{\param_1}$\\$\styleclock{\clock}:=0$}   (B)

 	(A)  edge node  [swap,align=center, above]   {$\styleclock{\clock}:=0$\\$\styleclock{\clockk}:=0$}  (B)

 	(B)  edge node  [swap,align=center, above]   {$\styleclock{\clock} = \styleparam{\param_2}\wedge\styleclock{\clockk} = \styleparam{\param_3}$}   (C);
 \end{tikzpicture}
 }
 }
 \caption{PTA of~\cite[Fig.\ 1]{AHV93}.}%
 \label{figure:PTAAHV}
 \end{figure*}
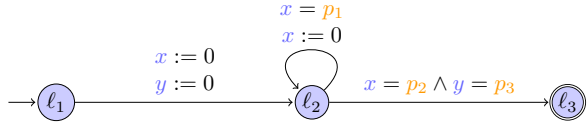

\end{rem}

\section{Conclusion and perspectives}\label{section:conclusionRtoPPTA}

Our class of \RPGRtoPPTAs{} is one of the few subclasses of PTAs (actually even extended with parametric updates) to enjoy decidability of EF-emptiness.
In addition, \RPGRtoPPTAs{} are the first ``subclass'' of PTAs to allow exact synthesis of bounded \emph{rational}-valued parameters.

In terms of future work, beyond reachability emptiness, we aim at studying unavoidability-emptiness and language preservation emptiness (``given a reference parameter valuation, does there exist another parameter valuation with the same untimed language''~\cite{ALM20}), as well as their synthesis.

We would also study the possibility to stop and start stopwatches without full updates; it would change the \mPDBM{} structure by creating new clock constraints, but seems promising and the reachability emptiness problem might be decidable as well.

Time bounded reachability is also interesting to study:
given an \RPGRtoPPTA{}~$\A$, a parameter~$\ptotal$ and a goal location~$\loc$, is there a parameter valuation~$\pval$ and a run in~$\pval(\A)$ reaching~$\loc$ in less than~$\pval(\ptotal)$ time?

Finally, we would like to investigate whether our parametric updates can be applied to decidable hybrid extensions of TAs~\cite{HKPV98,BDGORW13}. For example, we shall find a subclass of hybrid automata that can be reduced to \RPGRtoPPTA{} as done with initialized rectangular hybrid automata and TAs in~\cite{HKPV98}.

  \newcommand{\ENTCS}{Electronic Notes in Theoretical Computer Science}
  \newcommand{\FMSD}{Formal Methods in System Design}
  \newcommand{\IEEETSE}{IEEE Transactions on Software Engineering}
  \newcommand{\IJFCS}{International Journal of Foundations of Computer Science}
  \newcommand{\IJSSE}{International Journal of Secure Software Engineering}
  \newcommand{\JLAP}{Journal of Logic and Algebraic Programming}
  \newcommand{\JLC}{Journal of Logic and Computation}
  \newcommand{\LMCS}{Logical Methods in Computer Science}
  \newcommand{\LNCS}{Lecture Notes in Computer Science}
  \newcommand{\RESS}{Reliability Engineering \& System Safety}
  \newcommand{\STTT}{International Journal on Software Tools for Technology Transfer}
  \newcommand{\TCS}{Theoretical Computer Science}
  \newcommand{\ToPNoC}{Transactions on Petri Nets and Other Models of Concurrency}
  \newcommand{\TSE}{IEEE Transactions on Software Engineering}
  \bibliographystyle{alpha} %
\bibliography{resetPTAarxiv}

\versionProofOut{
\newpage
\appendix

\section{Proof of \crefLemma{lemma:sumvalid}}\label{appendix:proof:lemmasumvalid}
\recallResult{lemma:sumvalid}{\lemmasumvalid}
\begin{preuve}
\prooflemmasumvalid{}
\end{preuve}

\section{Proof of \crefLemma{lemmaDijDjiGeq}}\label{appendix:proof:lemmaDijDjiGeq}
\recallResult{lemmaDijDjiGeq}{\lemmaDijDji}
\begin{preuve}
\prooflemmaDijDjiGeq{}
\end{preuve}

\section{Proof of \crefLemma{lemmaDiiDijGeq}}\label{appendix:proof:lemmaDiiDijGeq}
\recallResult{lemmaDiiDijGeq}{\lemmaDiiDij}
\begin{preuve}
\prooflemmaDiiDijGeq{}
\end{preuve}

\section{Proof of \crefLemma{lemmaStableReset}}\label{appendix:proof:lemmaStableReset}
\recallResult{lemmaStableReset}{\lemmaStableReset}
\begin{preuve}
\prooflemmaStableReset{}
\end{preuve}

\section{Proof of \crefLemma{ssiReset}}\label{appendix:proofssiReset}
\recallResult{ssiReset}{\lemmaResetLong}
\begin{preuve}
\proofssiReset{}
\end{preuve}

\section{Proof of \crefLemma{lemmaLargestFracClock}}\label{appendix:proof:lemmaLargestFracClock}
\recallResult{lemmaLargestFracClock}{\lemmaLargestClock}
\begin{preuve}
\prooflemmaLargestFracClock{}
\end{preuve}

\section{Proof of \crefLemma{lemmaTEstable}}\label{appendix:proofTEstable}
\recallResult{lemmaTEstable}{\lemmaTEstable}
\begin{preuve}
\prooflemmaTEstable{}
\end{preuve}

\section{Proof of \crefProp{ssiTimeElapsing}}\label{proofssiTimeElapsing}
\recallResult{ssiTimeElapsing}{\propTimeElapsingLong}
\begin{preuve}
\proofpropTimeElapsingLong{}
\end{preuve}

\section{Proof of \crefLemma{ssiNPguard}}\label{appendix:proof:ssiNPguard}
\recallResult{ssiNPguard}{\ssiNPguard}
\begin{preuve}
\proofssiNPguard{}
\end{preuve}

\section{Proof of \crefLemma{ssiPguard}}\label{appendix:proof:ssiPguard}
\recallResult{ssiPguard}{\ssiPguard}
\begin{preuve}
\proofssiPguard{}
\end{preuve}

\section{Proof of \crefProp{completguardRational}}\label{appendix:proofcompletguardRational}
\recallResult{completguardRational}{\theoremCompletGuardRationalLong}
\begin{preuve}
\proofcompletguardRational{}
\end{preuve}

}

\end{document}